# The specific heat of astro-materials:

# Review of theoretical concepts, materials and techniques

Jens Biele[(1)], Matthias Grott[(2)], Michael E. Zolensky[(3)], Artur Benisek[(4)], Edgar Dachs[(4)]

*[(1)] DLR – German Aerospace Center, RB-MUSC, 51147 Cologne, Germany, e-mail: Jens.Biele@dlr.de*

*[(2)] DLR – German Aerospace Center, Institute for Planetary Research, Berlin, Germany*

*[(3)] NASA Johnson Space Center (Houston, United States)*

*[(4)] Chemistry and Physics of Materials, University of Salzburg, Jakob-Haringer-Str. 2a, 5020 Salzburg, Austria.*

**Keywords**
specific heat; minerals; solid matter; rocks; meteorites; thermophysical properties

**Author contributions**
JB and MG had the idea for the article. AB and ED contributed their expertise in minerals, experimental measurement and theory of heat capacity; MZ contributed his vast knowledge about meteoritic minerals; AB helped with artwork and many clarifications; JB performed the literature search, data analysis and drafted the work; all authors critically revised it.

## Statements and Declarations

- **Competing Interests:** The authors have no financial or proprietary interests in any material discussed in this article.
- **Acknowledgments** Grant P33904 to E. Dachs by the Austrian Science Fund (FWF) is gratefully acknowledged. M. Zolensky acknowledges a grant by the NASA-Hayabusa 2 Participating Scientist Program. We are grateful for discussions and share of data with Robert J. Macke (Vatican observatory), Dan Britt (University of Central Florida, Orlando) and Cyril P. Opeil (Boston College). We would also like to thank the Engauge team for providing an excellent free digitizer software [1] which made data import from published graphs easy and accurate.

**Abstract**

We provide detailed background, theoretical and practical, on the specific heat of minerals and mixtures thereof, 'astro-materials', as well as background information on common minerals and other relevant solid substances found on the surfaces of solar system bodies. Furthermore, we demonstrate how to use specific heat and composition data for lunar samples and meteorites as well as a new database of endmember mineral heat capacities (the result of an extensive literature review) to construct reference models for the isobaric specific heat $c_P$ as a function of temperature for common solar system materials. Using a (generally linear) mixing model for the specific heat of minerals allows extrapolation of the available data to very low and very high temperatures, such that models cover the temperature range between 10 and 1000 K at least (and pressures from zero up to several kbars). We describe a procedure to estimate $c_P(T)$ for virtually any solid solar system material with a known mineral composition, e.g., model specific heat as a function of temperature for a number of typical meteorite classes with known mineralogical compositions. We present, as examples, the $c_P(T)$ curves of a number of well-described laboratory regolith analogues, as well as for planetary ices and 'tholins' in the outer solar system. Part II will review and present the heat capacity database for minerals and compounds and part III is going to cover applications, standard reference compositions, $c_P(T)$ curves and a comparison with new and literature experimental data.

*Nullus est liber tam malus, ut non aliqua parte prosit | Plinius minor*

## Introduction

Specific heat $c_P(T)$ is one of the parameters which determine a surface's temperature response to (solar) heating. Remote sensing in the mid-infrared is often used to estimate a parameter termed the thermal inertia of the surface material, which is defined as $\Gamma(T) = \sqrt{\rho(T)k(T)c_P(T)}$, where (in SI units) $T$ is absolute temperature in K, $k$ is thermal conductivity in W m$^{-1}$ K$^{-1}$, $\rho$ is bulk density in kg m$^{-3}$, and $c_P$ is specific heat at constant pressure in J/kg/K. Knowledge or an estimate of $c_P(T)$ is required to extract information on, e.g., thermal conductivity $k$ from the data, which in turn allows for an estimation of important surface properties like grain size [2-6] and porosity [7]. Furthermore, knowledge of thermophysical surface properties (including porosity) is essential to model the Yarkovsky [8-10] and YORP [10, 11] effects as well as the response of planetary surfaces to impact cratering [12, 13]. In comets, the surface material is a mixture of ices (water ices, CO, $CO_2$) and silicate dust, which in most of a comet's orbit is at very low temperatures – with a very different specific heat than commonly assumed for silicates near room temperature. Trans-Neptunian objects (TNOs) and icy moons likewise have a surface composition very different from, e.g., the Moon – thus we need to know specific heats of solar system ices and of the so-called 'tholins', the 'complex abiotic organic gunk' [14] on the very surface.

Piqueux et al. [15] have recently studied the effect of composition- and temperature dependent specific heat on thermal modelling of surfaces in the solar system. We agree with them that under non-cryogenic conditions, the composition is typically (excluding perhaps metal-rich worlds like M-type asteroids) not a significant factor controlling $c_P(T)$ and thermal inertia trends, and even the temperature dependence of specific heat has usually only a second-order influence on surface temperatures (although it must at least be considered in the error budget since the advent of high-resolution, high-precision thermal datasets).

However, we also agree with [15, 16] that surface temperature models could be impacted by the drastic decrease in $c_P(T)$ values toward low temperatures; thermal models generally assume lunar basalt calorimetric properties, which are not well known outside the data range 90 – 350 K. Indeed, 'knowledge of specific heat variability as a function of temperature and bulk material composition remains largely under-constrained for the need of planetary thermal modelers' [15]. In particular, the specific heat capacity of geological materials relevant to solar system body surfaces below room temperature is not particularly well constrained and the thermal modelling community only has a limited set of adequate ready-to-use $c_P(T)$ trends for planetary surface temperature modelling.

The goal of Piqueux et al. is to provide a reference for thermal models by providing experimental data on a wide range of materials - covering a wide range of compositions and temperatures relevant to planetary surfaces - from which thermal models can incorporate the most appropriate one.

Our approach is complementary: We provide the means to calculate synthetic $c_P(T)$ from a known bulk composition, and additionally a method to predict the specific heat curve beyond the temperature range measured, even if the composition is not (well) known.

- Unbeknownst maybe to most astronomers and planetary scientists, many precise heat capacity data exist for hundreds of minerals, over wide temperature ranges, yet in particular for temperatures below 25°C [17], they are scattered in the literature. Our motivation thus is also to collect, merge, critically review and tabulate these data for substances of interest, and to make this database readily available.

Around room temperature, the temperature dependence of $c_P$ is a second-order effect in the thermal inertia, and except for the mass fraction of meteoritic iron (FeNi), and to a lesser degree phyllosilicates, specific heat is not very strongly dependent on the specific material. However, at low temperatures $c_P$ shows a strong temperature and compositional dependence. Specific heat must approach 0 as temperature approaches absolute zero, and it is usually proportional to $T^3$ at very low temperatures. Specific heat furthermore shows a noticeable, about linear increase at very high temperatures, which is caused by anharmonicities of the lattice vibrations and by thermal expansion (only harmonic lattice $C_V$, heat capacity at constant volume, obeys the Dulong-Petit limit).

The range of temperatures relevant for this study is given by the minimum and maximum surface temperatures in the solar system, which span a large range from asteroids with smallest perihelia and Mercury to cold TNOs at the edge of the Edgeworth-Kuiper belt. While Mercury has maximum surface temperatures of up to 700 K and some asteroids even ~1000 K (e.g., (3200) Phaethon and (155140) 2005 UD [18]), TNOs have night time temperatures down to ~10-30 K, and even on the Moon, surface temperatures as low as 25 K have been measured in permanently shadowed craters in the vicinity of the south pole [19, 20]. Therefore, we aim for a description and parameterization of specific heat in the temperature range between 10 and 1000 K, while simultaneously allowing for a physical reasonable extrapolation to 0 K as well as to the respective melting temperatures. The latter are typically of the order of 1400 K for silicates, while the threshold temperature for sintering of silicates is close to 700 K [21].

Note that knowledge of specific heat is also necessary to calculate thermal conductivity from thermal diffusivity measurements (e.g., by the flash method [22]).

Data on the specific heat of extra-terrestrial material (apart from the Apollo lunar samples) is scarce, and only a handful of $c_P$ data of meteorites have been published (most of them measured at temperatures at or above 300 K) until quite recently; since about 2012, there has been a surge of new meteorite specific heat data [15, 23-31], . The only other extra-terrestrial material with known $c_P$ over a wide temperature range is lunar samples from the Apollo missions, and lunar $c_P(T)$ has widely, but not always wisely, been used as a representative standard in studies covering solar system bodies ranging from asteroids [e.g., 5] to planets like Mars [e.g., 32]. However, heat capacity can strongly depend on composition, thus the use of lunar data for, e.g., C- or M-class asteroids or objects containing frozen volatiles may give rise to large systematic errors. Furthermore, most available data cover only a limited temperature range, introducing further uncertainty when extrapolating to lower or higher temperatures. In the next years, however, it is expected that the first specific heat data of asteroid material will become available, e.g., from the Bennu samples acquired by the OSIRIS-REx mission [33].

$c_P(T)$ data for rocks (in general, 'astro-material', any solid material present on the surface of solar system bodies) can be calculated from the contributions of the constituent minerals (and mineraloids, i.e., amorphous substances). This is particularly important when studying the surfaces of outer solar system objects like icy moons, comets or TNOs, as the specific heat capacities of ices are dramatically different from those of silicates near room temperature. We will also demonstrate how $c_P(T)$ measurements over a limited temperature range (example: lunar regolith) can be meaningfully extrapolated.

One of the problems that has to be considered when calculating specific heat of astro-material is that the minerals are usually neither perfectly mechanically mixed nor do they show solid solutions of the same composition throughout the sample of interest. This is clear from the study of meteorites, which show compositional zoning and obvious inhomogeneities in the form of chondrules (of mostly <1 mm diameter), embedded in a fine-grained matrix, but also from brecciated meteorites like siderolites (stony iron meteorites). A linear mixing model for $c_P$ is only valid at spatial scales larger than the intrinsic spatial inhomogeneities. Note that natural polycrystalline minerals often exhibit a range of solid solution compositions (i.e., characteristic zoning patterns) at length scales of the order of the grain size, indicating changes in pressure and temperature conditions during crystallization (e.g., [34] [35] and references therein). This has implications for any composition-dependent transition peaks in the $c_P$ –curve (FeNi is possibly the most important example, but only at temperatures between ~600 and ~1000 K), and appropriate averaging is necessary if utmost peak fidelity is sought.

More practically, if in a given sample volume a mineral with composition-dependent transition peaks is present with a significant mass fraction, and that mineral has a range of compositions within that sample volume, smearing out of the sample-averaged transition peaks is expected. We speculate that this effect will obscure the magnetic transition peaks in olivine (Fo-Fa) to a slight hump between ~20 and ~60 K, and in iron-bearing pyroxenes (Di-Hed, En-Fs; augites and pigeonites) between ~10 and ~40 K, possibly also in non-stoichiometric compounds like wüstite $Fe_{1-y}O$ and pyrrhotite $Fe_{1-x}S$, but not in minerals like magnetite.

When interpreting remote sensing observations of thermal emission, it is important to note that observations are sensitive to average thermal properties. First of all, averaging takes place horizontally over the size of the instrument's footprint, which can range from cm to km scales. Furthermore, thermal properties are also averaged vertically over the thermal skin depth[1] $s$, i.e., the e-folding length of the periodic surface temperature forcing. The diurnal skin depth of solar system bodies surfaces typically varies between 2 mm and 1 m [36] such that the specific heat of the observed surfaces can usually be regarded as homogeneous. However, laboratory specific heat measurements of meteorites often involve only tens of milligrams of material, and care has to be taken to grind and mix a representative volume of the specimen and all its constituents in unbiased proportions; this can be problematic with meteorites containing ductile FeNi metal besides brittle minerals (e.g., [37]).

---

1 $$s = \sqrt{\kappa P / \pi \rho c_p} = \frac{\Gamma}{\rho c_P}\sqrt{\frac{P}{\pi}}$$

$P$ is the diurnal period, $\kappa$ the thermal conductivity (note that erroneously, sometimes a factor of 2 in the denominator of the fraction under the root appears; also, it remains to be studied what the e-folding length is for surface temperature waves that are not sinusoidal, like the simple thermal model surface temperature (thermal inertia 0, $T$=0 at night) or for step T changes at the surface)

Note that $c_P$ of a homogeneous crystalline material is independent of particle size down approximately 50 nm, whereas nanoparticles show deviations from the bulk specific heat value due to surface effects and a strong discretization of possible lattice vibration modes, e.g., [38].

We will review the available data on lunar samples and meteorites as well as the specific heat capacities of the most abundant endmember minerals including iron-nickel metal. Furthermore, organic materials found in meteorites and frozen volatiles thought to exist on outer solar system bodies are also considered. From these data, we built up a computerized database to calculate the specific heat of approximately 100 minerals and compounds for temperatures between absolute zero and close to melting (or decomposition temperature) by use of tables and correlation equations apt for convenient but accurate interpolation.

The paper is organized as follows: in section 1, we first summarize the relevant background on heat capacity, its temperature and pressure dependence as well as useful approximations, and discuss the various transitions and effects of crystallinity and particle size. The concept of endmember minerals and mechanical mixtures versus solid solutions is introduced, and polymorphs as well as phase transitions are discussed.

Section 2 gives background on minerals and compounds reviewed in this work (Table 4). We also investigate which minerals are compositionally likely to be important on the terrestrial planets (other than Earth) and the moon, since otherwise we have focused on minerals known to be important in meteorite samples. This section also presents a table with an overview of our database. We then briefly summarize textbook descriptions of the most common and important mineral groups that occur in solar system materials and which are part of the $c_P$ database. Note the newly introduced sections on carbon-rich/organic matter, on solar system ices and on tholins. For each material, if necessary, important aspects of the specific heat like the influence of composition, (adsorbed/hydrate) water content, transitions, solid solutions, isomorphs, and thermal alteration at elevated temperatures are emphasized. The detailed description of methods and used input data, for each mineral and compound covered, will be given in paper II.

Section 3 gives some examples what can be done with the methodology presented here, using the database; section 4 summarizes the paper and section 5 gives an outlook.

In the appendix (Supplementary Information) we describe methods like our accurate Padé approximant to the 3D-Debye function and details on the results shown in section 3, list all data known to us on measured meteorite heat capacities and lunar samples, and present the reference $c_P(T)$ for lunar regolith and some (mostly commercial) laboratory regolith simulants along with the mineral compositions of the latter.

# 1 Background – heat capacity of solids

Heat capacity is a bulk thermodynamic quantity; at constant pressure, we have $C_P = \left(\partial H / dT\right)_P$ and at constant volume, $C_V = \left(\partial U / \partial T\right)_V$ where $H$ is enthalpy, $U$ internal energy. While strictly an extensive property, it is always made intensive. Molar heat capacity, $C$, is conventionally just called 'heat capacity' of a compound while the 'specific heat', $c$, refers to unit mass. In calorimetry, the temperature range 0-340±40 K is traditionally called 'low temperature' and the range 340±40 //K to melting (or decomposition) temperature 'high temperature'. Also traditionally, and somewhat arbitrarily, temperatures below 90±10 K are called 'cryogenic'. Experimentally, $c_P$ is measured, and $C_P$ can only be given for substances of known chemical composition: $C_P = c_P M$ where $M$ is the molar mass. In this paper, we use $C$, $c$ where necessary (e.g., in equations 1, 2). Wherever it does not matter, we use 'heat capacity', $C_P$ and 'specific heat', $c_P$, interchangeably.

$c_V$ is very difficult to measure directly, but can be calculated from $c_P$ (see below). The heat capacity of solids depends mainly on temperature, especially at low temperatures; the pressure dependence is negligible, so data measured at 1 bar can be used in a wide pressure range, from 0 to several kilobars. Yet in many substances, we see signals in the $C_P(T)$ curve from magnetic and substitutional order/disorder transitions leading to transition peaks (some obvious examples are shown in Figure 1), but sometimes (especially at very low temperatures) only to minor 'bumps' and 'shoulders. $C_P$ also depends to a lesser extent on vacancy defects, dislocations and effects of crystallinity. The effect of particle size is normally negligible (see below).

The seminal work of Cezairliyan et al. [39] is still a very good reference on the theory of specific heat of solids and their measurement (calorimetry).

$C_P(T)$ is important for thermodynamics and mineralogy/petrology; thus, there is abundant and precise data in the literature for endmember minerals. However, these data are scattered in the literature and often reflect different temperature ranges, methods and accuracies. There exist excellent collections of $C_P$ and other thermodynamic mineral data (some of them internally consistent) [40-48], but these collections typically only give polynomial $C_P(T)$ interpolation equations for high temperatures ≥298.15 K and include either none or rather crude descriptions of transition peaks; they do, however, give citations of the original (i.e., including the low temperature[2]) data. Therefore, we have undertaken to revise, combine, smooth and electronically tabulate $C_P(T)$ data for the most important endmember minerals, for a temperature range as wide as possible.

What about the *ab initio-based* prediction of thermodynamic properties like specific heat? This is indeed possible, with state-of-the-art theoretical techniques like density-functional theory (DFT), density-functional perturbation theory (DFPT) in quasi-harmonic approximation (QHA), combining, for magnetic contributions, with methods like the spin quantum Monte Carlo approach (QMC) for solving the quantum Heisenberg model (suitably mapped), e.g., [49-51]. A number of compounds (elements, oxides, simple minerals) have been calculated with satisfying accuracy (that is, systematic deviations to experimental data less than a few %). Benisek & Dachs [52] provide information about the uncertainties in DFT-calculated $C_P$´s on a number of well-known minerals; the uncertainties range from less than one % to a few %. For other minerals, see [53]; complex minerals are no problem in principle, just the computing time gets impractical if *Z*, the number of atoms in a unit cell, is larger than about 100. Note that there is an issue to transform $C_V$ into $C_P$: the quasi-harmonic approximation can calculate $C_P$ (not the anharmonic contributions though!) but it is a very time-consuming task. Also, we have little experience concerning the accuracy of magnetic (spin) contributions using QMC and it is unclear how to calculate $C_P$ contributions from other phase transitions. However, for minerals for which no $C_P$-data exist, DFT-calculations are really helpful and far better than estimation methods (see section 1.2.1).

[2] Indispensable to obtain 0$^{th}$-law entropy

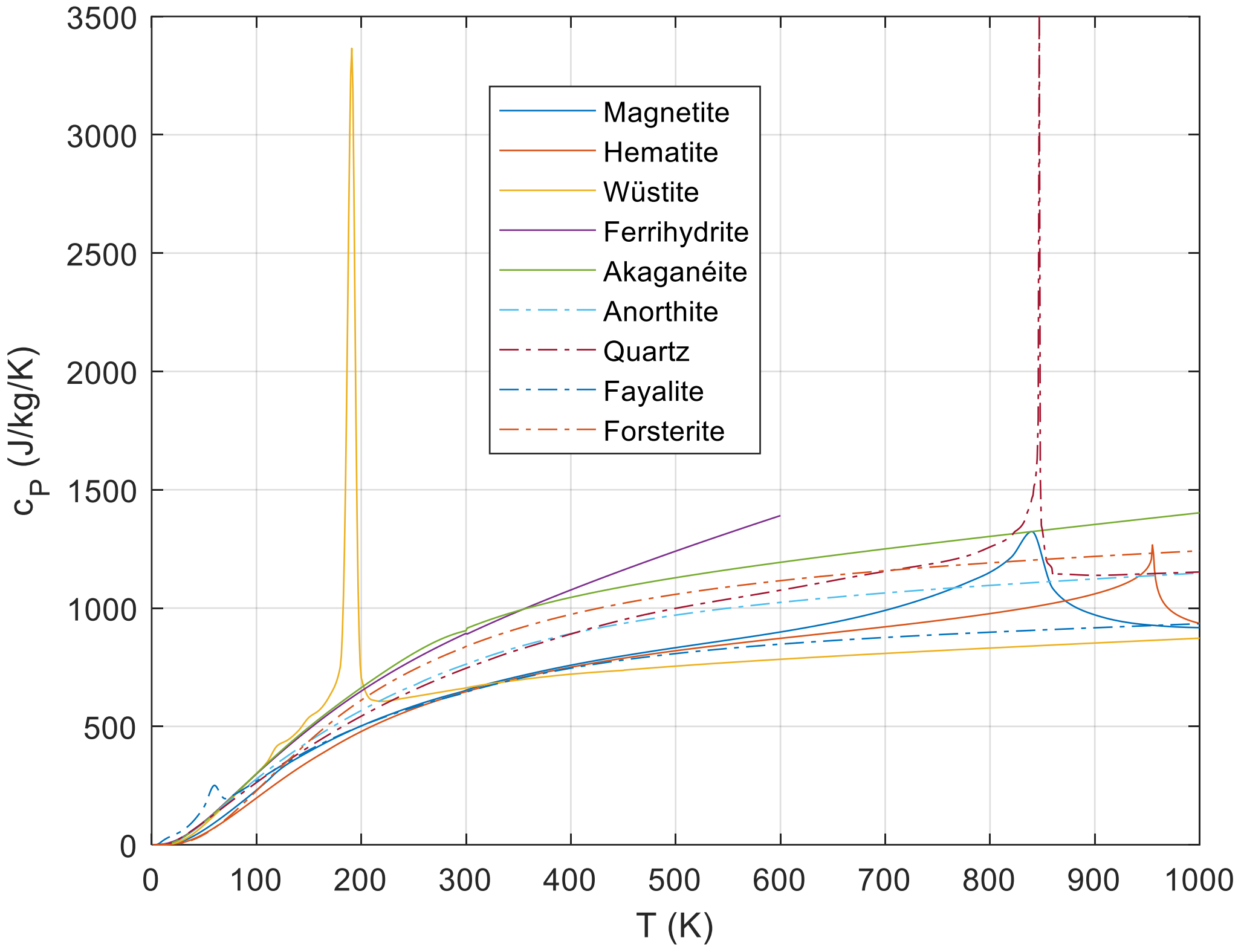

**Figure 1 Example $c_P$-curves, (magnetic) transition peaks in some iron oxides, quartz with the λ transition (α-β) at 843 K, fayalite with its low temperature magnetic transition, forsterite and anorthite with no anomalies. Note that magnetite has a small broad Verwey peak at ~124 K which here shows only as a 'bump'. Akaganéite here is β-FeOOH·0.65$H_2O$ and the ferrihydrite is 2-line.**

## 1.1 Theory

Heat capacity can be written as the sum of terms: lattice vibrational, Schottky, electronic and magnetic (ferromagnetic and ferromagnetic) contributions, order/disorder, activation (vacancy) and anharmonic contributions [54]. 'Lattice heat capacity' is the conventional, but rather colloquial term for the phonon heat capacity (the lattice as a mathematical construct has no heat capacity of its own).

In general, the heat capacity due to lattice vibrations (phonons) can be written

$$C_{V,lattice}(T) = \frac{\partial}{\partial T} \int_0^{\omega_{\max}} g(\omega) E\left(\frac{\hbar\omega}{k_B T}\right) d\omega \qquad (1)$$

with g(ω) the DOS (distribution of vibrational states function), ω the circular frequency, $k_B$ the Boltzmann constant, $\hbar$ is Planck's constant, and $E$ the Einstein oscillator function (oscillator energy times Bose-Einstein distribution with degeneracies 1)

$$E = \frac{\hbar\omega}{\exp\left(\frac{\hbar\omega}{k_B T}\right) - 1}$$

Due to the integral, heat capacity is not very sensitive to details in the DOS; only at very low temperatures the DOS becomes decisive, while at very high temperatures it has no influence at all.

<u>Lattice heat capacity.</u> The theory of Debye [55] is a reasonable approximation for simple, monoatomic, isotropic crystals (Pb, lead is a famous example). For polyatomic solids, it is only applicable if the following conditions hold (which they usually do not) [56]:

1. The various atoms have nearly equal masses
2. The coordination environments of the different atoms are nearly identical
3. The environments are essentially isotropic
4. The various near-neighbour interatomic force constants are nearly equal.

Still, one can use the simple and elegant Debye theory and calculate an effective ('calorimetric') Debye temperature $\theta_D$ that depends on temperature but much less than $C_P$ itself (we discuss this in detail below).

$$C_V = 3nR\,D(\theta_D/T) \quad [J/mol/K]$$
$$D(\theta_D/T) = 3\left(\frac{T}{\theta_D}\right)^3 \int_0^{\theta_D/T} \frac{x^4 e^x}{(e^x-1)^2}\,dx = 3\left(\frac{T}{\theta_D}\right)^3 \int_0^{\theta_D/T} \frac{x^4 e^{-x}}{(e^{-x}-1)^2}\,dx$$
$$\theta_D = \frac{h}{2\pi k_B}\left(\frac{6\pi^2 N_A}{ZV}\right)^{1/3} v_m \qquad (2)$$
$$v_m \cong \sqrt{\frac{3}{\left(1/v_P^2 + 2/v_S^2\right)}}$$

Nomenclature:

| | |
|---|---|
| $n$ | Number of atoms in a formula unit |
| $V$ | Molar volume |
| $Z$ | Number of formula units in the unit cell |
| $V_L$ | Volume of the primitive unit cell |
| $N_A$ | Avogadro constant |
| $N$ | Number of atoms in 1 mol of crystal |
| $k_B$ | Boltzmann constant |
| $R$ | $R= N_A\, k_B$, molar gas constant |
| $\theta_D$ | Deybe temperature (‚effective', ‚calorimetric') |
| $h$ | Planck constant |
| $v_P$ | Acoustic longitudinal wave velocity |
| $v_S$ | Acoustic shear wave velocity |

$v_m$  Mean sound speed

$M_r$  Molecular mass of the formula unit

$\alpha$  Isobaric coefficient of thermal volume expansion

$B$  Isothermal bulk elastic modulus $=1/\beta$

For isotropic or cubic crystals,

$$v_P = \sqrt{\frac{K+4G/3}{\rho}} = \sqrt{\frac{E(1-\nu)}{\rho(1+\nu)(1-2\nu)}}$$
$$v_S = \sqrt{\frac{G}{\rho}} \qquad (3)$$

where $K$ is the (isentropic) bulk modulus, $G$ the shear modulus, $E$ the Young's modulus, $\rho$ the density and ν the Poisson's ratio. Note that for anisotropic crystals, the relationship between sound velocities (in a given direction) and elastic constants (many more than 2) is much more complicated, cf. [57].

Note the second form of the Debye integral $D(\theta_D/T)$ in (2) with $\exp(-x)$ is equivalent but numerically much more robust (avoids overflow).

The low-temperature approximation of the Debye model is of course the famous $\sim T^3$ law:

$$C_V = \frac{12\pi^4}{5} n N_A k_B \left(\frac{T}{\theta_D}\right)^3 \qquad (4),$$

and the high-temperature approximation, Dulong-Petit's law:

$$C_V = 3nR$$

Actually, the series (Taylor) expansions of the Debye function are, for $T \to 0$

$$C_V = c_3\left(\frac{T}{\theta_D}\right)^3 - c_5\left(\frac{T}{\theta_D}\right)^5 + c_7\left(\frac{T}{\theta_D}\right)^7 - c_9\left(\frac{T}{\theta_D}\right)^9 + \ldots \qquad (5)$$

And for $T \to \infty$

$$C_V = 3nR\left[1 - \frac{1}{20}\left(\frac{\theta_D}{T}\right)^2 + \frac{1}{560}\left(\frac{\theta_D}{T}\right)^4 - \ldots\right] \qquad (6)$$

Equation (6) has rather bad convergence properties; modifications have traditionally been used (see chapter 1.2.3), and recently a novel, fast-converging series representation of the Debye function for high temperatures has been proposed [58], where the reciprocal square-root of the Debye function is written as 1 + (polynomial with only even powers of $T$).

The point of inflection of the Debye-curve, $C_V$ vs. $T$, is at $T \approx \theta_D/6.1$. At this point, $C_V$=0.7713675$nR$. The maximum of the curve $C_V/T$ is at $0.27985645\times\vartheta_D$ which is useful to quickly estimate the Debye temperature of a solid if $\vartheta_D$ is constant (which is, unfortunately, almost never the case for minerals).

The Einstein model [59] is given by

$$C_V(T) = 3nR \cdot \left(\frac{\Theta_E}{T}\right)^2 \cdot \frac{\exp\left(\frac{\Theta_E}{T}\right)}{\left[\exp\left(\frac{\Theta_E}{T}\right)-1\right]^2} \tag{7}$$

with $\theta_E$ the Einstein temperature. The Einstein model is unphysical for low temperatures $T\rightarrow 0$, but useful as a reasonable approximation for the lattice heat capacity of optical vibration modes at high temperatures.

In some cases, other vibrational $c_P$ contributions are observed, for example by hindered rotations, inversion vibrations etc. (e.g. in ammonia $NH_3$); often in molecular solids, polymers and complex organic substances, see, for example, [60]. Note that in polymeric science, where often linear chains of molecules dominate the vibrational modes, the 1-dimensional Debye function is often used [e.g., 61].

Coming back to silicate minerals, Kieffer [62-64] developed a more sophisticated theory which captures the main features of the vibrational spectra encountered in non-simple solids. It proposes a vibrational spectrum consisting of three acoustic branches, an optical continuum and optional Einstein oscillator(s). This theory contains up to 25 parameters; it is, however, independent of calorimetric data and not obtained by any fitting procedure. Kieffer's theory is useful – if measurements are not available – for the prediction of lattice heat capacities of structurally complex rock-forming minerals from their elastic constants and spectroscopic data. The parameters are given by elastic, crystallographic and spectroscopic (infra-red and Raman) data only, which are used to define upper and lower limits of the various vibrational branches. Its accuracy, if compared to accurate experimental data, is typically 30-50% below 50 K, 5% at 300 K and 1% at 700 K; fitting of ill-determined spectroscopic parameters by calorimetric data can improve the low-$T$ accuracy significantly. The theory however cannot model any anomalies (Schottky anomalies, electronic and magnetic contributions, transitions) and neglects the effects of thermal expansion (the spectrum is referred to the volume V at 0 K), defect/domain/surface contributions and, perhaps most significantly, anharmonic effects. All these effects are usually small in the temperature range 10 K<$T$<500 K. At high temperatures, when the details of the lattice vibration spectrum are not so important, often a single Einstein oscillator term (corresponding to the Si-O stretching mode) suffices to fit silicate $c_P$-data (to the order of 1% at 700 K).

A variant of Kieffer's lattice dynamics model using vibrational density of states for constructing thermodynamic databases is given by [65]. This model is computationally much simpler and faster than the Kieffer model, it models the vibrational density of states by the sum of (a large number, ~60) monochromatic Einstein frequencies and adds models for the dependence on volume of the Einstein temperatures, an equation of state for the static lattice contributions and a free electron gas model for the electronic contribution. It allows to predict also thermal expansion and anharmonicity [66] of minerals; the main input are data (infrared, Raman, inelastic neutron scattering) on the vibrational DOS.

There is an established alternative theoretical model for the lattice heat capacity, that of Komada and Westrum [67, 68] which is somewhat complex mathematically (discussed in [69]). This model needs also a number of input parameters from chemical and crystallographic data, besides a (nicely constant) characteristic temperature $\theta_{KW}$, and similarly to the Kieffer model does not describe any peaks and anomalies.

<u>The relation between $C_V$ and $C_P$</u> from thermodynamics is

$$C_P - C_V = TV\alpha^2 B = TV\alpha^2 / \beta \qquad (8),$$

where $\alpha = \frac{1}{V}\left(\frac{\partial V}{\partial T}\right)_P$ is the isobaric coefficient of thermal volume expansion, $V(T)$ the molar volume, $B$ the isothermal bulk modulus, $\frac{1}{B} = \beta = -\frac{1}{V}\left(\frac{\partial V}{\partial P}\right)_T$ the isothermal compressibility.

All quantities are temperature-dependent.

<u>The pressure dependence</u> $dC_P/dp$ is negligible for most minerals at pressures up to thousands of bars. As an example, for periclase (MgO), the maximum relative sensitivity $dC_P/dp/C_P$ , at ~70 K, is about 3E-6/bar, thus reaching 1% at pressures of 3000 bar or more. For forsterite, Chopelas [70] finds $dC_V/dp$ of 4.98E-5 J/mol/K/bar at 298 K or, in relative terms, 4e-7/bar or reaching 1% at 23 kbar. See [71, 72] for extensive information on the pressure dependence where it matters (e.g., in the Earth's mantle).

<u>Anharmonicity</u>. On top of the effects of thermal expansion ($C_V \rightarrow C_P$) the anharmonicity of lattice vibrations typically increases even $C_V$ beyond the Dulong-Petit limit at high temperatures; the anharmonicity of forsterite, fayalite, and periclase has been discussed by Anderson and Suzuki [73]. [73].

Anharmonicity in general is covered in [39, 66, 74-76]. For example, $C_P$ of feldspars ($n$=13) at 1400 K [43] is between 330 and 346 J/mol/K, where $3Rn$=324.26 J/mol/K would be the predicted limit for $C_V$. Forsterite has a high intrinsic anharmonicity, where even $C_V$ exceeds the Dulong-Petit limit for $T$>1550 K [77].

<u>Electronic heat capacity</u> occurs in conductors with free electrons, thus mostly in metals (Fe, Ni) but also in, e.g., in graphite and pentlandite $(Fe,Ni)_9S_8$ [78, 79]. It is a small effect only relevant at low temperatures. The usual low-temperature limit [80] is given in the free-electron approximation by ($T_f$: Fermi temperature, calculated with the number density and effective mass of the valence electrons)

$$C_{V,el} = \frac{\pi^2}{2} k_B \left( \frac{k_B T}{E_f} \right) = \frac{\pi^2}{2} k_B^2 \left( \frac{T}{T_f} \right) = \gamma T \qquad (9)$$

Various refinements valid for higher temperatures exist, e.g. [80] but deviations of a simple linear $T$-dependence are usually negligible. A different electronic heat capacity stems from electronic excitation from the ground state (energy set to 0, degeneracy $g_0$) to higher energy levels (degeneracy $g_i$; $T_\Delta$ is the energy difference expressed in Kelvin) and is usually called <u>Schottky-type heat capacity</u>. It has the form of a very broad asymmetric peak [54, 81, 82] which falls off $\propto 1/T^2$ at temperatures higher than the peak temperature.

For a two-level system the Schottky heat capacity is

$$C_{e,sh} = R(g_0 / g_1) \left( \frac{T_\Delta}{T} \right)^2 \frac{\exp\left(T_\Delta / T\right)}{\left[1 + (g_0 / g_1) \exp\left(T_\Delta / T\right)\right]^2}$$

Realistic systems often involve several transitions with various degeneracies, usually at very low temperatures, e.g., [83] for fayalite. Note that the peak temperature of the Schottky bump is of the order of (0.3 –0.4)$T_\Delta$ and its magnitude is of order ~0.2$R$ to ~0.8$R$, depending on the degeneracies, not on temperature; at very low temperatures, this can be a significant or even the dominating (in case of nuclear terms) contribution to heat capacity.

Note that a linear term in $c_P$ (at low $T \ll 100$ K) not necessarily stems from conduction electrons, but could also be caused by lattice vacancies [84].

It is customary to plot low-temperature $c_P/T$ *vs.* $T^2$; obviously, cubic (Debye-) and linear (electronic or glass anomaly) terms can then be easily determined from extrapolating a linear fit to zero K, see the example in Figure 2.

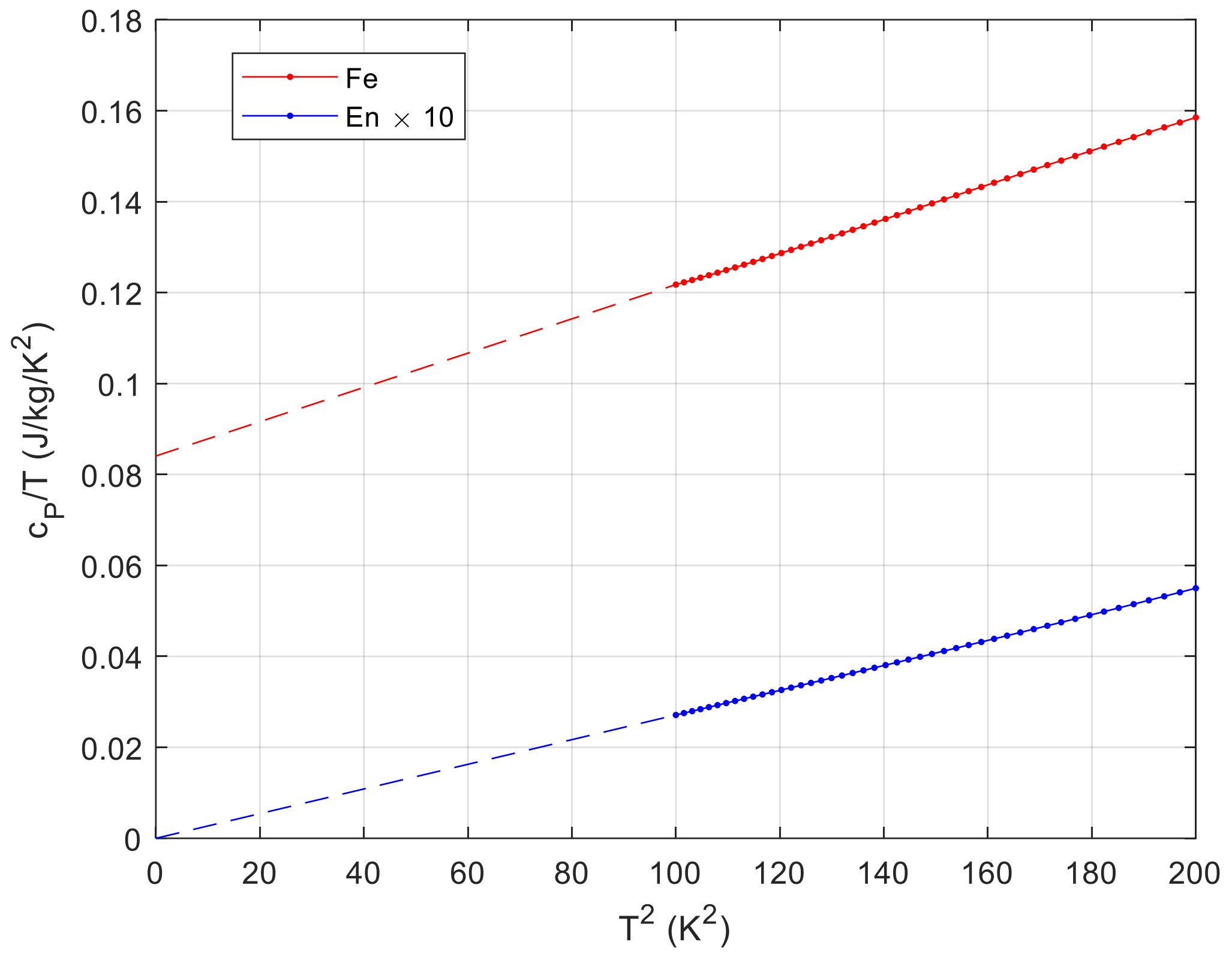

**Figure 2 Plotting $c_P/T$ versus $T^2$ for low temperatures, less than about 15 K, gives straight lines for most solids; the slope is ∝ $1/\theta_D^3$ , and extrapolation to 0 K gives directly γ, the electronic heat capacity term, while for Debye solids it is zero. Low-temperature anomalies (e.g., Schottky) also show up clearly. Smoothed $c_P$ data of our database have been used.**

Anomalies in glasses and gels. Glasses have a $c_P$ anomaly at low temperatures (and a glass transition $c_P$ anomaly, basically a step, at the high temperature glass transition temperature $T_g$ which is typically at a much lower temperature than the melting temperature of the crystalline phase). $T_g$ for silicate minerals depends strongly on water content [85]).

The low-temperature anomaly of glasses consists of an extra $c_P$ contribution, about linear in $T$ (∝ $T^{(1+\delta)}$, but vanishing at high $T$ > about 30 K (*au contraire* to electronic heat capacity). See the glass section 2.15 below for details.

Activation Heat Capacity. At high temperatures, especially for substances with a high melting point, lattice monovacancies can have a marked effect on $c_P$, e.g., for tungsten >1000 K [86]. This effect can be mixed with the 'premelting' increase in heat capacity caused by impurities (see below).

Magnetic (ferromagnetic and ferromagnetic) and order/disorder transitions are discussed in more detail below. They are generally very difficult to model precisely. They usually produce transition peaks in the $C_P(T)$ curve that can be very dominant (compare Figure 1). For the magnetic contributions, at least limiting cases for $T \rightarrow 0$ can be given (Table 1).

Nuclear contribution to the specific heat can become significant below ~1 K in certain compounds, depending on isotopic composition and dependent on external magnetic fields, e.g., [39, 87]. It's typically a Schottky peak at ~0.01 K; below this peak temperature, nuclear contribution tends to 0, at high temperatures it varies as ~1/$T^2$.

**Table 1 Limiting cases at low temperatures $T \rightarrow 0$ K, from [39]**

| | |
|---|---|
| $C_V = \beta T^3$ | lattice vibrations only, isolators |
| $C_V = \beta T^3 + \gamma \mathrm{T}$ | nonmagnetic conductors, glasses (approx.) |
| $C_V = \beta T^3 + \gamma \mathrm{T} + \delta \mathrm{T}^{3/2}$ | ferromagnetic and ferrimagnetic |
| $C_V = \beta T^3 + \delta \mathrm{T}^{3/2}$ | ferrimagnetic |
| $C_V = \beta T^3 + \gamma T + \delta \mathrm{T}^3$ | antiferromagnetic |
| $C_V = \beta T^3 + \eta / T^2$ | nuclear (2-level, >$10^{-3} \ldots 10^{-2}$ K) |

More theoretical background, in particular for low temperatures and 'heat capacity anomalies' can be found in [88, 89].

### 1.1.1 Mixing model

Except (presently) for Olivine (see below), we use a simple mechanical mixing model for astro-materials composed of endmember minerals:

$$C_P = \sum_i X_i C_P^{(i)} ,$$

$$c_P(T) = \sum_i w_i c_P^{(i)}(T) \qquad (10)$$

with $X_i$ the mole and $w_i$ the mass fractions of the constituents, $\sum X_i = 1$, $\sum w_i = 1$, and $C_P^{(i)}$ are the heat capacities of the endmembers. This model is exact for mechanical mixtures and for ideal solid solutions of endmember minerals (without interactions); deviations for solid solutions are discussed next.

#### *1.1.1.1 Solid solutions and the excess heat capacity of mixing*

Many minerals form solid solution series (‘joins’, in the jargon). Their $c_P$ is only ideally given by the linear combination of end-member $c_P$ with the end-member mass fractions as coefficients (mole fractions of end-members for $C_P$). For detailed background, theory and experimental, see e.g. [90].

However, non-idealities exist. The definition [90] of the non-ideality of $C_P$ , called *excess heat capacity* of mixing, is (‘real minus ideal’)

$$\Delta C_P^{ex} = C_P^{solid\ sol.} - \sum C_P^{(i)} X_i$$

where $C_P^{solid\ sol}$ is the heat capacity of the solid solution, $C_P^{(i)}$ are the heat capacities of the endmembers and $X_i$ are the corresponding mole fractions. Usually, measured excess heat capacities are used to compute the excess entropy $S^{ex}$ and modelled (at STP, 298.15 K) as function of composition. This done, the $\Delta C_P^{ex}(T)$ cannot be derived anymore. Rather, the measured data have to be used to calculate temperature-dependent Margueles parameters, e.g. for a binary mixture:

$C_P^{mix}(T) = (1 - X_2)C_P^1(T) + X_2 C_P^2(T) + X_2^2 W_{12}(T) + (1 - X_2)^2 W_{21}(T)$ or the Margueles formulation for an asymmetric ternary solution [91] which has 7 Margueles parameters, 6 $W_{ij}(T)$ parameters *and* $W_{123}(T)$.

Olivines, feldspars and pyroxenes are the most abundant rock-forming minerals, thus it is desirable to know the excess heat capacities for their solid solutions. At present we can do that only for olivine, a mixture of the two endmembers forsterite and fayalite, where the excess heat capacity is well characterized. For other minerals, there is a dearth of data on excess heat capacities, so we mostly ignore the deviations from ideal, (or mechanical) mixtures. This leads to uncertainties, which are negligible at high temperatures (>300 K), and possible systematic deviations from the mechanical mixing model in the low temperature range for some solid solution series. Maximum excess heat capacities found [90] are, e.g., ~25% at 40 K for grossular-pyrope, ~10% at 40 K for analbite-sanidine, ~3% at 400 K for annite-siderophyllite; ~50% at 10 K for bronzite (Fe-poor orthopyroxene) but negligible >65 K [92, 93], <2% for feldspars between 10 K and 800 K [91, 94, 95].

For olivine (Fo/Fa solid solutions), [96] measured a significant excess heat capacity, but only near the magnetic transition at 35 – 70 K. Since these data cover the whole composition range of olivines in sufficiently small increments (Figure 3), we are able to 2D-interpolate the $c_P$ of the solid solutions accurately. Thus, for olivines of known Fo/Fa composition, the database gives the accurate $c_P$ directly. A 2-D interpolation (table lookup) is used to obtain $c_P$ values for fayalite concentrations $0<X_{Fa}<1$ and *T*<300 K and a mechanical mixing model for 300<*T*<1400 K where the excess heat capacity in olivines is negligible.

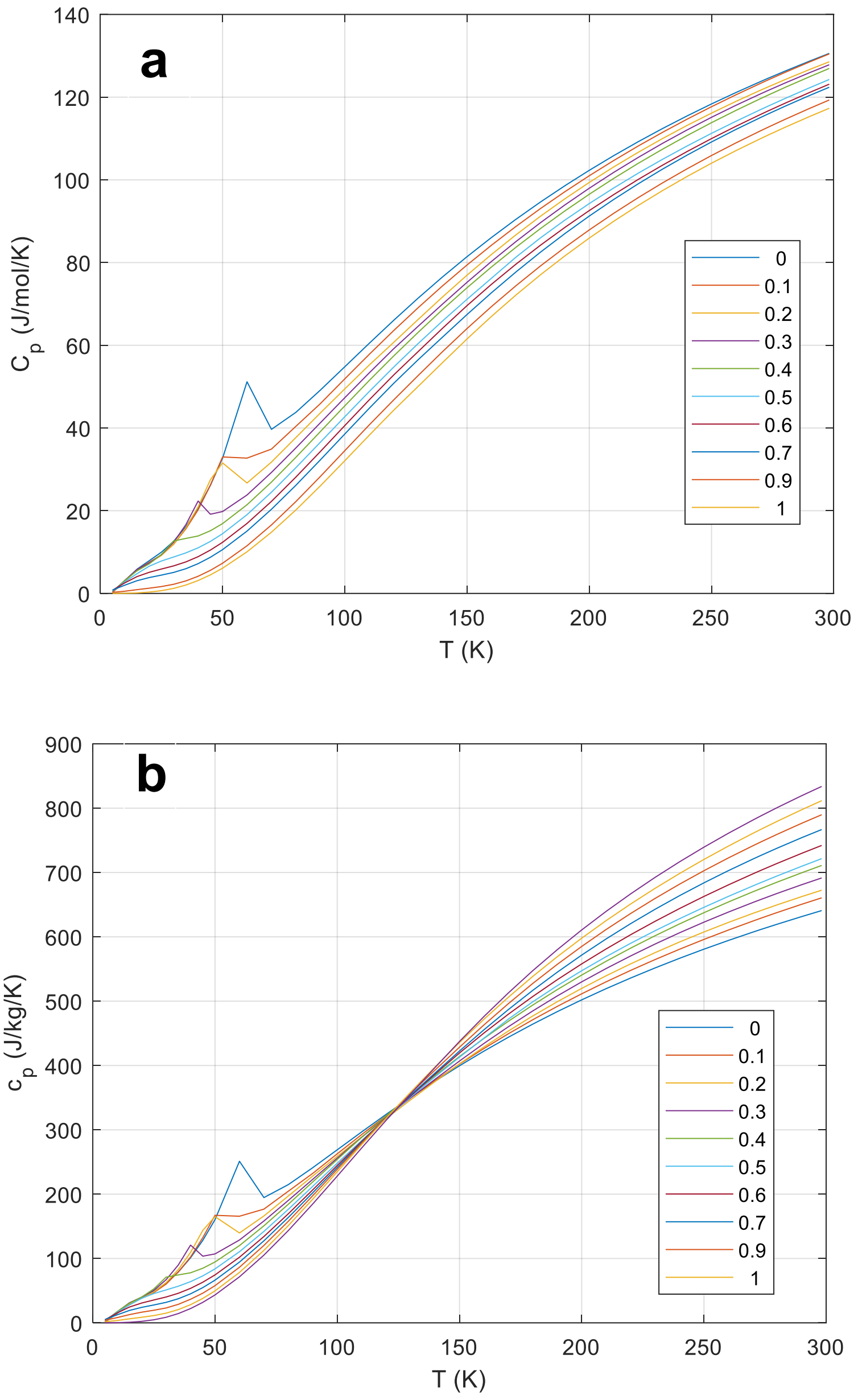

**Figure 3 $C_P$ (upper panel, a) and $c_P$ (lower panel, b) of olivines, after [96]. Note the X-point at ~125 K, where all compositions have about the same mass-based specific heat, which is not the case in the molar $C_P$. This is a quite natural effect of the vastly different formula weights of fayalite (203.778) and forsterite (140.693).**
**Parameter in legend: $x_{Fo}$, mole fraction forsterite ($w_{Fo}=x_{Fo}$*140.693/(203.778-63.085*$x_{Fo}$))**
**Higher-resolutions data around the transition peaks not shown for clarity.**

There are data on the excess heat capacities of feldspars and pyroxenes [91, 94, 95, 97-99], but they are presently difficult to model due to the up to 4-dimensional compositional range. For Fe-Ni alloys, the enthalpy of mixing is small since the two metals are very similar. However, the temperatures and amplitudes of the magnetic and structural transition lambda peaks >600 K change drastically with composition (see Figure 10). Our database currently employs the curve for a standard Fe/Ni ratio for all temperatures and a real mixing model for olivine; it is planned to give at least approximate real mixing models for idealized feldspars and pyroxenes in the future, i.e., for idealized anorthoclase (alkali) Ab-Or and plagioclase Ab-An feldspars and for idealized orthopyroxenes En-Fs and clinopyroxenes Di-Hed.

So, what accuracy can be expected for the $c_P(T)$ of an astro-material of given mineral composition, if accurate endmember mineral's $c_P(T)$ are in the database? We have indications that the remaining uncertainty is very low at high temperatures, e.g., [100] could reproduce the measured $c_P$ of 4 'standard rock samples' from 300 to 1000 K with a standard deviation of about 1% if calculated from mineral compositions. For very low temperatures, if there are solid solutions (not olivine, which we already treat as non-ideal mixture) with a high excess heat capacity the few examples given above suggest a *maximum* relative deviation, outside or near $c_P$ anomalies, of ~25% [90] or ~50% at 10 to 40 K, decreasing rapidly for $T$>65 K, and for transition peaks (if seen at all in the $c_P(T)$ curve) a possibly significant change in peak temperature and amplitude.

Note that even for laboratory samples, the uncertainties of chemical analysis[3] (for normative mineral composition) and especially the uncertainties of modal analysis[4] are typically of the order of a few % even for major constituents; of the order of 10% or more for minor constituents. This translates, already, into a few % uncertainty in $c_P$ on average; for less well known astro-material, it follows that the uncertainties stemming from compositional uncertainty are usually more significant than those from the non-ideality effects of $c_P(T)$ in solid solutions.

An example: bronzite

Bronzite is Fe-poor orthopyroxene (hypersthene) and its $c_P$ should thus be a linear combination of En and Fs. Krupka et al. [92, 93] have measured its $c_P$ from 5 K to 1000 K. The sample is a natural crystal of idealized composition $Mg_{0.85}Fe_{0.15}SiO_3$. (x=0.15 Fe, 1-x=0.85 Mg).

It turns out that the nominal $c_P$ calculated with x=0.15 and the ideal molar mass , M=105.120 g/mol (corresponding to $Mg_{0.85}Fe_{0.15}SiO_3$) already matches very closely (better than ±1%) the data except below ~100 K where the broad Fs transition peak occurs at 38 K, but the corresponding Bronzite peak (actually only a Schottky bump) is shifted to ~12 K (see Figure 4 and Figure 5). A free fit of the composition (Mg and Fe only), with temperatures < 100 K excluded from the fit, results in a slightly better agreement of measured and calculated $C_P$ (0.25% less bias, overall agreement mostly better ±0.5%) and returns compositions entirely consistent with the chemical analysis and its inherent uncertainties. For supporting data and figures see Appendix, section 6.

[3] The relative 1-σ uncertainty of a mass fraction $p$ in chemical analysis is about $\sigma(p)/p \approx 0.25 \exp(-5\sqrt{p})$, correlating data of [101-104]. This reflects the state-of-the-art ca. 1950-1985, combination of atomic absorption spectroscopy and standard wet-chemical techniques [105]. Nowadays, uncertainties are much smaller.

[4] Was generally point counting of thin sections, with significant statistical uncertainties of the order of 20% (relative) [106] depending much on component abundance, number of points counted. Potentially much more depending on the homogeneity of the sample, i.e., the representativeness of that thin section. No one does it anymore that way so those old analyses will stand forever. Newer SEM X-ray mapping techniques are probably not as good.

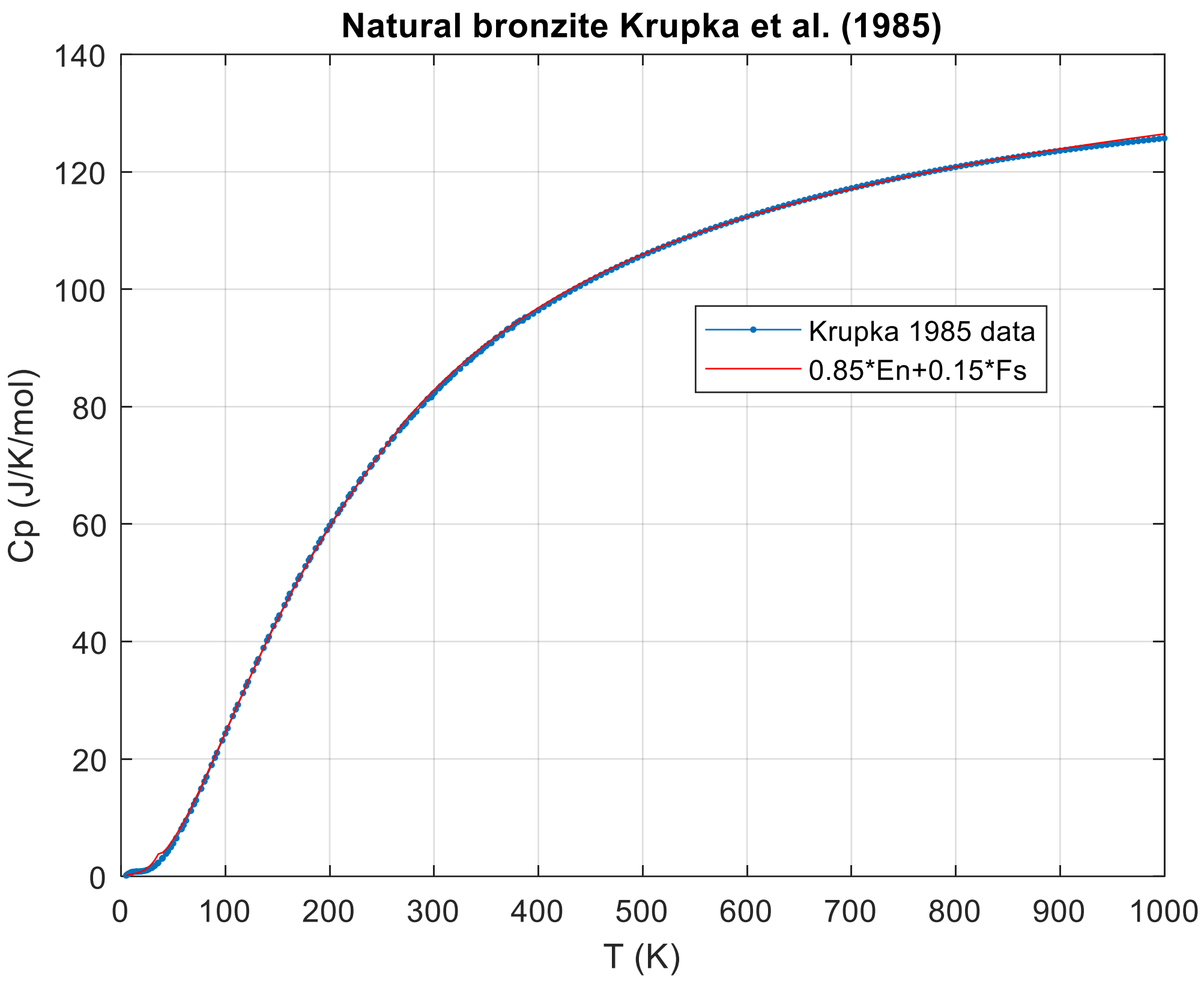

**Figure 4 Bronzite [93], data and ideal $c_P$ calculated for ideal composition.**

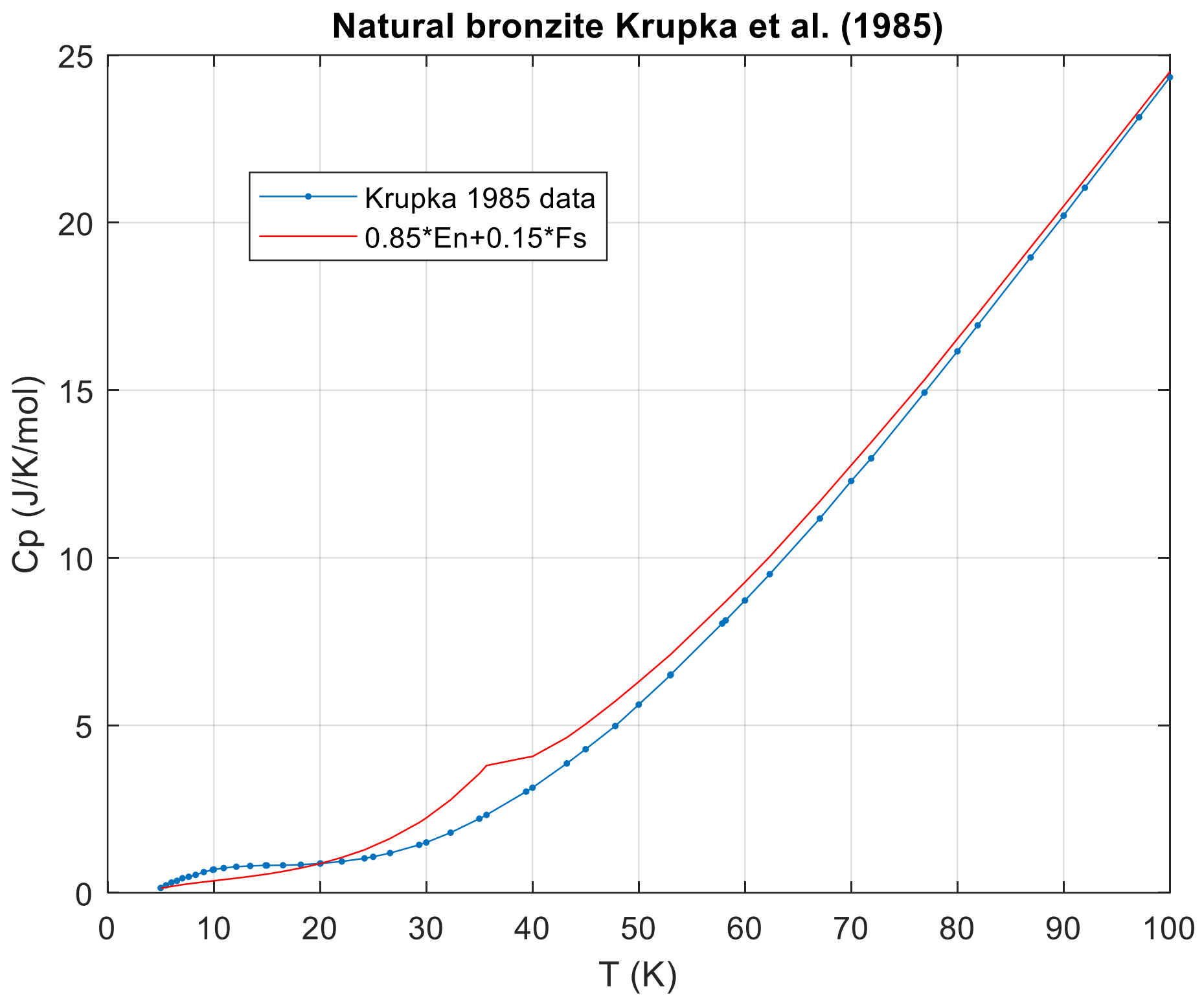

**Figure 5 Bronzite [93], low T, data and ideal $c_P$ calculated for the ideal composition. The Fs transition at 38 K and the Schottky peak of bronzite near 12 K do not scale linearly.**

We conclude that nonideal mixing is negligible (<1 % effect similar to experimental uncertainties) here for *T*>100 K but significant (up to ~60%) at certain very low temperatures, near 12 K and near 38 K, but only due to the change of the magnetic/Schottky-transition peaks at low temperatures with composition.

### 1.1.2 Polymorphs and phase transitions

Polymorphism is the ability of a mineral to exist in more than one form or crystal structure. Different polymorphs can have slightly differing $C_P(T)$ and there may be a peak in $C_P$ at the phase transition temperature, where the low-temperature form transforms into the high-temperature structure.

There are three main types of *structural* phase transitions [e.g., 107]

- Reconstructive (metastable at low *T*, since they require diffusion)
- Order-disorder (metastable at low *T*, since they require diffusion)
- Displacive (instantaneous, since they only require a distortion of the lattice)

The rate of solid diffusion required for reconstructive and order-order phase transitions follows approximately an Arrhenius equation, $\propto \exp(-E/RT)$ with $E$ the activation energy. To give an example, for the Al/Si disorder rate in albite and microcline, activation energies between 280 and 360 kJ/mol have been determined and a 50% transformation time of 5 days at 1050°C [108]. From this it can be estimated that below ~400°C, the phases are 'frozen in' over timescales comparable with the age of the solar system (4.5 Ga).

For other atoms in solids, much lower activation energies of the order of ~60 kJ/mol have been determined. The atomic migrations of Al and Si in feldspars are probably slower than those of any of the other major ions, including oxygen, at least when water is present [109]; hence the migrations of these species may be rate limiting for a number of processes in feldspars. There are no data for Al or Si diffusion in feldspar because the rates are so slow, but studies of AI-Si order-disorder kinetics are one way to get at this problem [108]. For the coupled substitution (Na,K)+Si = Ca+Al, where the tetraeder system is involved, a diffusion coefficient of $10^{-22}$ cm²/s at 800°C has been determined [110, 111].

There are also high-pressure polymorphs of minerals, e.g., for forsterite[5], wadsleyite and ringwoodite which have been reported from shocked meteorites. Depending on the cooling history, both high-temperature and high-pressure modifications, although metastable, can remain 'frozen in' for billions of years.

Phase transitions can also be caused by (or coupled to) magnetic effects [113, 114]; other transitional behavior includes Verwey, Jahn-Teller, metal-insulator, superconductivity, electrical, plastic and 'crystalline liquid' phenomena that show up as anomalies in the $C_P(T)$ curve [39, 115-118].

#### *1.1.2.1 Modeling of phase transitions*

The most commonly adopted thermodynamic classification of phase transitions still follows the Ehrenfest [119, 120] terminology, by assigning the order of the transition appropriate to the order of the derivative of the Gibbs function (with $P$ or $T$) showing a finite discontinuity. A 1st-order phase transition (1O) is characterized by a latent heat (= energy is absorbed or released by a substance during a change in its physical state without changing its temperature). The fact that the temperature is not changed during the 1O phase transition causes $C_P$ to go to infinity (theoretically). The explanation for the presence of the latent heat is that chemical bonds are broken during the (heating) transition (melting of a crystal, vaporizing of a liquid) and this is responsible for the absorption of energy without increasing the temperature, thus, a vertical jump of enthalpy at the transition temperature $Tc$, thus $C_P(T_c)=\infty$ (but with a finite integral = phase change enthalpy *ΔH*). A simple 1-O phase transition ideally produces a δ-peak in $C_P$; experimentally, due to non-zero thermal homogeneity, the peak has always a finite width ε, order of 1 K. This can be described by the Gaussian approximation to the Dirac δ-function,

$$\Delta C_P(T) = \frac{\Delta H}{\sqrt{2\pi\varepsilon}} \exp\left(-[T - T_0]^2 / 2\varepsilon\right).$$

[5] $Mg_2[SiO_4]$ is trimorph, α = forsterit and high-pressure phases (metastable at low pressure; usually shocked but not melted forsterite), ringwoodite (γ-$(Mg,Fe)_2[SiO_4]$) and wadsleyite (β-$(Mg,Fe)_2[SiO_4]$). Small differences of the $C_P(T)$ curves (α,β,γ) have been measured by [112]

The Ehrenfest higher-order transitions have received much evolution since 1933 [119]. It became clear that not only the existence of discontinuities in thermodynamic derivatives but also the actual nature of the discontinuity of the *m*th derivative of the Gibbs free energy at the transition point is important, whether, for example, $c_P$ appears to go to infinity at the transition point or is merely one which is finite but very large [121-126]. See Figure 6 for a schematic overview. 1$^{st}$ order transitions (other than melting) are rare in minerals (Quartz probably) as are strictly 2$^{nd}$ order transitions; most are 'in-between'.

Summarizing, phase transitions other than first order (called second order (2O) or, maybe better, continuous) including tricritical phase transitions [127] are less well understood and rarely analytically tractable. A strictly 2$^{nd}$-order phase transition has no latent heat and hence $C_P$ does <u>not</u> go to infinity. It describes displacive phase transitions without breaking chemical bonds. This is also true for magnetic phase transitions (magnetic ordering gives rise to a distortion of the lattice).

Transitions with finite discontinuities in specific heat at a definite transition temperature (classical 2-O) are extremely rare [107]. Phase transitions which are not 1$^{st}$ order, yet which show (probably) infinite heat capacity, are called λ-transitions, with no or small first order break at $T_c$ [128], see also the provocative papers by Mnyukh [129, 130]. The heat capacity of the system increases (coming from $T \ll T_c$) long before the critical temperature $T_c$ and typically falls off much faster. Examples are order/disorder transitions in alloys or solid solutions, ferromagnetism and the transition from liquid to superfluid helium. A famous example is ammonium chloride, $NH_4Cl$. Lambda (λ) transitions are very common and may be distinguished from classical second-order (2-O) phase transitions in that heat capacity $C_P$ (not $C_V$) tends toward infinity as the transition temperature is approached. Some transitions are mixed' or 'superimposed' as, for example, the ferroelectric transition in $KH_2PO_4$ (KDP) at about 122 K is mixed displacive and order-disorder with one transition triggering the other [107].

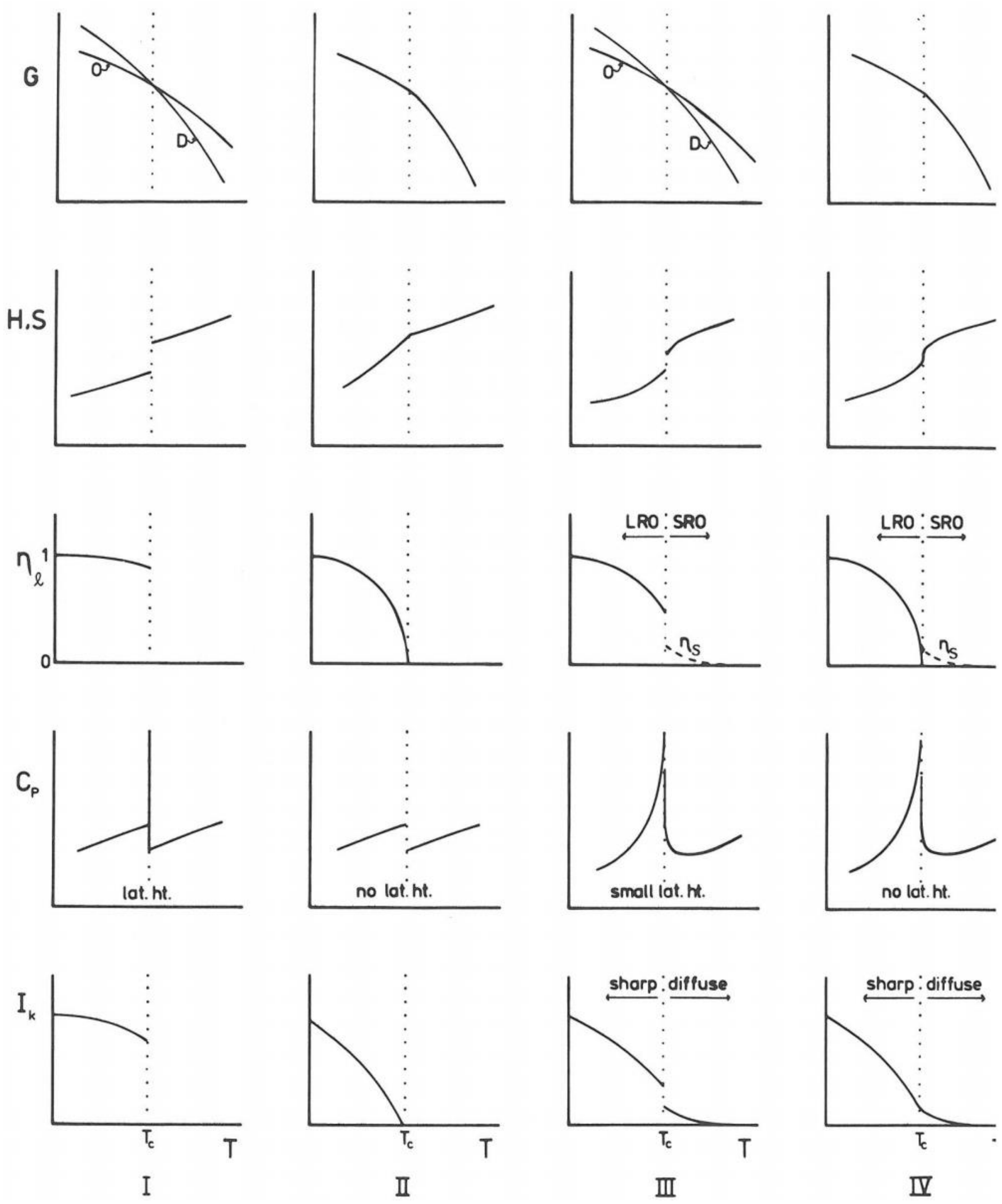

**Figure 6 After [128] Schematic form of the principal thermodynamic parameters through a phase transformation at $T_c$. Column I = first order; column II =second order; column III = λ transformation with a small first order break at $T_c$ ; column IV = λ transformation with no first order break. G = free energy, H = enthalpy, S = entropy, $\eta_l$ = long range order parameter $\eta_s$= short range order parameter describing precursor ordering above $T_c$; $C_P$ = specific heat; $I_k$ = integrated intensity of a superlattice reflection. D = disordered state, O= ordered state. LRO = long range order, SRO = short range order. Volume is not shown, but must be continuous or discontinuous in some manner analogous to H and S**

A very careful analysis of the lambda-transition in quartz is given by [131].

Lambda transitions can often be treated in the framework of the Landau theory [132, 133]:

$$C_P(T) = C_{P,L}(T) + \begin{cases} \dfrac{a^{3/2}}{4c^{1/2}} \dfrac{T}{\sqrt{T_d - T}}; T < T_c \\ 0; T > T_c \end{cases}$$

where $C_{P,L}$ is the lattice heat capacity, $T_c$ is the temperature for which the experimental specific heat curve has the maximum value and $T_d$ is the metastability limit on cooling*, a, c* are constants.

Long-range correlations and fluctuation effects can be semi-empirically modelled ('critical exponents'), by an additive term $C_\lambda$ [96, 134, 135],

$$C_\lambda(\varepsilon) = \begin{cases} A'\varepsilon^{-\alpha'} & ,T < T_c \\ A\varepsilon^{-\alpha} & ,T > T_c \end{cases}$$
$$\varepsilon = |T - T_c| / T_c$$

where the critical exponents α, α′ can be slowly varying functions of (reduced) temperature ε or log(ε). Note that fits usually give slightly different $T_c$, $T_c'$ for the portions above and below the peak. Dachs et al. [136] applied the $C_\lambda$ model successfully to the magnetic phase transition of almandine. Some compilations of mineral thermodynamic data (e.g., [41]) represent transition peaks either with Landau parameters or with the parameters of the Bragg-Williams theory (see [137, 138]). Improved theories for ferro- or antiferromagnetic transitions are available [139, 140].

The heat capacity behaviour related to a phase transition depends on the degree of crystallinity of the crystal (e.g., the concentration of imperfections), is rate-dependent and has hysteresis.

Thus the shape of the corresponding peak is very likely sample dependent (impurity content, grain size) and whether the temperature is raised or lowered through $T_c$ and how fast [141]. The so-called Verwey transition in magnetite near 125 K is an example of a displacive structural transition coupled to a magnetic phase transition. The temperature and shape of the Verwey transition peak are highly sensitive to the stress state of magnetite and to the stoichiometry; non-stoichiometry in the form of metal cation substitution or partial oxidation can lower the transition temperature or suppress it entirely**.** Similarly, in wüstite, $Fe_{1-x}O$ (a classical example of a non-stoichiometric phase) the antiferromagnetic/displacive lambda peak near 190 K is strongly composition dependent [142, 143].

## 1.2 Useful approximations

### 1.2.1 The Neumann-Kopp rule and estimation models

The so-called Neumann-Kopp rule is just stating the obvious, namely that heat capacities of mixtures are additive and that $C_V$ scales with $n$, the number of atoms per formula unit, and $c_V$ (and $c_P$) scales with $n/M_r$, thus with $1/\bar{A}_r$; $\bar{A}_r$ is the average atomic mass of the elements involved, $\bar{A}_r = M_r/n$. The $\bar{A}_r$ of silicates is of the order of 20 g/mol, which explains why meteoritic iron ($\bar{A}_r$ ≈56 g/mol) has a $c_P$ about half of the $C_P$ of silicates. It also explains that per unit of mass, heat capacity is rather similar in all rocks, while the molar heat capacity can assume rather high values if $n$ is high.

The approximate scaling of $c_P$ with $1/\bar{A}_r$ is very useful to estimate the specific heat of a mineral that is not in the database, but an isostructural mineral with similar composition is (extension: principle of corresponding states (see, e.g., in [93])

The additivity is used for the subtraction of impurities (secondary phases) from experimental $c_P$ data [e.g.,144] and for deviation from end-member stoichiometry [145, 146]:

$$c_{P,\mathrm{miner}} = \frac{c_{P,\mathrm{sample}} - \sum_i x_i c_{P,i}}{x_{\mathrm{miner}}}$$

where $x_{miner}$ is the mass fraction of the mineral, $c_{P,sample}$ the heat capacity of the sample, $x_i$ the mass fraction of impurity $i$, and $c_{P,i}$ the heat capacity of impurity $i$ , all in J/kg/K.

The Neumann-Kopp rule is also invoked to roughly estimate the heat capacity of compounds from known heat capacities of constituent compounds [115], e.g., $C_P(MgAl_2O_4) \approx C_P(MgO) + C_P(Al_2O_3)$. Leitner et al. [74] discuss the extensions to the empirical Neumann-Kopp rule, a combination of an additive and a contribution method to estimate the heat capacity of complex compounds. See also [147] and [148].

Indeed, several schemes have been devised to estimate the thermodynamic properties of minerals for which they are unknown. These models are all based on the premise that the thermodynamic properties of minerals can be described as a stoichiometric combination of the fractional properties of their constituents: $X = \sum n_i x_i$ where $X$ is the property of interest, $x_i$ are the fractional properties of each constituent and $n_i$ are the stoichiometric amounts of that constituent in the mineral. Different building blocks are used in the models ranging from elements [149], oxides [46, 150], iso-structural minerals [151] to elements in their respective crystallographic coordination (the polyhedron method, [152-154]) and other schemes [155]. Up to now, all these models have been made only for the high temperature range, ≥298 K (interesting for terrestrial geophysics), the temperature dependence conveniently cast into one of the usual polynomial representations (see section 1.2.4). However, it should also be possible to extract a $C_P(T)$ polyeder-'kit' from the many existing low-$T$ data, or the $C_P(T)$ of 'exchange vectors', i.e., the change of $C_P$ by substitution from a well-known (or DFT-calculated) endmember. To the best of our knowledge, this has not been done (or tried) yet.

If there is no experimental $c_P$ data for a particular mineral, all these estimation methods are certainly better than nothing; the crux is that the *a priori* uncertainty of the predicted $c_P$ values is rather unpredictable.

### 1.2.2 Determination of lattice heat capacity

In order to isolate a heat capacity anomaly, e.g. a transition peak, it is necessary to estimate the pure vibrational ('lattice') heat capacity in the complete temperature range where the anomaly has a significant effect. Various methods, more or less empirical, exist:

the procedure described by Robie et al. (1982b) Robie RA, Hemingway BS, Takei H (1982b) Heat capacities and entropies of Mg2SiO4, Mn2SiO4 and Co2SiO4 between 5 and 380 K. Am Mineral 67:470–482 was used. [156].

Komada-Westrum model fit to temperature regions not affected by the anomaly, extrapolation assuming a constant KW-temperature.

principle of corresponding states (Lewis and Randall, 1961; McQuarrie, 1973) with respect to tremolite.That is, the ratio of the low-temperature Cp per gram of tremolite (Robie and Stout, 1963) to the Cp; per gram of pure magnesioanthophyllite was used for a smooth extrapolation to zero Kelvin [93]

### 1.2.3 Modeling $C_P$ - $C_V$

The relative difference of $C_P$ and $C_V$ is usually, for silicates, of the order of <1% below room temperature, and of the order of 4% at 700 K. The part due to thermal expansion can be modelled [77, 157, 158], eqn. (1.8), if the thermal volume coefficient of expansion α, bulk modulus $B$=1/isothermal compressibility are known as a function of temperature.
A perfect knowledge of α and $B$ is still not sufficient to calculate $C_P$-$C_V$ exactly, because of the additional contribution of the anharmonicity which has the approximate form $c_{p,anh} = c_{p,harm}(1+aT)$ (crude average over all vibrational modes, see [75, 159]).

In practice, after [160, 161], [162-165], we write

$$C_p = C_v(1+\alpha(T)\gamma_G T) \quad \text{or, since } \alpha\gamma_G \propto C_P,$$

$$C_p - C_v = AC_p^2 T, \qquad (11)$$

$$C_p = \frac{1-\sqrt{1-4ATC_v}}{2AT}$$

with $\gamma_G$ the thermodynamic Grüneisen parameter [166]. $A$ and $\gamma_G$ can be taken as approximately constant, over a wide range of temperatures $T>\theta_D$ (for $T<\theta_D$ $C_P$-$C_V$ is, fortunately, usually small). The second formula is known as the Nernst-Lindemann relation. The unit of $A$ is mol/J (or kg/J, for $c_P$).

Parameter $A$ may be crudely estimated from melting temperature $T_m$ by [167] $A \simeq 1\cdot 10^{-10}[mol/J/K]\,T_m\,[K]$ . If data on the thermal expansion coefficient and compressibility at one temperature $T_0$ are available, $A$ may be calculated as

$$A \cong \frac{V(T_0)\alpha(T_0)^2 B(T_0)}{C_P^2(T_0)}$$

Alternatively, $A$ can be estimated from high temperature $C_P$ -data alone, by invoking the empirical constraint that the effective Debye-temperature ≈ constant at the highest temperatures $\gg \theta_D$. This $A$ then also includes the effects of anharmonicity in an approximate way.

### 1.2.4 Polynomial expressions for $c_P$ at high temperatures

Various empirical polynomials are in use (see Table 2, below), they have been discussed by [46], see Table 3. They all diverge for $T\rightarrow 0$ and are only useful for $T >$ (100–300) K and only if no transition peaks appear in the fitted range.

[168] recommend a semi-empirical expression

$$C_P = 3Rn(1+k_1T^{-1}+k_2T^{-2}+k_3T^{-3})+(A+BT)+\Delta C_P$$

$R$ and $n$ are the gas constant and the number of atoms in the chemical formulae, respectively. $A$ and $B$ are calculated from thermal expansion coefficient and isothermal bulk modulus data. The $k_i$ are determined by fitting the measured low-temperature heat capacity data. $\Delta C_P$ is the departures from the $3Rn$ limit for some substances due to anharmonicity, and possibly electronic contributions or cation disordering.

**Table 2 Common high-T polynomial fit equations for $C_P$. The various empirical equations are commonly known by the author/date-citations in the table. The numbered references for each are given in the table footer**

| | | |
|---|---|---|
| $C_p = a + bT + \frac{c}{T^2}$ | (1) | Maier and Kelley (1932) |
| $C_p = a + bT + \frac{d}{T^{1/2}}$ | (2) | Chipman&Fontana (1935) |
| $C_p = a + bT + \frac{c}{T^2} + \frac{d}{T^{1/2}} + eT^2$ | (3) | Haas & Fisher (1976), Robie (1978) |
| $C_p = k_0 + \frac{k_{0.5}}{T^{1/2}} + \frac{k_2}{T^2} + \frac{k_3}{T^3}$ | (4) | Berman & Brown (1985) |
| $C_p = k_0 + bT + \frac{k_1}{T} + \frac{k_2}{T^2} + \frac{k_3}{T^3}$ | (5) | Fei and Saxena (1987) |
| $C_p = k_0 + k_{\ln} \ln T + \frac{k_1}{T} + \frac{k_2}{T^2} + \frac{k_3}{T^3}$ | (6) | Richet and Fiquet (1991) |
| $C_p = a + bT + \frac{c}{T^2} + \frac{d}{T^{1/2}}$ | (7) | Holland (1981) |

(1) [169]
(2) [170]
(3) [43, 171]
(4) [46]
(5) [168]
(6) [172]
(7) [173]

**Table 3 Summary of the merits of $C_P$ equations. [76, 77]**

| | **Eq. (1)** | **Eq. (3)** | **Eq. (4)** | **Eq. (5)** | **Eq. (6)** |
|---|---|---|---|---|---|
| Reference | [169] | [43],[171] | [46] | [168] | [172] |
| **Representation of measurements** | mediocre | excellent | good | excellent | excellent |
| **Low-temperature extrapolation [a]** | good | mediocre | bad | bad | bad |
| **High-temperature extrapolation of DSC measurements** | bad | bad | mediocre | mediocre | mediocre |
| **Drop calorimetry data up to 1800 K** | mediocre | bad | good | good | excellent |

[a] Extrapolation to lower temperatures for phases that are stable at high temperatures only.

Equation (7), by Holland [173], retains the extrapolatory merits of the Maier-Kelley equation while allowing superior representation of the measured heat capacities. However, the added flexibility of such a polynomial requires that one or two dummy data points at high temperatures (~linear extrapolation, low weight) be used in the fitting procedure (or constraint $b \geq 0$).

At low temperatures, $c_p \propto T^3$ (Debye limit, without effects like magnetic, spin, electronic contributions, see Table 1 for that); $c_p(0) \equiv 0$ in any case. We find that the type of equation best suited for a particular data set depends on the data and their accuracy and the temperature range. In practice, a case-by-case approach is best. One can start with the Maier & Kelley equation and add terms (all possible permutations) until (with the smallest number of terms) the fit does not improve any more (but does not start to oscillate, either), for example measured by a minimum in the Akaike information criterion [174] *AIC* for *n* data points, data uncertainties $\sigma_i$ and *k* parameters:

$$AIC = n\left(\ln(2\pi\chi^2 / n) + 1\right) + 2k + 2k(k+1)/(n-k-1)$$

$$\chi^2 = \sum_i \frac{c_{p,fitted,i} - c_{P,observed,i}}{\sigma_i}$$

### 1.2.5 Debye function approximation

The Debye-integral can be evaluated by numerical quadrature. It is not generally known that it is tractable analytically in terms of a finite sum of polylogarithms [175-177], see Appendix A.1. We found, however, that the evaluation of polylogarithms is computationally even more inefficient than quadrature. There are also rational approximants [178] and an analytic expression by [179], the first one being only accurate for $T/\theta_D$>0.1 and the latter deviating up to 6.5% at $T/\theta_D$<0.2. Padé approximations provide a convenient and very fast alternative [180, 181]. The Padé approximant that fits both the high- and low-*T* power law asymptotes of $C_v(T/\theta_D)$ and has additional terms in powers of 1/*T* in the numerator and denominator to fit the intermediate *T* range is:

$$C_V / R = \frac{\sum_m N_n / x^n}{\sum_m D_m / x^m}, \quad x = \frac{T}{\theta_D}, \quad m = n + 3$$

The approximant by Goetsch, Anand et al. [181] does not deviate from the normalized Debye function by more than $2 \times 10^{-4}$ at any *T*. By construction, the deviation goes to zero at both low and high *T*. The relative error has its maximum magnitude of 0.3% at low *T*. We have constructed an even more accurate Padé approximant (n=8, m=11; 17 independent coefficients; maximum relative deviation to true Debye = $5.755 \cdot 10^{-6}$ at $T/\theta_D$ ~0.1); full information is given in the appendix.

.

### 1.2.6 ‘Calorimetric’ Debye temperatures and their fit

The calorimetric Debye temperature $\theta_D$ *by definition* leads to the same $C_P$ (actually $C_V$) that was measured calorimetrically. Note that one can also define, after Grimvall [182] an 'entropy' Debye temperature, which leads to the measured $S(T)$; it is different from the calorimetric (heat capacity) Debye temperature we will discuss here.

Since we fit a function $C_{exp}$ that depends on $3nZ$ vibrational degrees of freedom to a model having a single free parameter $\theta_D$, it is obvious that we must pay a price, *i.e.*, $\theta_D$ will vary with the particular temperature at which the fit is done. Typical $\theta_D(T)$ curves for minerals are shown in Figure 5.

What is observed, for calorimetric Debye temperatures (from $C_V$!) is typically (e.g., [58], see Figure 7)

(i) Rapid fall from their $T \to 0$ limiting $\theta_D(0)$ plateau starting at a few K, to a minimum, $\theta_{D,min}$ at a temperature of the order of $\theta_D(0)/24$

(ii) Subsequent **rise** to $\theta_D(\infty)$ = *const.*, if anharmonic effects are negligible or have been removed.

Empirically, we found that $\theta_{cal}(T)$ can often be fitted very well with

$$\theta_{cal}(T) = a_1 \exp(-b_1 T) + a_2 \left[1 - \exp(-b_2 T^n)\right] + c \tag{12}$$

(Example Anorthite: $a_1$ = 440.2 K, $a_2$ = 647 K, $b_1$ = 0.1384 $K^{-1}$, $b_2$ = 0.002616 $K^{-n}$, c = 330.7 K, n = 1.19).

Often, the calorimetric Debye temperature shows a plateau for $T$<5 – 10 K, with a limiting value of $\theta_{D,0K}$ which can be estimated from the Debye temperature calculated from elastic constants (mean sound speed) and molar volume measured at room temperature[6], [183] see equation (1.2).

[6] Since $v_p$ and $v_s$, decrease with increasing T, $\theta_D$ also decreases with increasing T. For typical minerals, $\theta_D$ (acoustic, 0 K) is about 22 K higher than at 300 K (-0.07 K/K at 300 K , [183] Fig.1 p. 81.)

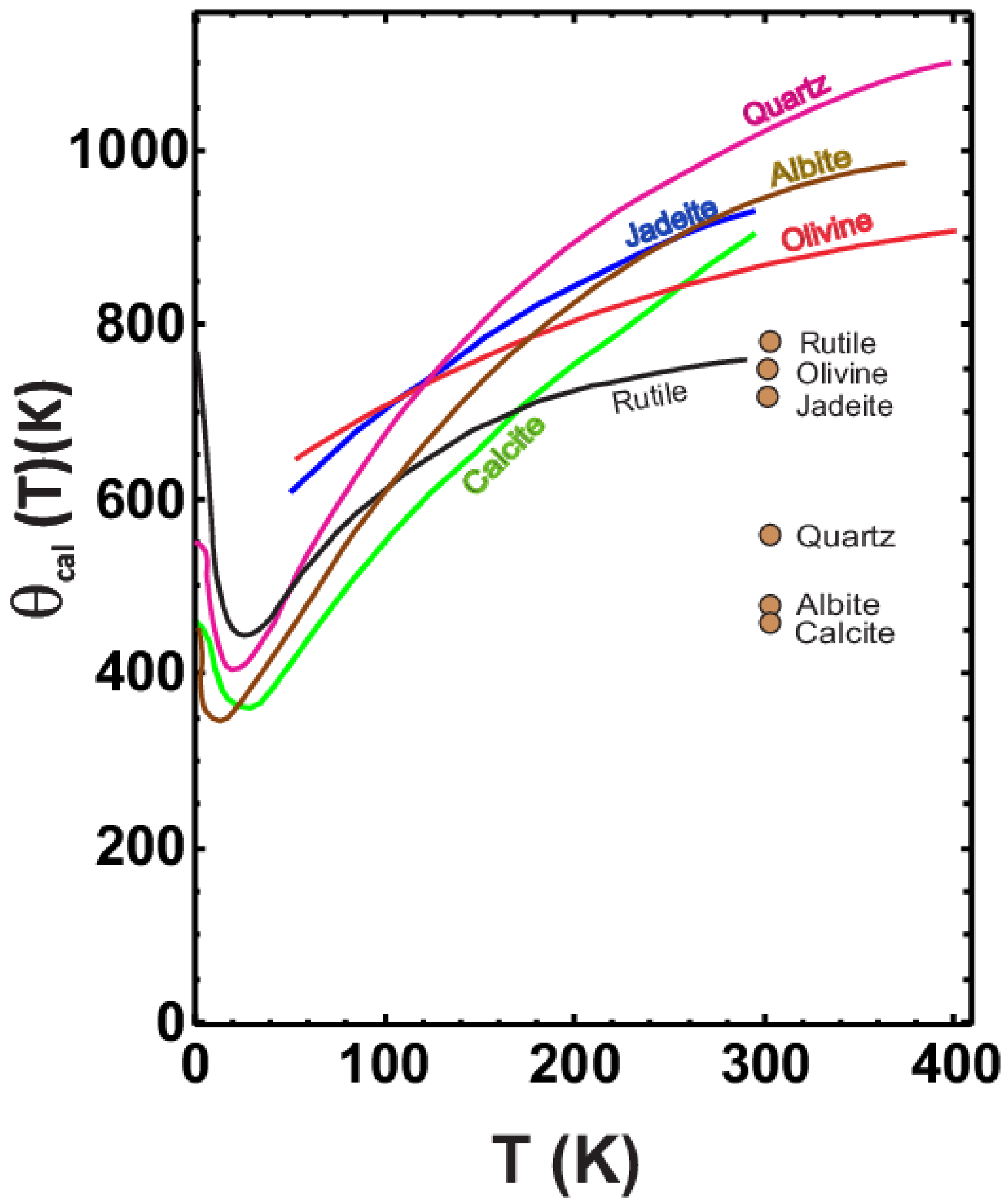

**Figure 7 Values of $\theta_D$ (*T*) for representative minerals. Room temperature elastic values $\theta_D$ are shown by circles at 300 K; they are assumed to apply, approximately, at low temperatures, T→0, as well. After [62].**

Our extended analytical model can capture this (paper II); however, this is usually only relevant for $T$<10 K.

### 1.2.7 Practical fitting of $C_P$ and *H* data, a brief review

Gurevich et al. [184] describe the practical fitting of $C_P$ and *H* data with a combination of Debye, Einstein and Kieffer functions plus an additive $b_0TC_V^2$ term for expansion and anharmonicity. Something similar is also advocated by Boerio-Goates et al. [185] and Yong et al. [186], who propose the 'DES-function'

$$C_P = nD(\theta_D / T) + mE(\theta_E / T) + n_S S(\theta_S / T)$$

where $D$ is the (3D) Debye-function, $E$ the Einstein-function, and a dramatic improvement of the fit could be achieved by including a two-level Schottky-function $S$

$$S = \frac{(\theta_S / T)^2 \exp(-\theta_S / T)}{\left[1 + \exp(-\theta_S / T)\right]^2} \quad \text{(in units of } R\text{)}$$

with the degeneracies of both levels set to one. No physical significance is attributed to the Schottky function. $n+m$ should approximate the number $Z$ of atoms per unit cell, and for many silicates the Debye temperature is of the order of 400 K, the Einstein temperature of the order of 500-1300 K, and the Schottky temperature $\theta_S$ is about 90 K . Maybe it is useful to add an expansion ($C_P$-$C_V$) multiplicative term like (1+A*$T$+B*$T^2$) which must be positive. Most experimental data can only be fit well if the temperature range is broken into at least a low-$T$ and a high-$T$ range, with individual fits joined at the best overlap point, typically around 150 K.

### 1.2.8 Adopted procedure to represent experimental data

We can determine, by weighted nonlinear least-squares, the parameters of $\theta_{cal}(T)$, equation (12) which allows to calculate $C_V$, and with the Ernst-Lindemann relation (parameter A), $C_P$ to which the experimental data are fitted.

We find that our model of $\theta_{cal}(T)$, evaluating the Debye function with a fast high-precision Padé approximant and with $C_P(C_V)$ calculated by (11) can very well fit experimental (lattice) heat capacities from 0 K to melting (decomposition) temperature to usually << 1% systematic accuracy. Of course, any $\Delta C_P$ from anomalies, lambda peaks etc. have to be smoothed/fitted/represented on a case-by-case basis, and this is the reason why we store the final $c_P(T)$ curves, at least for low temperatures, in our database in tabular form, not as coefficients of some correlation equation; for datasets with anomalies, we often smooth the merged, weighted experimental data with orthogonal polynomials (in log $c_P$ vs. log $T$). Since there is no good fit function or theoretical description for the $c_P(T)$ for all minerals over all temperatures, it is best to represent smoothed $c_P(T)$ data in (electronic) tables and to use 1D interpolation on these tables. We will show that a suitable (and very fast) interpolation of the tabular data is able to reproduce any original fitted or smoothed (merged) data set to very high accuracy in the complete temperature range. Usually though we join the low-temperature tabulated data with high-temperature polynomial correlations at some temperature close to 300 K where there is no jump in $c_P$ and at most a very small change of slope d$c_P$/d$T$.

**<u>Representation of $c_P$ in tables, temperature- and pressure sensitivity</u>**

We have performed numerous numerical experiments on the best temperature spacing and interpolation method (see paper II) – suffice to say here that $c_P$ is not very sensitive to temperature except of course near transition peaks, such that the effects of different temperature scales (e.g. ITS-90 vs. IPTS-68/48) are negligible and that the pressure dependence is negligible– so we can use $C_P$ measured at 1 bar for the surfaces of atmosphere-less bodies as well as for a rocky subsurface down to many km on a terrestrial planet.

## 1.3 Practicalities: atomic masses, mineralogical composition, conversion of mass, volume, mole fractions

Note that **atomic masses** are not constant in natural samples and have vastly different uncertainties [187, 188]. Since experimental $C_P$ determinations actually measure sample mass, thus, $c_P$, it is best to use the assumed molecular weight used in the original paper to convert back molar heat capacities $C_P$ to specific heats $c_P$. Otherwise, we use the IUPAC (2013) atomic masses of elements common in minerals [187]. Where IUPAC gives ranges, the most likely value for rocks and minerals has been used.

**Mineral composition**

The formulae of mineral can either be written as simplified formulas, e.g., $(Ca,Mg,Fe)(Mg,Fe)Si_2O_6$ indicating the possible substitutions (and vacancies □), the number of atoms in the substitution brackets not being specified, or as empirical formula. The latter can have fractional subscripts, but cations and anions must be charge-balanced; this is common for solid solutions. Example: $Ca_{0.25}Mg_2FeAl_{0.5}Si_{3.5}O_{10}(OH)_2$ (a saponite). We prefer, wherever possible, ideal formulas of the endmembers, even though these ideal compositions rarely exist in nature. For complex minerals, Hawthorne [189] discusses the correct and recommended endmember formula syntax.

The empirical chemical formula can be calculated from the elemental (or oxide) composition (mass fractions), which is rather straightforward for one mineral. For mixtures of minerals, normative analysis estimates the *idealised mineralogy* of a rock based on a quantitative chemical analysis according to the principles of geochemistry (i.e., likely reactions during formation). Normative mineral calculations can be done via the CIPW Norm [190] or other schemes [191]. Note that normative mineralogy is merely a calculation scheme based on predefined chemical entities (not all of which have mineralogical analogues) and thus provides an estimate of the hypothetical mineralogy of an igneous rock (a rock that crystallized from a melt). Its merit lies in the geochemical comparison of various igneous rocks suites, but it usually differs from the visually observable mineralogy (modal analysis).

Quantitative modal analysis, which we prefer, is used to determine the volumetric proportions of the minerals that make up the sample; it is estimated by identification and fractional area count of distinct minerals in thin sections, and gives volume fractions of these minerals in the sample. Densities of minerals $\rho_i$ need to be known to convert volume fractions $\varphi_i$ into mass fractions $w_i$.

To relate atomic percent $x$ and weight percent $w$, aka mole (atomic) fraction and mass fraction:

Mean molecular mass of mixture: $M = \sum x_j M_j = \left(\sum w_j / M_j\right)^{-1}$ (13)

Mole fraction (at-%) $x_j = (w_j / M_j) / \sum w_j / M_j = w_j \dfrac{M}{M_j}$ (14)

Mass fraction (weight-%) $w_j = x_j M_j / M$ (15)

Relationship between volume fraction $\varphi$ (vol-%) and mass fraction $w$ (mass-%):

$$w_i = \frac{\rho_i}{\rho^0}\varphi_i \quad \text{with} \quad \rho^0 = \sum \varphi_i \rho_i \qquad (16)$$

Note that there is a occasionally confusion and inaccurate use of various concentration units, in particular where volume and mass quantities are mixed. Volume quantities depend on temperature (negligible for solid minerals, though) and on whether ideal or non-ideal mixtures are assumed (i.e., whether the volume of the solution after mixing is used as the reference or the sum of volumes of constituents prior to mixing).

We will use [$p,T$ = const]:

Volume fraction $\varphi_j = \dfrac{V_j}{V_0}, \quad V_0 = \sum V_j, \quad \sum \varphi_j = 1$

Volume concentration $\sigma_j = \dfrac{V_j}{V}$ , $V$ is the volume of the mixture. $V_j$ : Volume of solute *prior* to mixing. $\rho$ is the density of the mixture.

Then ($n$: total number of mol):

$V^E = V - V_0$ excess volume (usually given as molar excess volume $V^E / n$, can be up to about ±1 cm³/mol), $V_m = M / \rho$ (m³/mol).

$\sigma_j \neq \varphi_j$ in general (equality only for ideal solutions).

$$\sigma_j = w_j \frac{\rho}{\rho_j}, \quad w_j = \sigma_j \frac{\rho_j}{\rho}$$
$$\varphi_j = \frac{w_j / \rho_j}{\sum w_j / \rho_j}, \quad w_j = \varphi_j \frac{\rho_j}{\sum w_j / \rho_j} = \varphi_j \frac{\rho_j}{\sum \varphi_j \cdot \rho_j} \qquad (17)$$

Mass concentration $\gamma$ is defined as the mass of a constituent $m_j$ divided by the volume of the mixture $V$:

$$\gamma_j = \frac{m_j}{V}, \quad w_j = \frac{\rho_j}{\rho}$$

For the automation of stoichiometric calculations (e.g., for thermal alteration decomposition reactions or composition from oxide content), see Anderson and Bjedov [192]; we use the convenient MATLAB® tool, *stoichtool* [193].

## 1.4 Experimental methods and their accuracies

Background: mineralogists are interested in low-temperature $C_P$ data, because these are needed to calculate the zero-point (i.e., third-law) entropy $S^0$ : $S^0 - S^{T=0K} = S^0 = \int_0^{298.15K} \frac{C_P}{T} dT$ .

In terms of silicate minerals, it has been standard practice prior to about 2005, to measure their low temperature $C_P(T)$ behaviour (often only once) using adiabatic calorimetry. Nowadays, a number of different devices is available that allow $C_P$ coverage from ~0 K to roughly 2,000 K.

It is usually sufficiently accurate, for $S^0$, to measure $C_P$ down to about 5-20 K and extrapolate $C_P/T$ vs. $T^2$ (approximately straight line for most minerals and metals) to perform the $C_P$ integration for the range 0 K to lowest measured temperature. Lower temperatures than 5 or 10 K are sometimes needed, if there are magnetic transitions in this range. The bulk of mineral $c_P$ data is available down to 10 K or (with lesser accuracy) 5 K since ca. 1955, but often only to ~15 K. Before ca. 1950, the limit was typically 50 K. Accurate data for minerals appear since about 1935. Many mineral $c_P$ data have been measured or re-measured since 1985. A good survey, reference and recommendations for the various calorimetric techniques for solids can be found in [39, 194, 195]. There are various calorimetric standard substances to calibrate calorimeters, such as corundum ($Al_2O_3$, *aka* alumina, synthetic sapphire, SRM720), benzoic acid and copper; we have re-analysed and re-fitted all available data on these standard substances (paper II). Their relative $c_P$ accuracy can be as low as 0.05% (corundum at medium temperatures, [196]). In the following, we briefly present the main methods for accurate $c_P$ measurements, in particular to give the reader a feeling on the typical experimental uncertainties. See also [17, 115] for a more detailed overview.

### 1.4.1 Low-temperature adiabatic calorimetry ('low-TAC') for up to ~340±40 K

Low-temperature adiabatic calorimetry, which is typically carried out between about 5 and 400 K, is capable of delivering an experimental precision of about 0.1% in the heat capacity. It requires rather large samples (order of 10-30 g) and is complicated and time-consuming [197]. This method has been used extensively to measure the heat capacity of silicates and oxides and the compilation of Robie & Hemingway [44] summarizes results obtained over many years of study. Adiabatic calorimeters are not available commercially and there are only a few laboratories world-wide that are capable of making such measurements. Many of the data found in [44] derive from investigations made at the U.S. Bureau of Mines (U.S.B.M.) from 1940 to about 1970. Real accuracies, including systematic errors and non-reproducibility of samples, tend to be rather 5% at 10 K, 2% at 15 K, 1% at 20 K and ~0.2% above 40 K [198].

### 1.4.2 Heat-pulse relaxation calorimetry (PPMS) for 2-400 K

New techniques and devices for small sample calorimetry (in the mg range) were developed in the 1970s. Based on this and later work [199] Quantum Design® [200] constructed a commercial relaxation calorimeter (available since ca. 1998 – 2003), implemented as the heat capacity option of the Physical Properties Measurement System (PPMS) [200]. Technical details of the instrument, as well its measuring procedures and performance, have been described in detail [201, 202]. PPMS measurements can be automated to a large extent; the accuracy is comparable to that of DSC. Older PPMS measurements (before 2005) often have higher uncertainties at the low temperatures (e.g., 1% above 10 K, 5% at 10 K) and failed at 1$^{st}$ order phase transitions (due to automatic evaluation of the raw data with 2 relaxation constants). Kennedy et al., 2007 [203] showed that the accuracy of heat capacity determinations using the QD PPMS can be within 1% for 5 K < $T$ < 300 K and 5% for 0.7 K < $T$ < 5 K under ideal conditions. Otherwise, significantly higher uncertainties were quoted [203]. Dachs and Benisek [204] found that the accuracy of $C_P$ data obtained from powder measurements using the PPMS is generally lower compared to single crystal measurements. It is 1–2% for not too low temperatures and critically depends on sample geometry and sample mass, similarly to what Kennedy et al. [203] found. At best, the accuracy that could be obtained for powders calibrated to DSC at RT: 10% @ 20 K, 3% @ 40 K, 1% for >60 K.
However, the methods continues to improve [17], e.g., improvements in accuracy on loose powders can be achieved by undertaking measurements on powder samples wrapped and pressed into thin and light Al-foil holders weighing ~5.5 mg (see also [205]) such that nowadays the accuracy of low-TAC can be similar, PPMS even better for T<15 K [17].

Since ca. 2000, more and more data appear down to 1.9 K (lowest temperature for PPMS; recently 0.4 K with $^3$He cooling). The required sample amount is typically 1–100 mg. Very careful handling is required for good accuracy <60 K [206]. For further details of PPMS techniques, see [69, 201, 203].

### 1.4.3 Differential scanning calorimetry (DSC)

There are two basic types of DSC methods: heat flux and power compensation DSC. In heat flux DSC calorimeters, the sample and the reference are heated in the same furnace while measuring the temperature difference between sample and reference. The temperature difference is converted to a difference in power using a calibration. Such calorimeters can be operated between ca. 100 and 1800 K. In spite of this large temperature range, these calorimeters are not often used for measuring the heat capacity, because of their only moderate accuracy. In power compensation DSC, sample and reference are heated separately by micro furnaces. These are maintained at the same temperature during heating while measuring the difference in heating power (heat flow). A power compensation DSC [207-210] can be operated between ca. 100 K and 1000 K with better accuracy (~1-2%), enabling rather precise heat capacity measurements; commercial DSCs are widely used in industry and science and are often very conveniently automated. DSC techniques for very high (e.g., up to 1500 K [61, 211]) and low (down to ca. 1 K [61]) have been developed, but most measurements with commercial instruments are conducted within the range 100 – 700 K.
A principal disadvantage of DSC is [39] that because it is so easy to use it is also very easy to abuse: It has been said [194] that the very ease of obtaining data by DSC can lead to work which is of questionable accuracy if the operator fails to observe many necessary and rigorous principles. For accurate measurements, the 3-curve method with (not necessarily overlapping) temperature scans (step-scanning) and baseline postprocessing is recommended, see [204, 212, 213].

Note that DSC measurements are inherently dynamic, as the sample temperature is a (linear) function of time; thermodynamic equilibrium is never attained, meaning that in practice (due to finite thermal conductivities) the sample is never at a uniform temperature. With typical instruments, heating rates of ~10 K/min are used; with a sample mass of ~30 mg in powdered form, typical thermal inhomogeneities in the sample are of the order of ~1 K, thus broadening features (like an actually sharp peak) accordingly.

### 1.4.4 Drop calorimetry

Drop calorimetry (e.g., [157]) is used to measure the heat capacity at temperatures higher than ca. 900 K. In this calorimetric technique, a sample (equilibrated to e.g., room temperature, $T_1$) is dropped into the calorimeter, whose high temperature ($T_2$) is controlled by a surrounding furnace. The small temporary temperature decrease generated by dropping the sample into the calorimeter is recorded as a function of time. Integrating these data and applying a calibration yields the absolute heat content (enthalpy $H$ change) of the sample when heated from $T_1$ to $T_2$. Heat capacity is then calculated from differentiating the $H(T)$ curve; the difficulty here is to estimate reasonable uncertainties of $C_P$.

A simple and elegant but not very accurate variation of drop calorimetry for the non-destructive measurement of the $c_P$ (at ~180 K) of meteorites has been devised by Consolmagno et al. [24] using liquid nitrogen vaporization; basically, the enthalpy difference between 77 K and 'room temperature' is determined, with random and systematic uncertainties of ~2% and ~4%, respectively.

### 1.4.5 Important notes for experimental $c_P$ data

**Homogeneity.** PPMS and DSC require only ~3-30 milligram of sample. Consequently, high purity and homogeneity of the sample are required for the measurement to be representative of the whole sample.

**Curvature correction.** The true heat capacity at temperature $T$ is given by

$$C_P = \lim_{\Delta T \to 0} \Delta H / \Delta T = dH / dT$$

The result of classical, stepwise measurements is the mean heat capacity, $C_{P,mean} = \Delta H / (T_2 - T_1)$, associated with the mean temperature of the interval, $T_m = (T_1 + T_2)/2$. Deviation from linearity in the $C_P$ versus $T$ curve will therefore require adjustment of the mean heat capacity by a curvature correction [198] to yield the true heat capacity at $T_m$, or, equivalently, a correction to $T_m$. The curvature correction can often be neglected if $\Delta T$ is only a few K and if there is no transition peak.

**Sample preparation, humidity control.** Handling of the sample in humid laboratory air can change the (sorption or crystal water) content of a sample. Dehydrated phyllosilicates can adsorb of the order of 10% terrestrial water rapidly, which changes specific heat significantly. Drying the sample to a defined state is mandatory, and frost depositing on cold sample or calorimeter surfaces must be avoided.

**Premelting.** Most substances studied today by accurate calorimetric methods are pure enough to render the effect of impurities on the observed heat capacity data negligible except in the region just below the melting point. In some minerals, an abnormal increase in enthalpy and $C_P$ well below the melting point has been observed [172], which can be caused by structural changes, by Frenkel thermal vacancies [214] or by classical impurity premelting.

**The effect of temperature uncertainty** (including the temperature scale used, ITPS-27, -48, -68, ITS-90, and various low temperature extensions for example) is significant only at low temperatures (or in the vicinity of first-order or sharp lambda peak). The absolute $T$ uncertainty leads to typical $C_P$ uncertainties of the order of 1-2% at 1 K, 0.3% at 10 K, and ~0.01% at $T$>100 K. Note that for commercial DSC at high temperatures, the temperature uncertainty including non-isothermality of the sample can be up to 1 K [215] – the relative error introduced by this is <1% for $T$>100 K and <0.2% for $T$>300 K. This leads us to

**Temperature resolution of $C_P(T)$ curves**: a temperature range in the sample volume is inherent for the dynamic (PPMS and DSC) techniques. Conversely, $C_P(T)$ is averaged, in the sample volume, over a finite temperature range $\Delta T$ in low-TAC or drop-calorimetry; $\Delta T$ can be chosen by the experimenter, but of course there is a trade-off with noise/accuracy. This means that sharp $C_P(T)$ peaks invariably get distorted (lower peak heights and broader peaks, shift of peak temperature), cf. [210, 216-218].

**Differences between natural and synthetic crystals -** It is possible that slight differences in $C_P(T)$ behaviour could exist between some natural and synthetic crystals. The latter could be more structurally 'disordered' than their natural analogues due to the shorter times and often higher temperatures associated with their crystallization. Grossular $Ca_3Al_2Si_3O_{12}$ has been well studied [17]; the relative $C_P$ difference at low-$T$ (<100 or 200 K) for various natural grossular versus synthetic grossular reaches is small but measurable, reaching ~10% at 40 K and ~20% at 20 K. Internal stresses and strains in natural materials are also known to influence transition features. Not to be confused with the effect of compositional zoning in a sample volume. For olivines, the $C_P(T)$ behavior of the natural, well crystallized forsterite $Fo_{0.894}Fa0_{.106}$ and a crystalline synthetic $Fo_{90}Fa_{10}$ sample are in excellent agreement between about 7 and 300 K [219]. It would appear that any phonon difference arising from possible variations in $Fe^{2+}$-Mg order-disorder are minimal to nil despite their contrasting crystallization histories and small differences in chemistry.

# 2 Background - Description of minerals and compounds

A mineral or mineral species is a naturally occurring, macroscopically homogeneous solid chemical compound with a fairly well-defined chemical composition and a specific crystal structure. Traditionally amorphous substances which fulfil the other criteria are included but called mineraloids. Minerals can be elements, organic or inorganic compounds – some minerals in fact are among the most complex inorganic compounds known. Of the ~6000 mineral species recognized today, most are inorganic and are silicates. Many minerals form solid solution (substitution) series, 'joins', and usually we seek the physical properties of the endmember minerals of a series. Note that e.g., 'olivine' is not a mineral, but a mineral group; but an olivine with a specified composition, like $Fo_{90}Fa_{10}$, is a mineral. Solid solutions can be considered in terms of three categories: complete solid solutions without structural ordering, solid solutions with structural ordering, and partial solid-solutions. The recommended mineral nomenclature in each of these categories is discussed in [220].

A solid solution may be defined as a homogeneous phase composed of different chemical substances whose concentrations may be varied without the precipitation of a new phase. This variation can be classified in three types: substitutional (by far the most important, includes multicomponent and coupled substitutions), interstitial (example: tridymite, $SiO_2$, towards nepheline, $NaAlSiO_4$) and omissional (example: wüstite $Fe_{1-x}O$).

In this paper, we include condensed gases like methane and carbon dioxide because on some cold outer solar system bodies they are believed to form the bulk of the surface material. We also include the enigmatic tholins, 'complex abiotic organic gunk' [14] for the same reason.

Two common classifications, ‘Dana’ and ‘Strunz’ in short, are used for minerals; both rely on composition, specifically with regards to important chemical groups, and structure. Dana, as of 1997, is in its eighth edition [221]. The less commonly used (Nickel-)Strunz classification [222] is based on the Dana system, but combines both chemical and structural criteria, the latter with regards to distribution of chemical bonds.

The International Mineralogical Association (IMA) is the generally recognized standard body for the definition and nomenclature of mineral species. The IMA Database of Mineral Properties https://rruff.info/ima/**,** is representing the ‘official’ IMA list of minerals on the web. For detailed mineral descriptions and properties, it is linked to the following useful websites:

- Handbook of Mineralogy, pdf online version, http://www.handbookofmineralogy.org/ [223]
- American Mineralogist Crystal Structure Database, https://rruff.geo.arizona.edu/AMS/amcsd.php [224]
- RRUFF™ database, https://rruff.info/, https://rruff.geo.arizona.edu/doclib/hom for pdf summaries [225]
- mindat.org page, the world's largest open database of minerals, rocks, meteorites and the localities they come from. Mindat.org is run by the not-for-profit Hudson Institute of Mineralogy [226]
- webmineral.com, © 1997-2014 by David Barthelmy [227]

Crystallographic, thermal expansion and elastic data for many minerals have been compiled in [183].

In this chapter we describe the most common/important mineral groups that, to our current knowledge, occur in extra-terrestrial regoliths. What we present here, in an order only loosely resembling the IMA classification, is (with some exceptions) not new but rather a condensed and simplified textbook (e.g., [228]) knowledge. In fact we often employ the same sources as *Wikipedia, The Free Encyclopedia*, https://en.wikipedia.org/ (which is a usually a high-quality source if it comes to minerals). We believe that this condensed background is useful for the non-mineralogist, since the terminology of mineral species can be quite complex (in particular for the so-called clay minerals, or phyllosilicates), and there are often conventional/historical names in addition to the official ones. Even the official mineral names can be quite exotic! For minerals with variable compositions caused by solid solutions we will identify and treat normally only the (idealized) endmembers. We will focus, besides definition and composition, on describing the properties relevant for specific heat (polymorphs, phase transitions, or and dehydration/decomposition/melting at elevated temperatures).

In literature describing the mineralogy of, e.g., meteorites, the reader will often encounter component’s mineral names which are not in our database (and won’t ever be). One the one hand, this has historical reasons, some minerals have had different names until the IMA came to an official recommendation; yet mostly, analytic methods do not well resolve composition within solid solution series, and there are numerous names for minerals of intermediate composition between (2, 3 or more) endmembers, some of them obsolete, some very common. We mention, for example, feldspars (plagioclase, anorthoclase, oligoclase, andesine, labradorite, bytownite), pyroxenes (augite, pigeonite, hypersthene, bronzite), olivine, hornblende and many others[7].

Phyllosilicates belong to the most complex inorganic compounds. It is therefore not surprising, that for example in modal analysis, the exact empirical formulas of the observed phyllosilicates very often cannot be given, and only a broad categorization into ‘saponite’ and ‘serpentine’ is done. What exactly these terms mean is not uniform in literature and depends also on context information, like iron content.

Table 4 below, an abridged version of a much more detailed ‘master table’, given an overview of our minerals & compounds database to date. The entries are alphabetically sorted alphabetically (1) after mineral group, (2) after sub-group (if any; not shown), (3) after name.

[7] The old name ‘**hypersthene**’ is an orthopyroxene with ~30% En and rest Fs. **Bronzite** is also a member of the pyroxene group of minerals, belonging with enstatite and hypersthene to the orthorhombic series of the group. Rather than a distinct species, it is really a ferriferous variety of enstatite. $(Mg,Fe)SiO_3$, the iron(II) oxide ranges from about 12 to 30%. **Hornblende** is an IMA-CNMNC defined amphibole root-name with strict chemical boundaries but is often confused with the more loosely defined ‘hornblende’, which is a term used by petrologists and mineral collectors for any black undefined amphibole in the calcium (and sometimes in the sodium-calcium) subgroups [226]

**Table 4 Abridged database-overview. Here, we list all minerals and substances currently contained in the $c_P$ database (full version, incl. molar mass, molar volume, theoretical density, references for $c_P$ (low-T, high-T), peaks if any (temperature, type, enthalpy), melting and/or decomposition temperature (or triple point), melting (or sublimation) enthalpy, comments) see paper II SOM (and, for future updates, in the data repository). The formula given here is the actually used formula for the $c_P$ in the database if different from the ideal endmember formula. The abbreviation, which is also used for the software and data file names, follows [229] wherever possible; in some cases, we use the chemical formula or invented a new abbreviation. The densities are important for the conversion of modal mineralogy into mass fractions and are taken from various sources ([226, 227, 230]); they refer to a temperature of 25°C, except in the case of ices, where they refer to a temperature midway between 0 K and triple point temperature.**

| # | Abbr | Name | Group | Formula | ρ (g/cm³, 25°C) |
|---|---|---|---|---|---|
| 1 | Fact | Ferro-actinolite | Amphibole | $Ca_2Fe_5Si_8O_{22}(OH)_2$ | 3.34 |
| 2 | Tr | Tremolite | Amphibole | $Ca_2Mg_5Si_8O_{22}(OH)_2$ | 2.98 |
| 3 | Cum | (Magnesio-)Cummingtonite | Amphibole | $\square Mg_7Si_8O_{22}(OH)_2$ | 2.97 |
| 4 | Gru | Grunerite | Amphibole | $\square(Fe^{2+})_7Si_8O_{22}(OH)_2$ | 3.53 |
| 5 | Ath | Anthophyllite | Amphibole | $\square Mg_7Si_8O_{22}(OH)_2$ | 2.86 |
| 6 | Fath | Ferro-anthophyllite | Amphibole | $\square(Fe^{2+})_7Si_8O_{22}(OH)_2$ | 3.59 |
| 7 | Cal | Calcite | Carbonates | $CaCO_3$ | 2.71 |
| 8 | Dol | Dolomite | Carbonates | $CaMg(CO_3)_2$ | 2.87 |
| 9 | Mgs | Magnesite | Carbonates | $MgCO_3$ | 3.01 |
| 11 | Sd | Siderite | Carbonates | $FeCO_3$ | 3.94 |
| 10 | Na2CO3 | Sodium carbonate, anh. | Carbonates | $Na_2CO_3$ | 2.54 |
| 12 | Ab | Albite | Feldspars | $NaAlSi_3O_8$ | 2.62 |
| 13 | An | Anorthite | Feldspars | $CaAl_2Si_2O_8$ | 2.76 |
| 14 | Or | Orthoclase | Feldspars | $KAlSi_3O_8$ | 2.56 |
| 15 | Nph | Nepheline | feldspathoid | $(Na_3K)Al_4Si_4O_{16}$ | 2.59 |
| 16 | Alm | Almandine | Garnets | $Fe_3Al_2(SiO_4)_3$ | 4.32 |
| 17 | Adr | Andradite | Garnets | $Ca_3Fe_2Si_3O_{12}$ | 3.86 |
| 18 | Grs | Grossular(ite) | Garnets | $Ca_3Al_2(SiO_4)_3$ | 3.59 |
| 19 | Prp | Pyrope | Garnets | $Mg_3Al_2(SiO_4)_3$ | 3.58 |
| 20 | Sps | Spessartine | Garnets | $Mn^{2+}{}_3Al_2(SiO_4)_3$ | 4.19 |
| 21 | Uv | Uvarovite | Garnets | $Ca_3Cr_2(SiO_4)_3$ | 3.83 |
| 22 | Hl | Halite | Halides | NaCl | 2.16 |
| 23 | Syl | Sylvite | Halides | KCl | 1.99 |
| 24 | CO2 | Carbon dioxide | Ices | $CO_2$ | 1.10 |
| 25 | CO | Carbon monoxide | Ices | CO | 0.92 |
| 26 | C2H6 | Ethane | Ices | $C_2H_6$ | 0.60 |
| 27 | C2H5OH | Ethanol | Ices | $C_2H_5OH$ | 1.03 |
| 28 | Ice | Ice Ih | Ices | $H_2O$ | 0.92 |
| 29 | CH4 | Methane | Ices | $CH_4$ | 0.50 |
| 30 | CH3OH | Methanol | Ices | $CH_3OH$ | 0.70 |
| 31 | N2 | Nitrogen | Ices | $N_2$ | 0.95 |
| 32 | Al | Aluminium | Metals | Al | 2.70 |
| 33 | Cu | Copper | Metals | Cu | 8.92 |
| 34 | Fe | Iron | Metals | Fe | 7.88 |
| 35 | Kamc | Kamacite | Metals | α FeNi | 7.90 |
| 36 | FeNi | Meteoritic iron | Metals | FeNi | 8.00 |
| 37 | Ni | Nickel | Metals | Ni | 8.91 |
| 38 | Tae | Taenite | Metals | γ FeNi | 8.30 |
| 39 | Tta | Tetrataenite | Metals | L10-FeNi | 8.30 |
| 42 | Fa | Fayalite | Olivines | $Fe_2SiO_4$ | 4.39 |
| 43 | Fo | Forsterite | Olivines | $Mg_2SiO_4$ | 3.21 |
| 44 | Hem | Hematite | Ox-/Hydroxides | $Fe_2O_3$ | 5.28 |
| 45 | Ilm | Ilmenite | Ox-/Hydroxides | $FeTiO_3$ | 4.79 |
| 46 | Mgh | Maghemite, Fe(II)-deficient magnetite | Ox-/Hydroxides | $\gamma$-$Fe_2O_3$ | 4.86 |
| 47 | Mag | Magnetite | Ox-/Hydroxides | $Fe_3O_4$, $Fe^{2+}(Fe^{3+})_2O_4$ | 5.20 |
| 48 | Wüs | Wüstite | Ox-/Hydroxides | FeO | 5.99 |
| 49 | Fhy | 2-line ferrihydrite | Ox-/Hydroxides | FeOOH • $0.027H_2O$ | 3.80 |
| 50 | Aka | Akaganéite | Ox-/Hydroxides | β FeOOH • $0.652H_2O$ | 3.52 |
| 51 | Cor | Corundum | Ox-/Hydroxides | $Al_2O_3$ | 3.99 |
| 52 | Bse | Bunsenite | Ox-/Hydroxides | NiO | 6.67 |
| 53 | Lm | Lime, calcium oxide anh. | Ox-/Hydroxides | $CaO$ | 3.34 |
| 54 | Per | Periclase | Ox-/Hydroxides | $MgO$ | 3.58 |
| 55 | Rt | Rutile | Ox-/Hydroxides | $TiO_2$ | 4.24 |
| 56 | Fap | Fluorapatite | Phosphates | $Ca_{10}(PO_4)_6F_2$ | 3.20 |
| 57 | Hap | Hydroxyapatite | Phosphates | $Ca_{10}(PO_4)_6(OH)_2$ | 3.16 |
| 58 | Chm | Chamosite | Phyllosilicates | $(Fe^{2+},Mg)_5Al(AlSi_3O_{10})(OH)_8$ | 3.13 |
| 59 | Clc_L | Chlinochlore (Mg-Chl) | Phyllosilicates | $(Mg_{0.097}Fe^{2+}{}_{0.903})_5Al(Si_3Al)O_{10}(OH)_8$ | 2.72 |
| 60 | Clc_M | Clinochlore (Fe-Chl M) | Phyllosilicates | $(Mg_{0.594}Fe^{2+}{}_{0.406})_5Al(Si_3Al)O_{10}(OH)_8$ | 3.32 |
| 61 | Clc_W | Clinochlore (Fe-Chl W) | Phyllosilicates | $(Mg_{0.589}Fe^{2+}{}_{0.411})_5Al(Si_3Al)O_{10}(OH)_8$ | 3.05 |
| 62 | Kln | Kaolinite | Phyllosilicates | $Al_2Si_2O_5(OH)_4$ | 2.62 |
| 63 | Prl | Pyrophillite | Phyllosilicates | $Al_2Si_4O_{10}(OH)_2$ | 2.78 |
| 64 | Tlc | Talc | Phyllosilicates | $Mg_3Si_4O_{10}(OH)_2$ | 2.78 |
| 65 | Ilt | Illite (group) | Phyllosilicates | $K_{0.65}Al_{2.0}Al_{0.65}Si_{3.35}O_{10}(OH)_2 \bullet n(H_2O)$ | 2.80 |
| 66 | Ms | Muscovite 2M1 | Phyllosilicates | $KAl_2(AlSi_3O_{10})(OH)_2$ | 2.83 |
| 67 | Ann | Annite | Phyllosilicates | $KFe_3^{2+}AlSi_3O_{10}(OH)_2$ | 3.36 |
| 68 | Eas | Eastonite | Phyllosilicates | $KAlMg_2(Si_2Al_2)O_{10}(OH)_2$ | 2.59 |
| 69 | Phl | Phlogopite | Phyllosilicates | $KMg_3AlSi_3O_{10}(OH)_2$ | 2.79 |
| 70 | Sid | Siderophyllite | Phyllosilicates | $KFe^{2+}{}_2Al(Al_2Si_2O_{10})(OH)_2$ | 3.18 |
| 71 | Atg | Antigorite/Lizardite, Chrys-otile | Phyllosilicates | $Mg_3Si_2O_5(OH)_4$ | 2.59 |
| 72 | Plg | Attapulgite = Palygorskite | Phyllosilicates | $Mg_{1.5}Al_{0.5}Si_4O_{10}(OH) \bullet 4(H_2O)$ | 2.40 |
| 73 | Brh | Bertherine | Phyllosilicates | $(Fe_{2.5}Al_{0.5})[Si_{1.5}Al_{0.5}O_5](OH)_4$ | 3.00 |
| 74 | Vrm | Vermiculite | Phyllosilicates | $Mg_{0.7}(Mg,Fe,Al)_6(SiAl)_8O_{20}(OH)_4 \cdot 8H_2O$ | 2.26 |
| 75 | Bei | Beidellite | Phyllosilicates | $Na_{0.5}Al_{2.5}Si_{3.5}O_{10}(OH)_2 \bullet (H_2O)$ | 2.00 |
| 76 | Mnt | Montmorillonite | Phyllosilicates | $(Na,Ca)_{0.33}(Al,Mg)_2(Si_4O_{10})(OH)_2 \cdot n(H_2O)$ | 1.85 |
| 77 | Non | Nontronite | Phyllosilicates | $Na_{0.3}Fe_2((Si,Al)_4O_{10})(OH)_2 \cdot n(H_2O)$ | 2.25 |
| 78 | Sap | Saponite | Phyllosilicates | $Ca_{0.25}(Mg,Fe)_3((Si,Al)_4O_{10})(OH)_2 \cdot n(H_2O)$ | 2.27 |
| 79 | Ks | Kuishiroite (Ca-Tschermak) | Pyroxenes | $CaAlAlSiO_6$ | 3.43 |
| 80 | Aeg | Aegirine (Acmite) | Pyroxenes | $(NaFe^{3+})[Si_2O_6]$ | 3.58 |
| 81 | Wo | Wollastonite | Pyroxenes | $Ca_2Si_2O_6$ | 2.91 |
| 82 | Di | Diopside | Pyroxenes | $CaMgSi_2O_6$ | 3.28 |
| 83 | En | Enstatite | Pyroxenes | $Mg_2Si_2O_6$ | 3.19 |
| 84 | Fs | Ferrosilite | Pyroxenes | $Fe_2Si_2O_6$ | 4.00 |
| 85 | Hd | Hedenbergite | Pyroxenes | $CaFeSi_2O_6$ | 3.65 |
| 86 | Lch | Lechatelierite | Silicates | $SiO_2$, amorphous | 2.20 |
| 87 | Qz | Quartz | Silicates | $SiO_2$ | 2.65 |
| 89 | Nams | Sodium-Metasilicate | Silicates | $Na_2SiO_3$ | 2.61 |
| 91 | Spl | (Magnesio-)Spinel | Spinels | $MgAl_2O4$ | 3.58 |
| 92 | Chr | Chromite | Spinels | $FeCr_2O_4$ | 5.10 |
| 93 | Hc | Hercynite | Spinels | $FeAl_2O_4$ | 4.34 |
| 94 | Anh | Anhydrite | Sulfates | $CaSO_4$ | 2.96 |
| 95 | Esm | Epsomite | Sulfates | $MgSO4 \bullet 6.868H_2O$ | 1.68 |
| 96 | Gp | Gypsum | Sulfates | $CaSO_4 \bullet 2H_2O$ | 2.31 |
| 97 | FeSO4 | Iron(II)sulfate, anh. | Sulfates | $FeSO_4$ | 3.65 |
| 98 | MgSO4 | Magnesiumsulfate anh. | Sulfates | $MgSO_4$ | 2.66 |
| 99 | Pn | Pentlandite | Sulfides | $(Fe,Ni)_9S_8$ | 4.80 |
| 100 | Py | Pyrite | Sulfides | $FeS_2$ | 5.01 |
| 101 | Po | Pyrrhotite | Sulfides | $Fe_{0.9}S$ ; $Fe_{1-x}S$, x=0 – 0.2 | 4.63 |
| 102 | Tro | Troilite | Sulfides | FeS | 4.83 |
| 40 | Dia | Diamond | C-rich matter | C (cubic) | 3.53 |
| 41 | Gr | Graphite | C-rich matter | C (hexagonale) | 2.23 |
| 103 | ICOM | Complex organic matter | C-rich matter | C,H,O,N, ... | 0.90 |
| 104 | Coal | sub-bitumous coal | C-rich matter | C,H,O,N, ... | 1.35 |
| 105 | ANG | Apiezon® N grease | Other | $C_{111}H_{208}$ | 0.91 |
| 106 | Bza | Benzoic acid | Other | $C_7H_6O_2$ | 1.27 |
| 107 | PE | Polyethylen | Other | $(C_2H_4)_n$, polymeric | 0.92 |
| 108 | Adh | Ammonia dihydrate | Other | $NH_3 \bullet 2H_2O$ | 0.99 |
| 109 | Cem | Cementite, iron carbide | Other | $Fe_3C$ | 7.69 |
| 110 | Coh | Cohenite, FeNi carbide | Other | $(Fe,Ni)C_3$ | 7.65 |
| 111 | WIh | Hydrate water/icelike | Other | $H_2O$ | 0.92 |
| 112 | WZeo | Hydrate water/zeolithic | Other | $H_2O$ | 0.92 |

**Abbreviations**

| | |
|---|---|
| tr | triple point |
| anh | anhydrous |
| P | density |
| ☐ | (atom) vacancy |

## 2.1 Minerals in the solar system

Here, we briefly have a look which mineral could be important in the solar system, apart from those known to be common (at a mass fraction >1% or so) in meteorites. For the latter, see e.g., the excellent review on meteoritic minerals by Rubin & Ma (2017) [231]; about 435 mineral species have been identified in meteorites, of which only a few are significant for $c_P$.

**Table 14 Probable rocks and minerals on the surface of Mercury, Venus, and Mars, compilation from various references**

<table>
<tr><th>Body</th><th>Rocks and minerals</th></tr>
<tr><td>Mercury</td><td>
Fe, Ti-rich anorthosites [228];
Plagioclase, forsterite, enstatite, some graphite and sulfides. Poor in $Fe^{2+}$ [232]
Fe, Ti-rich anorthosites. 38–58 wt% albite, up to 37 wt% enstatite, up to 22 wt% diopside, up to 33 wt% forsterite, and up to 8 wt% quartz along with some graphite and minor sulfide (plausibly oldhamite, niningerite, keilite, troilite, and wassonite). Olivine and pyroxene on Mercury must be very magnesian. It is plausible that mercurian rocks contain high-pressure phases formed from common minerals: e.g., ringwoodite from olivine, majorite from low-Ca pyroxene, coesite, stishovite (and perhaps other high-density polymorphs) from silica, and lingunite ($NaAlSi_3O_8$) from plagioclase. [233]
</td></tr>
<tr><td>Venus</td><td>
K-rich basalts, olivine gabbro-norite (calcium-rich plagioclase labradorite, orthopyroxene, and olivine) [228];
Probably mostly felsic; common felsic minerals include quartz, muscovite, orthoclase, and the sodium-rich plagioclase feldspars (albite-rich). Possibly dehydrated saponite, montmorillonite; ferric oxide; likely sulfides, sulfates (anhydrite!). ~11% Fa in olivines, ~22% Fs in pyroxene.<br>+ likely carbonates (less calcite because this converts mostly to anhydrite, but dolomite, magnesite). [234]
Normative compositions based on Venera probes [235]:
<table>
<tr><th>Mineral</th><th>Venera 13</th><th>Venera 14</th><th>Vega 2</th></tr>
<tr><td>Orthopyroxene</td><td>-</td><td>18.2</td><td>25.4$^*$</td></tr>
<tr><td>Clinopyroxene</td><td>-</td><td>-</td><td>2.5$^†$</td></tr>
<tr><td>Diopside</td><td>10.2</td><td>9.9</td><td>-</td></tr>
<tr><td>Olivine</td><td>26.6</td><td>9.1</td><td>13.9$^‡$</td></tr>
<tr><td>Anorthite</td><td>24.2</td><td>38.6</td><td>38.3</td></tr>
<tr><td>Albite</td><td>3.0</td><td>20.7</td><td>18.9</td></tr>
<tr><td>Orthoclase</td><td>25.0</td><td>1.2</td><td>0.5</td></tr>
<tr><td>Nepheline</td><td>8.0</td><td>-</td><td>-</td></tr>
<tr><td>Ilmenite</td><td>3.0</td><td>2.3</td><td>0.5</td></tr>
<tr><td>Total</td><td>100.0</td><td></td><td>100.0</td></tr>
</table>
$^*$ 75 mol% En<br>$^†$ 1.2 mol% Wo, 0.9 mol% En, 0.4 mol% Fs<br>$^‡$ 75 mol% Fo
Plains units ( ~80 percent of the surface): basaltic lava flows, alkaline basalt (Venera 13) and tholeiitic basalt (Venera 14 and Vega 2); we can infer that major minerals in the crust include calcium-rich plagioclase, orthopyroxene, Ca-pyroxene and olivine.<br>Some regions of Venus (the tessera) are assumed to be more felsic (based on their relatively low ~1 µm emissivity values) and might contain silica minerals.<br>The rest is guesswork: anhydrite ($CaSO_4$) is stable at the surface. Pyrite ($FeS_2$) may also occur; moderately ferroan olivine and low-Ca pyroxene are also stable. Weathering of phases containing ferrous iron could produce magnetite, hematite, and/or maghemite, hematite being probably the
</td></tr>
</table>

| | |
|---|---|
| | dominant (if not exclusive) iron oxide on Venus, consistent with the visible and near-infrared spectra obtained by the landers Venera 9, Venera 10. [233] |
| Mars | $H_2O$ ice, $CO_2$ ice (polar caps).<br>*Andesites and basalts* with plagioclase, pigeonite, augite, enstatite, olivine, magnetite<br>*Secondary rocks* with pyrrhotite, phyllosilicates (smectite), goethite, jarosite, gypsum, montmorillonite [228]<br><br>In situ (rovers): Mineralogy seems highly variable with location. The most abundant minerals in the *igneous rocks, andesites and basalts*, are plagioclase, olivine, pigeonite, augite, and magnetite. The *sedimentary rocks and soils* contain phyllosilicates (smectites like montmorillonite), gypsum, anhydrite, pyrrhotite, magnetite, goethite, hematite, jarosite, opal, Mg-rich carbonate (probably dolomite), siderite, epsomite, halides, tridymite [233].<br><br>Martian meteorites: the most abundant variety of martian meteorite is the group of *shergottites* and related rocks. The most abundant phases are compositionally zoned pyroxene (augite, subcalcic augite, pigeonite and/or orthopyroxene), olivine (Fa24–40), and maskelynite and/or plagioclase (Ab30–50). [233] |
| Moon | Most lunar samples are mare material, i.e., *basaltic*, with a few samples from highland material, which is mostly *anorthosite* (end member mineral: anorthite $CaAl_2Si_2O_8$). Mare *basalts* are further distinguished as 'Low-Ti', 1.5-9% of $TiO_2$, and 'high-Ti', >9% $TiO_2$. The *mare basalts* are richer in $TiO_2$ than the *highland rocks* (0-5%). Ilmenite ($FeTiO_3$) is one of the minerals which have been detected widespread on the surface of the Moon. The abundances of ilmenite in high-Ti basaltic lava are higher (9-19%) than in high-Ti mare soil (<10%) [236].<br><br>Lunar minerals in order of importance [236]<br>• Pyroxene $(Ca,Fe,Mg)_2Si_2O_6$<br>• Plagioclase feldspar $(Ca,Na)(Al,Si)_4O_8$, mostly anorthite (90mol-%), rest albite<br>• Olivine $(Mg,Fe)_2SiO_4$; most mare basalt olivines have compositions in the range Fa20-Fa70<br>• Potassium feldspar (orthoclase/microcline) ($KAlSi_3O_8$)<br>• Mineraloids are present, up to ~30 vol%, as various glasses, either 'highland glasses', mostly anorthosite plagioclase glass with some iron oxide or 'mare (lowland) glasses', mostly basaltic (pyroxene glass).<br>• Ilmenite $(Fe,Mg)TiO_3$<br>• Spinel (various compositions, e.g. $MgAl_2O_4$)<br>• Rare: silica minerals $SiO_2$, only cristobalite, quartz, tridymite), <1% usually; the most common silica mineral in mare basalt lavas is not quartz but cristobalite, which can constitute up to 5 vol.% of some basalts.<br>• There are no $Fe^{3+}$ compounds, no hydrated minerals like clays, micas, amphiboles<br>• Armalcolite $(Fe,Mg)Ti_2O_5$ only in Ti-rich basalts<br>• Native iron (Fe, ~0.3% by mass) and troilite (FeS) <1% |

## 2.2 Feldspars (Framework (Tecto-)silicates)

Framework silicates comprise the feldspar group, the quartz family (treated separately in this paper), the feldspathoids like leucite, nepheline, sodalite , the scapolite group and the zeolite family (not found yet in astro-materials).

Feldspars proper are the most common rock-forming minerals. 'It is an understatement to claim that feldspar structures are complicated' [228] and this is why we spend some effort to explain the feldspar polymorphs in this section.

There are three main feldspar endmembers (Figure 8):

- Orthoclase, potassium feldspar (K-feldspar) endmember $KAlSi_3O_8$, and polymorphs
- Albite, sodium feldspar (Na-feldspar), endmember $NaAlSi_3O_8$ and polymorphs
- Anorthite, calcium endmember $CaAl_2Si_2O_8$.

Only limited solid solution occurs between K-feldspar and anorthite, and in the two other solid solutions, immiscibility occurs at temperatures common in the crust of the Earth; solid solutions between K-feldspar and Na-feldspar are called 'alkali feldspars' (anorthoclase) and solid solutions between albite and anorthite are called 'plagioclases' and have traditional names according to Ca mole fraction *x* (see Figure 8).

In extra-terrestrial materials plagioclase is by far the most abundant feldspar. Sanidine is present but much less abundant.

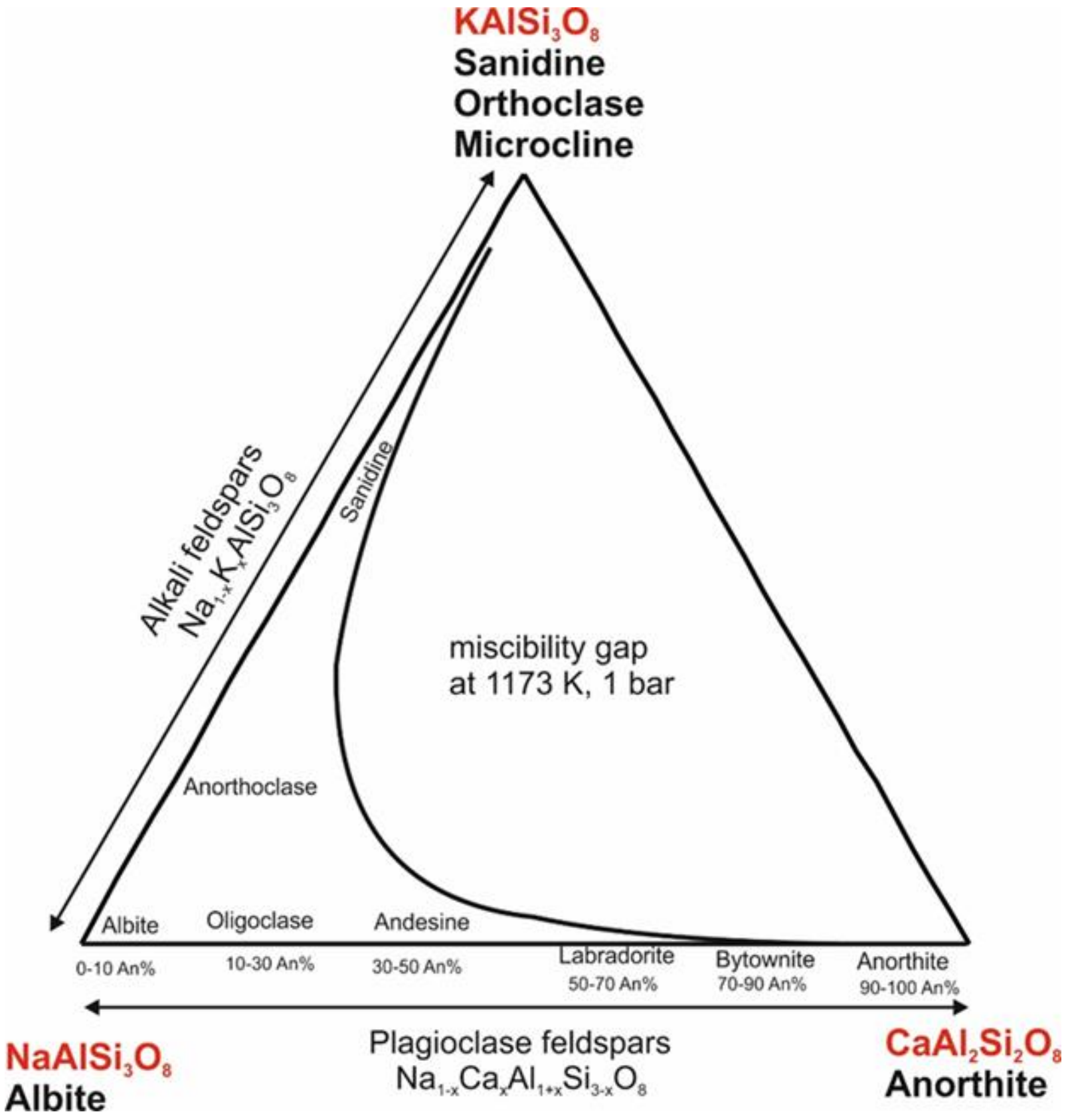

**Figure 8 Compositional phase diagram of the different minerals that constitute the feldspar solid solution. Ternary phase diagram of the feldspars (at 900 °C). Miscibility gap line after (Benisek, Dachs et al. 2010c)**

### 2.2.1 The feldspar polymorphs

We discuss the feldspar polymorphs mainly to clarify the complicated nomenclature. The heat capacity difference $\Delta c_P$ of the various polymorphs of a feldspar endmember is small, and we will neglect it (using the arithmetic average of $c_P$ data of the polymorphs of a given endmember, where available).

Ordering/disordering reactions of Al and Si on the tetrahedral sites results in the different feldspar polymorphs. For the potassium feldspars, sanidine, monoclinic, is the high temperature form with a disordered Al/Si distribution on the tetrahedral sites. It is found most typically in felsic volcanic rocks such as obsidian, rhyolite and trachyte. Orthoclase is a monoclinic polymorph stable at lower temperatures. Slowly cooled K-feldspar gives microcline with a triclinic structure and stable at yet lower temperatures.

For the Na-feldspars, it is generally accepted that there are two stable and one metastable modifications [237]:

(a) Stable modifications monalbite and albite

*monalbite [238]*: disordered; topochemical and actual symmetry: monoclinic (*C2/m*), corresponding to high sanidine; stable above 1290°C. Below this temperature it transforms by a displacive transition to a triclinic (Cl) albite structure. The most important feature of this transition is that it is very strongly coupled to the degree of Al,Si order. The Al,Si ordering transition is extremely slow compared with the unquenchable displacive transition. The effect on $c_P$ of these transitions is low: the heat capacity difference between ordered albite and analbite is 1.5% at most, the order-disorder transition at 416 K produces a $c_P$-step (low>high) of ~1% and the predicted $\Delta c_P$ peak of albite in thermal equilibrium, originating from the structural phase transition (~17% at 950 K, [239]), is so slow that it is unobservable. Thus, we will use average $c_P$ values for all polymorphs of albite.

*albite*: topochemical and actual symmetry: triclinic $C\bar{1}$; low albite, corresponding to low microcline, ordered form, stable below ≈950°C; high albite, corresponding to high microcline, disordered form, stable between ≈950°C and 1251°C [239].

(b) Metastable modification analbite

If monalbite is rapidly quenched, it undergoes a rapid displacive transformation to triclinic analbite C $C\bar{1}$ at $T_{displ}$ ≈ 930-980 °C (range of literature data). The diffusive transition (ordering of Al/Si-distribution) needs time. Because analbite is topochemically monoclinic (with a disordered Al/Si distribution), but metrically triclinic, it is unstable at any temperature.

Summarizing, monalbite and high sanidine are the high-temperature, disordered polymorphs; low albite and low microcline the low-temperature, fully ordered polymorphs; high albite and high microcline the low-temperature, polymorphs with a slightly disordered Al/Si distribution; and analbite the low-temperature metastable polymorph with disordered Al/Si distribution.

In contrast to Na-feldspar, a K-feldspar that cooled fast from high temperatures (volcanic) will preserve its Al/Si distribution as well as its monoclinic structure because the larger K atom keeps the structure open; sanidine forms from really fast cooling, or later exsolution during metamorphism. Orthoclase (Or) forms from slow cooling.

There is a significant $I\bar{1}-P\bar{1}$ phase transition in $c_P(T)$ in anorthite at ~510 K, which is displacive (fast), but to occur at all, the precise composition and degree of Al/Si-ordering is important, it only happens in pure or almost pure anorthite [240]. In ordered anorthite $T_c$ = 510 K and the transition is tricritical; in slightly less well-ordered anorthite $T_c^*$ = 530 K and the transition is second order.

The effect of this phase transition can be relatively large (Figure 9, after [241]: The heat capacities of three different anorthite samples show large differences in the temperature range 400– 600 K. *Natural* An can show a ~500 K structural phase transition $\Delta C_P$ in the range 430 – 580 K peaking at ~23 J/mol/K (~8.5%), see Figure 9. No peak was observed in synthetic An by [98]. The 500 K peak is not included in our database for anorthite.

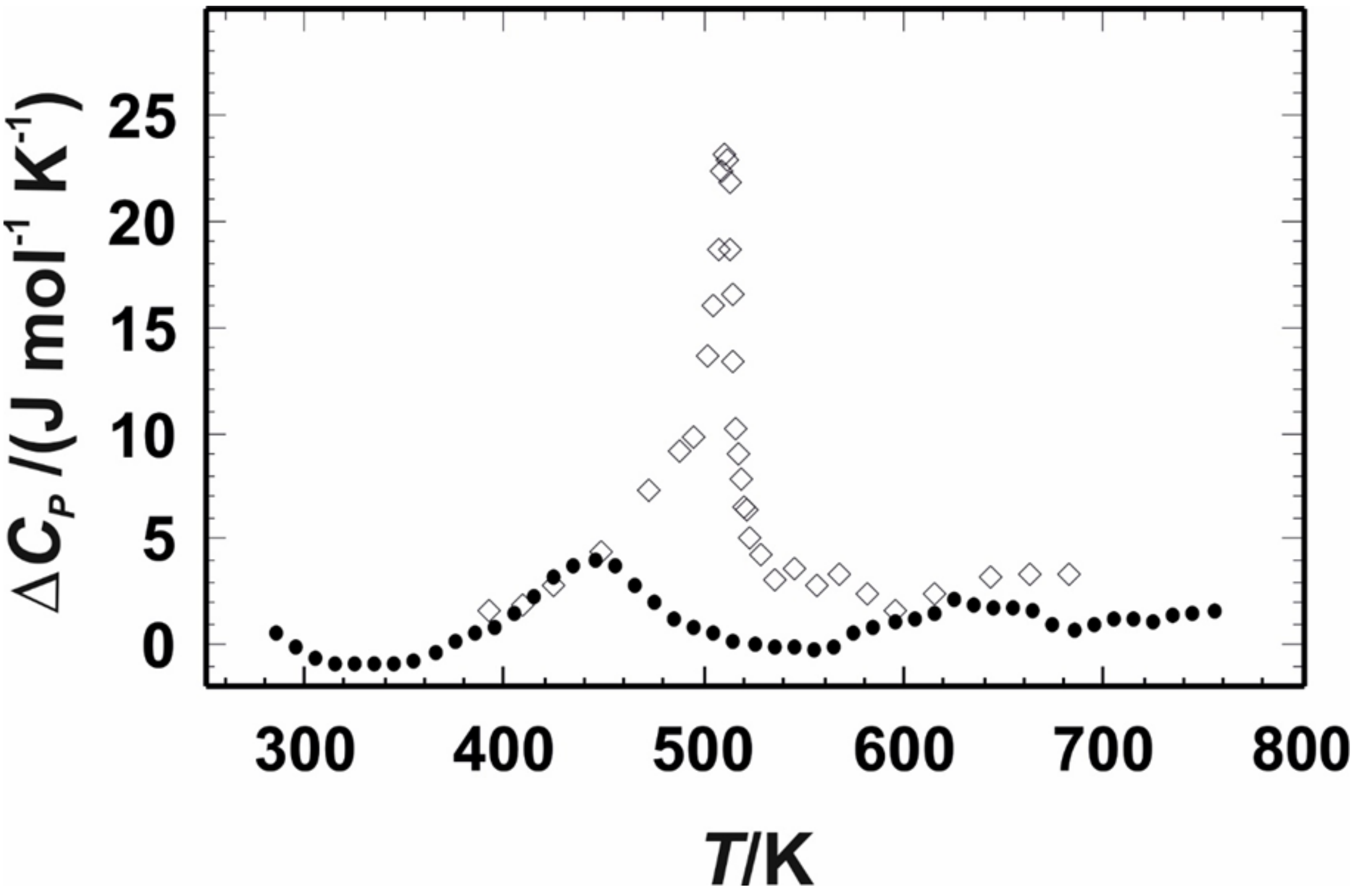

**Figure 9 excess heat capacity (ΔCp) of the ordering transition in anorthite, which was defined as $C_P$ MonteSomma – CpAn100 (solid circles) and $C_P$ Pasmeda –$C_P$ An100 (open symbols), respectively. An100 is a synthetic anorthite crystallised at 1573 K, Monte Somma is a volcanic anorthite (An98) and Pasmeda is a metamorphic anorthite (An100). From [241]. The average $C_P$ (without the peak) at 510 K is ~272 J/mol/K.**

### 2.2.2 Feldspathoids

Feldspathoids resemble feldspars but have a different structure and much lower silica content. They are a family of rock-forming minerals consisting of aluminosilicates of sodium, potassium, or calcium and having too little silica to form feldspar. There is considerable structural variation, so it is not a true group. We consider nepheline $Na_3K(Al_4Si_4O_{16})$, as a common feldspathoid, in the database.

## 2.3 Pyroxenes (Single chain inosilicates)

Pyroxenes are, besides olivines, the primary mineral phases in most primitive meteorites and in many types of non-chondritic meteorites.

Pyroxenes are a group of minerals that share the chemical formula (M2) (M1) (Si, Al)$_2$ O$_6$ [223, 225, 226]. Three pyroxene subgroups have been defined [242]  based on occupancy of the M2 site. In low-Ca pyroxenes, the M2 site is occupied by Fe or Mg, in high-Ca pyroxenes by Ca, and in the less common sodium pyroxenes by Na. Because high-Ca pyroxenes (solid solution series between endmembers diopside, $CaMgSi_2O_6$, and hedenbergite, $CaFeSi_2O_6$) have monoclinic symmetries, they are often referred to as clinopyroxenes. The term orthopyroxenes is commonly used for the orthorhombic low-Ca pyroxene solid solution series with endmembers enstatite ($Mg_2Si_2O_6$) and ferrosilite ($Fe_2Si_2O_6$). See Figure 10, the well-known ‘pyroxene quadrilateral’.

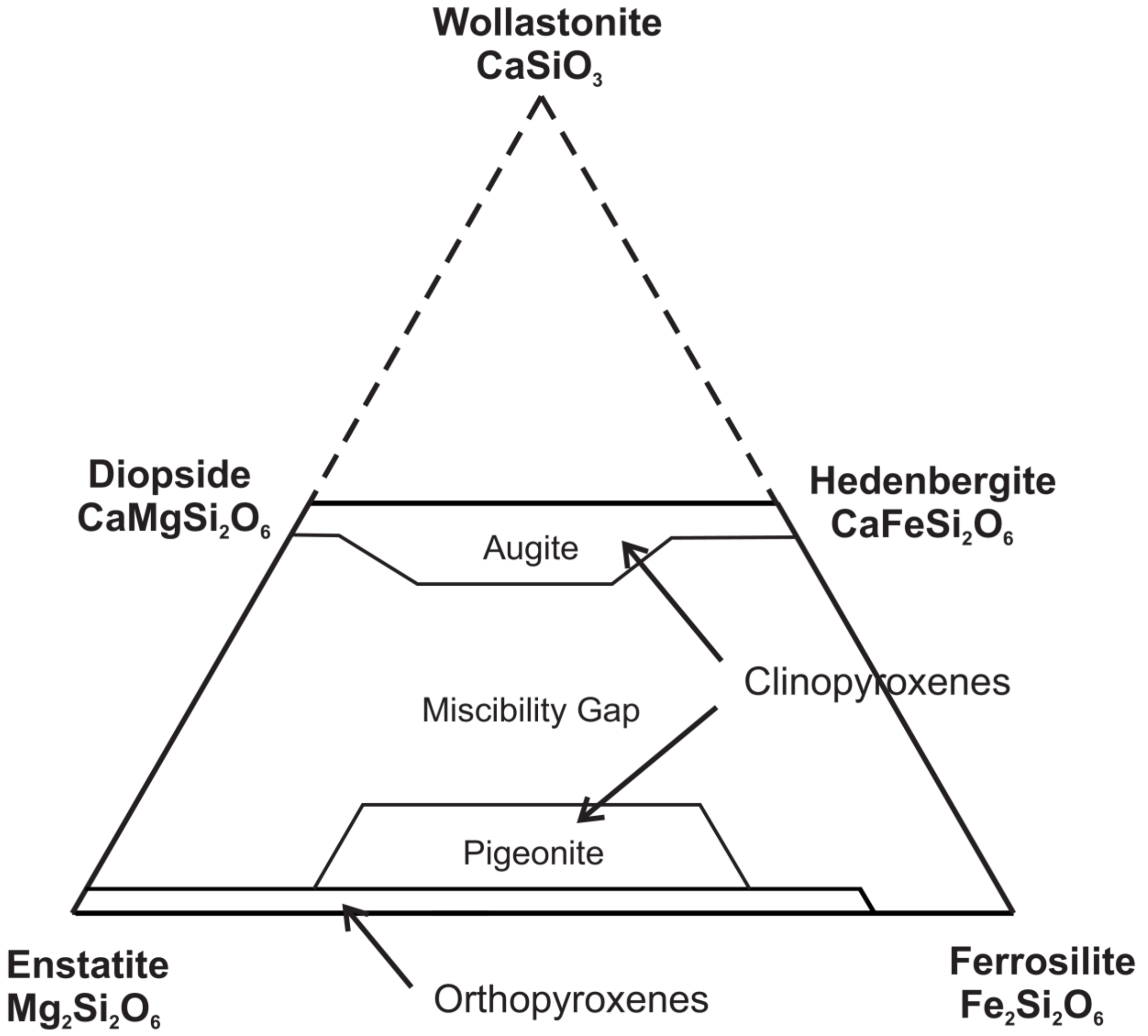

**Figure 10 pyroxene quadrilateral. Note that the old name ‘hypersthene’ is an orthopyroxene with ~30% En and rest Fs. Macke (2018 priv. comm.) takes diopside for hypersthene, which is incorrect. Bronzite is a member of the pyroxene group of minerals, belonging with enstatite and hypersthene to the orthorhombic series of the group. Rather than a distinct species, it is really a ferriferous (12 to 30% iron(II) oxide) variety of enstatite. The augites (Di-Hed) are monoclinic: ‘clinopyroxenes’. In natural orthopyroxenes , a small amount o  Ca (<2%) is always present in the structrure.**

The enstatite-ferrosilite series ([Mg,Fe]$SiO_3$) contains up to 5 mol.% calcium and exists in three polymorphs, orthorhombic orthoenstatite and (at high temperatures only protoenstatite) and monoclinic clinoenstatite (and the ferrosilite equivalent clinoferrosilite). Increasing the calcium content prevents the formation of the orthorhombic phases and pigeonite ([Mg,Fe,Ca][Mg,Fe]$Si_2O_6$) only crystallises in the monoclinic system.

Wollastonite $CaSiO_3$ and its high-temperature polymorph pseudowollastonite are not really pyroxenes but pyroxenoids.

Other pyroxene families exist; most importantly, the ones containing aluminium. Diopsidic pyroxene in many terrestrial rocks and meteorites commonly contains $Al_2O_3$, and the mineral was traditionally called fassaite [243]. Its fully aluminium endmember is Ca-Al-pyroxene $CaAlAlSiO_6$ = 'calcium Tschermak'; its official name is now Kushiroite. It is an important mineral in CAIs of carbonaceous chondrites [244]. It forms solid solutions with diopside [99].

Iron-bearing pyroxenes show (magnetic) $C_P$ peaks at cryogenic temperatures.

**Aegirine** is a member of the sodium-pyroxene family. It is rather rare. Monoclinic aegerine is the sodium-iron endmember of the jadeite-aegirine series and has the chemical formula $NaFeSi_2O_6$ in which the iron is present as $Fe^{3+}$. It is also known as **acmite.**

## 2.4 Olivines (Neo (Ortho-)silicates)

Olivine $(Mg^{2+}, Fe^{2+})_2SiO_4$ is a common mineral in the Earth's mantle but weathers quickly on the surface. Olivine rock is called Dunite (>90% olivine, $Fo_{90}$). Mg-rich olivine has also been discovered in meteorites (chondrites, pallasites), on the Moon and Mars, falling into infant stars, as well as on asteroid 25143 Itokawa [223, 225, 226].

The ratio of magnesium and iron varies between the two endmembers of the solid solution series: forsterite (Mg-endmember: $Mg_2SiO_4$) and fayalite (Fe-endmember: $Fe_2SiO_4$). Compositions of olivine are commonly expressed as molar percentages of forsterite (Fo) or fayalite (Fa) (e.g., $Fo_{70}Fa_{30}$). Forsterite has a high melting temperature at atmospheric pressure 2163 K, but the melting temperature of fayalite is much lower (about 1490 K). The melting temperature varies smoothly between the two endmembers, as do other properties. Olivine generally incorporates only minor amounts of elements other than oxygen, silicon, magnesium and iron; in extra-terrestrial materials Ca is more abundant than Mn or Cr and the Ca-rich kirschsteinite is occasionally present as a secondary phase in chondrites.

## 2.5 Amphiboles (Double chain inosilicates supergroup)

Amphiboles are found in some meteorites, including SNCs and some chondrites. Hornblende is the most commonly reported but others have also been seen [245].

Amphiboles crystallize into two crystal systems, monoclinic and orthorhombic. In chemical composition and general characteristics, they are similar to the pyroxenes. The chief differences from pyroxenes are that (i) amphiboles contain essential hydroxyl (OH) or halogen (F, Cl) and (ii) the basic structure is a double chain of tetrahedra (as opposed to the single chain structure of pyroxene). Amphiboles (as phyllosilicates) often have cation vacancies, symbolized by ☐ in chemical formulas [223, 225, 226].

Four of the amphibole minerals are among the minerals commonly called asbestos. These are: anthophyllite, riebeckite, the cummingtonite/grunerite series, and the important actinolite/tremolite series (see Figure 11) Those, however, are very rare to absent in known astro-materials, save for actinolite-tremolite (which is just ‘rare’). Note that another mineral commonly called ‘asbestos’ and common in C chondrites, *chrysotile* $Mg_3(Si_2O_5)(OH)_4$, is *not* an amphibole but a serpentine (Phyllosilicate/Kaolinite-serpentine group).

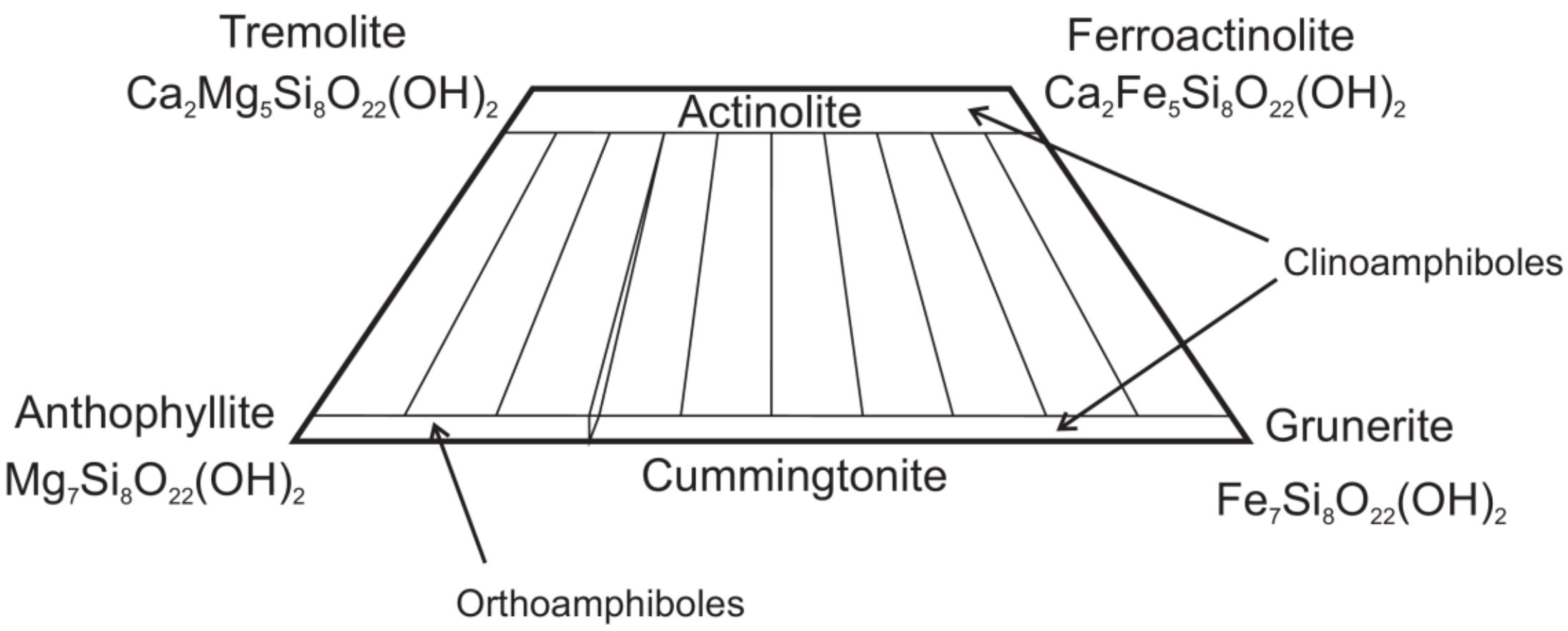

**Figure 11 Amphibole quadrilateral. The orthorhombic anthophyllites (low Ca, ≤3.8 at-%) extend up to ~30 at-% Fe, the monoclinic cummingtonite-grunerite (low Ca, ≤4.5 at-%) series from ~ 30 to 100 at-% Fe. The calcium content of the actinolite series is centered around 2/7 ≈ 29 at-% (referred to total metal cations).**

**Hornblende** is a complex monoclinic inosilicate series of minerals (ferrohornblende – magnesiohornblende). It is not a recognized mineral in its own right, but the name is used as a general or field term, to refer to a dark amphibole. It can usually be considered an isomorphous mixture of three molecules; a calcium-iron-magnesium silicate, an aluminium-iron-magnesium silicate, and an iron-magnesium silicate [223, 225, 226]. The general formula[8] can be given as $(K,Na)_{0-1}(Ca,Na,Fe,Mg)_2(Mg,Fe,Al)_5(Al,Si)_8O_{22}(OH)_2$ or $\square(Ca_2)(Z^{2+}{}_4Z^{3+}\}(AlSi_7O_{22})(OH,F,Cl)_2$, simpler as $Ca_2(Mg, Fe, Al)_5 (Al, Si)_8O_{22}(OH)_2$.

Simplifying, (no Na, no F); the physical properties of these hornblende endmembers are so similar that using two the following endmembers will be sufficient:

| | |
|---|---|
| Ferrohornblende | $\square\{Ca_2\}\{Fe^{2+}{}_4Al\}(AlSi_7O_{22})(OH)_2$ |
| Magnesio-hornblende | $\square\{Ca_2\}\{Mg_4Al\}(AlSi_7O_{22})(OH)_2$ |

[8] to be precise, a differentiation is necessary depending on normalization to 15 or 13 cations. For normalization to 13 cations: 13=Si+Ti+Al+Fe+Mn+Mg in the T and C (M1, M2, M3) sites. This method excludes Ca from the C sites and $Fe^{2+}$, Mn and Mg from the B (M4) site. In contrast, for normalization to 15 cations: 15=Si+Ti+Al+Fe+Mn+Mg+Ca in the T, C, and B sites [246].

## 2.6 Oxides and hydroxides

### 2.6.1 Simple oxides

Here we have first corundum $Al_2O_3$, periclase MgO, rutile $TiO_2$, quartz $SiO_2$ (see below), which are rather inert minerals with high melting points and simple stoichiometry. Lime, (anhydrous) CaO on the other hand, is quite reactive.

Note that $TiO_2$ occurs naturally in three phases: rutile, anatase, and brookite. Both rutile and anatase are accessory minerals that form small percentages of a vast array of rocks, soils, and sediments. Brookite is much rarer. Rutile is the most common phase in nature, and anatase transforms into rutile above 400–600 °C.

### 2.6.2 Quartz and its polymorphs

While quartz, $SiO_2$, a tectosilicate, oxide and silica mineral, is ubiquitous on Earth (think of common sand), it is, surprisingly, probably not an important phase anywhere else. In meteorites, in lunar regolith: almost no quartz. Apparently not on Mercury [232]. Only locally on Mars, due to aqueous alteration [247]. Maybe Venus has, as suggested by [234] a felsic crust that contains quartz, but the Venera probes have detected no evidence of it in-situ [235].

Quartz has many polymorphs with complicated transformation paths; under low or zero pressure, only α/β-quartz and christobalite are relevant. Low (α) quartz transforms instantly and reversibly at 843 K into high (β) quartz; above ~1143 K β-tridymite, above 1743 K β-cristobalite is stable, the latter melting at ~1978 K. Metastable cristobalite and tridymite can exist at $T$<<1000 K. High-pressure polymorphs (shocked quartz) are coesite and stishovite, which are metastable at low temperature and pressure.

Just for completeness (and comparison to terrestrial analogues that may contain significant amounts of quartz), the database contains the heat capacity of α-quartz (trigonal low-temperature form), β-quartz (hexagonal high-temperature form, >573°C, fast and reversible structural transition lambda peak), tridymite, cristobalite and amorphous $SiO_2$, (silica, lechatelierite) as well as the industrial compound sodium-metasilicate ($Na_2SiO_3$) which has been used in analogue materials (regolith simulants, see, e.g., appendix section 5).

### 2.6.3 Iron oxides and hydroxides

Iron forms rather complicated, but fascinating oxides and hydroxides. Iron oxides exist as several different polymorphs which can be divided into two groups: anhydrous (oxides) and hydrous (oxyhydroxides). The more common anhydrous forms include hematite (α-$Fe_2O_3$), maghemite (γ-$Fe_2O_3$), magnetite ($Fe_3O_4$), and wüstite ($Fe_{(1-x)}O$) with magnetite and wüstite containing both ferrous and ferric iron. Stoichiometric wüstite FeO is called ferrous oxide [248].

<u>Magnetite</u> = $FeO·Fe_2O_3$ = $Fe^{2+}Fe^{3+}{}_2O_4$, rather than $Fe_3O_4$. Mix of Fe-II and Fe-III. At low temperatures, magnetite undergoes a crystal structure phase transition (from a monoclinic structure to a cubic structure) known as the Verwey transition. The Verwey transition occurs around 124 K and the precise temperature, shape and magnitude of the lambda peak is dependent on grain size, domain state, residual stresses and the iron-oxygen stoichiometry in complicated, not fully understood ways [116, 142]. The Curie temperature of magnetite is 858 K, producing there a large λ peak in $C_P$.

<u>Hematite</u>, $Fe_2O_3$ can be obtained in various polymorphs. α-$Fe_2O_3$ has the rhombohedral, corundum (α-$Al_2O_3$) structure and is the most common form. It is antiferromagnetic below ~263 K (Morin or spin-flip transition temperature), and exhibits weak ferromagnetism between 263 K and the Néel temperature, 955 K. It shows three interesting $C_P$ features, an anomaly at ~ 10 K, visible in the effective Debye-*T* plot, a very broad bump around 500 K (< 4 %) and a strong λ -peak near 955 K. There is no λ-peak nor anomaly at the Morin temperature.
γ-$Fe_2O_3$, maghemite, is the ferromagnetic polymorph of hematite. It is Fe-II-deficient, has a cubic structure; it is metastable and converted from the α phase at high temperatures.

<u>Wüstite</u> = $Fe_{1-y}O$, ideally y=0, FeO, is particularly complicated. It is the most reduced variant, ferrous oxide, only $Fe^{2+}$(and metallic Fe if further reduced). It is typically iron deficient (classical example of a non-stoichiometric phase, y>0) with compositions ranging from $Fe_{0.84}O$ to $Fe_{0.95}O$ (eutectoid composition is $Fe_{0.932±0.004}O$). Wüstite forms from Fe and $Fe_3O_4$ (magnetite) at the eutectoid temperature of (847±7) K [143]. It is quenchable and remains metastable at ambient conditions for extended periods (tending to disproportionate to metal and $Fe_3O_4$: 4FeO → Fe + $Fe_3O_4$ but no transformation was observed at 200°C and lower [249].
Below 190 K antiferromagnetic ordering is observed in wüstite. It is accompanied by a slight rhombohedral deformation and a peak in $C_P$, which depends strongly on composition [142, 143].

**The oxide-hydroxides of iron** may occur in anhydrous (FeO(OH)) or hydrated (FeO(OH)·$n$$H_2O$) forms. The monohydrate (FeO(OH)·$H_2O$) might otherwise be described as iron(III) hydroxide ($Fe(OH)_3$), and is also known as hydrated iron oxide or yellow iron oxide.

Iron(III) oxide-hydroxide occurs naturally as four minerals, the polymorphs denoted by the Greek letters α, β, γ and δ. Goethite (α-FeOOH), lepidocrocite (γ-FeOOH), and akaganeite (β-FeOOH) comprise

the majority of the hydrous polymorphs with these materials often containing excess water. One of the most hydrated forms is semi-amorphous ferrihydrite $FeOOH \bullet nH_2O$ but with widely variable hydration [248].

Goethite, α-FeO(OH), is the main component of rust and bog iron ore.

Akaganéite is the β polymorph, formed by weathering and noted for its presence in some meteorites and the lunar surface. Cl is always present in akaganéite, serving to stabilize the molecular framework (e.g., 0.34% chlorine by mass, [248]). Decomposes > 230°C.

The γ polymorph lepidocrocite is commonly encountered as rust on the inside of steel water pipes and tanks. Feroxyhyte (δ) is formed under the high-pressure conditions of sea and ocean floors, being thermodynamically unstable with respect to the α polymorph (goethite) at surface conditions.

Ferrihydrite (Fh) $FeOOH \bullet nH_2O$, officially $Fe^{3+}{}_{10}O_{14}(OH)_2$, also written $(Fe^{3+})_2O_3 \bullet 0.5H_2O$, is a widespread hydrous ferric oxyhydroxide mineral at the Earth's surface, and a (weathering product?) constituent in extra-terrestrial materials. Ferrihydrite only exists as a fine grained and highly defective nanomaterial. The powder X-ray diffraction pattern of Fh contains two scattering bands in its most disordered state, and a maximum of six strong lines in its most crystalline state. The principal difference between these two diffractions end-members, commonly named *two-line* and *six-line* ferrihydrites, is the size of the constitutive crystallites. The two-line form is also called hydrous ferric oxides (HFO). Ferrihydrite is a metastable mineral.

Finally, we mention bunsenite, NiO, which is notable as being the only well-characterized oxide of nickel.

## 2.7 Carbonates, halides and brine salts

The well-known rock-forming carbonates are of course calcite $CaCO_3$, dolomite $CaMg[CO_3]_2$, and magnesite $MgCO_3$. Iron carbonate, siderite $FeCO_3$, is (on Earth) commonly found in hydrothermal veins.

Water-soluble carbonates form *evaporites*: Natrite, $Na_2CO_3$ (anhydrous; various hydrates have mineral names, e.g., $\cdot 1H_2O$ thermonatrite, $\cdot 10H_2O$ natron), and the bicarbonate nahcolite, $NaHCO_3$.

**Brine salts**

Ceres' most famous bright faculae in Occator Crater probably originated from the recent crystallization of brines that reached the surface from below; the brine composition is thought to be [250] a

mixture of $NaCl \cdot 2H_2O$ (hydrohalite) with smaller amounts of $NH_4Cl$ (ammonium chloride), $Na_2CO_3$ (natrite) and ammonium bicarbonate ($NH_4HCO_3$) .

Thus we include also the halides halite NaCl, and sylvite KCl, as well as the ammonia salts $NH_4Cl$ (ammonium chloride) and ammonium bicarbonate ($NH_4HCO_3$) in the database.

## 2.8 Phosphates, sulphates and related minerals

As common phosphate minerals, we include fluorapatite $Ca_{10}(PO_4)_6F_2$ and hydroxyapatite $Ca_{10}(PO_4)_6(OH)_2$: they form a solid solution series.

The most common sulphates are the ones of calcium, magnesium and iron: anhydrite $CaSO_4$, gypsum $CaSO_4 \cdot 2H_2O$, anhydrous magnesium sulphate $MgSO_4$ and epsomite $MgSO_4 \cdot 7H_2O$, and iron(II)sulphate $FeSO_4$. Note that epsomite is extremely soluble in water, and looses crystal water already slightly over room temperature.

Iron(II)sulphate $FeSO_4$ is a weathering product of FeS (or meteoritic iron with acid rain) in meteorite finds and is usually hydrated, $FeSO_4 \cdot n(H_2O)$. These compounds exist most commonly as the heptahydrate ($n$ = 7) but are known for several values of $n$ ($n$ = 1, 4, 5, 6, 7: Szomolnokit, Rozenite, Siderotil, Ferrohexahydrite, Melanterite) [223, 225, 226].

## 2.9 Sulfides and related minerals

Sulfides are an important accessory mineral in meteorites, we list troilite FeS, pyrite $FeS_2$, pyrrhotite $Fe_{1-x}S_x$ (x=0 … 0.2), and pentlandite $(Fe,Ni)_9S_8$.

### 2.9.1 Pyrite $FeS_2$

Iron (II) Disulphide $FeS_2$, is used as a replacement for troilite in meteorite analogues. It is dimorph, with pyrite (cubic) and marcasite (orthorhombic) phases; the latter is less stable than pyrite and decays in ambient air within a few years; heating marcasite >400°C produces pyrite.

Decomposition of pyrite into pyrrhotite and elemental sulfur starts at 813-843 K; at around 973 K, $p(S_2)$ is about 1 atm [226].

### 2.9.2 Pentlandite $Fe_{4.5}Ni_{4.5}S_8$

It is the most common terrestrial iron-nickel sulphide, compare troilite. It is non-magnetic.

Berezovskii et al. [78] measured the $c_P$ of pentlandite ($Fe_{4.60}Ni_{4.54}S_8$) between 6 and 306 K. There are no observable phase transitions in this temperature range, but a clear $\gamma T$ term indicating electronic heat capacity typical of conductors. The metal-sulphur ratio of pentlandite implies unusual valence of Fe and Ni atoms; metal bonding has been proposed. According to Warner et al. [251], pentlandite has a 2$^{nd}$ order lambda transition between 323 and 473 K. According to Sugaki and Kitakaze [252] there is a low phase <584°C, an order-order phase transition in the range 580 – 620°C; decomposition starts at 613°C, pentlandite melts over the range 865°C – 952°C.

### 2.9.3 Troilite FeS, pyrrhotite $Fe_{1-x}S$

Troilite FeS is a typical example for a non-stoichiometric compound; it is rarely found on Earth as $Fe_{1.00}S$ but rather pyrrhotite $Fe_{1-x}S$, while it is near- stoichiometric in iron meteorites (where it is in equilibrium with metallic Fe). Most troilite on Earth is of meteoritic origin. One iron meteorite, Mundrabilla, contains 25 to 35 volume percent troilite [253]. The most famous troilite-containing meteorite is Canyon Diablo. As troilite lacks the iron deficiency which gives pyrrhotite its characteristic magnetism, troilite is non-magnetic [223, 225, 226].

Iron-deficient pyrrhotite has the formula $Fe_{(1-x)}S$ (x = 0 to 0.2). Thermodynamic properties of the α/β phase transformation in terrestrial troilite vary systematically with prior thermal history of the troilite; both the transition temperature and enthalpy change for the α/β transformation decrease with increasing maximum temperature of prior heat treatment. DSC measurements on troilite from various meteorites indicate clear differences in the α/β thermodynamic properties that are consistent with differences in the natural thermal histories of the meteorites [254].

The heat capacity of troilite has been measured [255] for 5—1000 K. It exhibits transitions due to disappearance of the lower-temperature antiferromagnetic or ferromagnetic phase. Stoichiometric FeS shows three transitions in the temperature range 300 K to 1000 K, with heat-capacity maxima at 419.6 K, 440 K and 590 K; the Néel temperature is near 590 K. The low-temperature transition originates from structural changes, whereas the higher ones are mainly of magnetic origin. For $Fe_{0.98}S$ only one additional transition takes place, with maximum heat capacity at $T$ = 405 K. $Fe_{0.89}S$ exhibits a transition 30 K below the Néel temperature. The maximum heat capacity at $T$=560 K is due to a structural transition coupled to a magnetic-order-to-order transition. In addition, a smaller effect, related to a phase reaction, is observed in the range $T$ = 650 K to 760 K.

## 2.10 Meteoritic iron

Meteoritic iron FeNi is mainly composed of iron and nickel, with Ni content up to 65% and minor cobalt (0.25-0.77% Co ; Fe+Ni+Co make >95%). The bulk of meteoric iron consists of taenite and kamacite. Taenite is a face centered cubic and kamacite a body centered cubic iron-nickel-alloy (plessite is a fine-grained intergrowth of kamacite and taenite), see Table 5.

Meteoric iron can be distinguished from telluric iron by its high Ni content and by its microstructure, notably, the famous Widmanstätten patterns, interleaving of kamacite and taenite bands or ribbons called lamellae. They form when meteoric iron cools and kamacite is exsolved from taenite. They appear, however, only in octahedrites, with an average Ni content of 5 to 18%, not in hexahedrites which contain only kamacite with 4 – 7% average Ni content, nor in ataxites (only taenite) with average Ni content >15% [223].

Yang et al. [256] determined the Ni content and crystal structure of the various regions in meteoritic metal with <50 nm resolution, revealing a very clear compositional zoning.

**Table 5 Phases of meteoritic iron [223]**

| Mineral | Formula | Nickel (Mass-%) | Crystal structure | Notes |
|---|---|---|---|---|
| Antitaenite | $\gamma_{Low\ Spin}$-(Ni,Fe); $Fe_{\sim 3}Ni$ | 20-40 | fcc | Only approved as a variety of taenite by the IMA. Low magnetic moment |
| Kamacite | α-(Fe,Ni); $Fe_{\sim 0.9}Ni_{\sim 0.1}$ | 5-10 | bcc | Same structure as ferrite |
| Taenite | γ-(Ni,Fe); $Ni_{\sim 0.5}Fe_{\sim 0.5}$ | 20-65 | fcc | Same structure as austenite<br>High magnetic moment |
| Tetrataenite | (FeNi) | 48-57 | tetragonal | < 320°C |
| Awaruite | $Ni_3Fe$ | ~1/3 | cubic | |

The typical composition of meteoritic iron is [257]

- 64-98% kamacite [258], rest taenite/plessite and sometimes 5-8% cohenite and graphite, minor sulphides
- Ni content in kamacite: 6.8-8.2% (average 7.1±0.7), a bit lower (6.0%) near taenite borders

- Ni content in taenite: 29-60%

For example, the Canyon Diabolo meteorite has typically 87% kamacite with 6.8%Ni, 2.1% taenite with ~46%Ni, 1.1% plessite with ~26%Ni, 6.5% cohenite, thus on average 7.17% metallic Ni as measured. Often, meteoritic iron is associated with schreibersite $(Fe,Ni)_3P$.

Using a 'typical' meteoritic iron $c_P(T)$ curve for ~10% Ni is possible below ~400 K, since at low temperatures the composition dependence is small (see Figure 12). At higher temperatures, the magnetic transition peak depends strongly on composition (both peak temperature and shape).

Since the $c_P$ of FeNi is much smaller (factor 2) than that of silicates, meteorites (or regolith) containing FeNi haven also a lower $c_P$. However, weathering of meteorites can change the specific heat significantly, if the meteorite contains elemental iron-nickel (the $c_P$ of the oxides is significantly higher than the $c_P$ of the elemental metals), see [28]. Even without knowing the composition (the content of metal), the bulk or grain density $\rho$ correlates well with $c_P$, since FeNi is also much denser (7800 kg/m³ compared to ~3000 kg/m³ for silicates) and $c_P(\rho)$ may be approximated by the relation [259]: $c_P = a + b/\rho_b$, where $a$, and $b$ are constants (at 298.15 K we calculate $a$ = 262.81 J/(kg · K), and $b$ = 1.4616 · $10^6$ J/(K · $m^3$)).

There is a dearth of accurate experimental data on meteoritic iron, we only found Butler & Jenkins [260] who used an octahedrite sample of the Canyon Diablo meteorite. These data, obtained by a Xe lamp flash method, seem to be ~40% too high systematically (Figure 12).

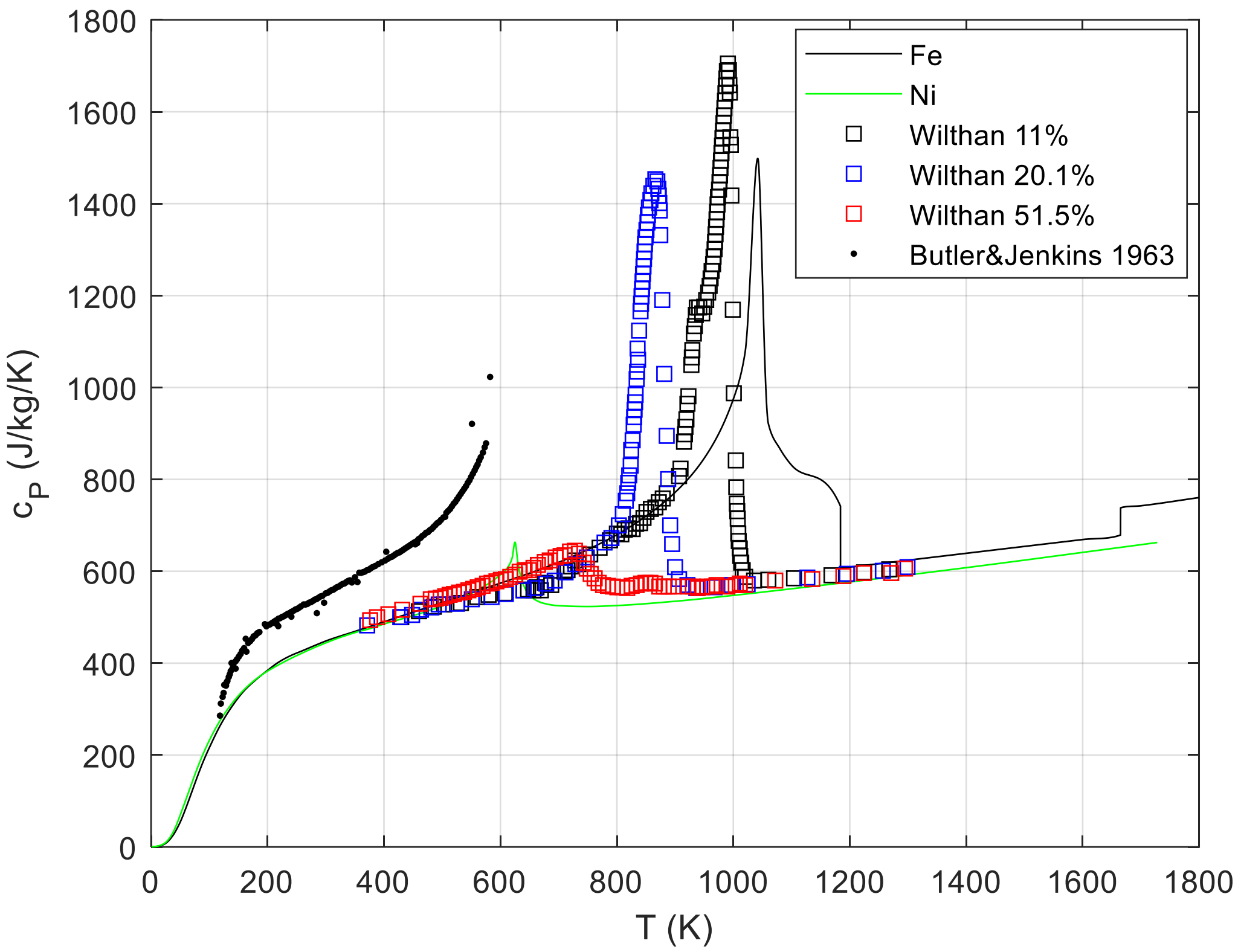


**Figure 12 Specific heat of FeNi alloys [261] together with the curves for pure Fe and pure Ni. In alloys, the iron α→γ transition at ~1190 K seems to vanish and the amplitude and position of the magnetic Ni transition varies systematically with composition. The black dots are the (digitized) data of [260], they seem systematically off.**

In the field of meteoritics, the equilibrium Fe-Ni phase diagram is of great importance. The phase relations between the alpha phase (kamacite) and the gamma phase (taenite) are best described by means of the Fe-Ni phase diagram [256, 262]. The low-temperature region of the phase diagram has been constructed with the help of meteoritic iron analyses, since this metal took ~$10^8$ years to cool [263] and is closest to equilibrium (even if actually metastable). At 300°C, it takes more than $10^4$ years for one atomic jump to occur; atomic diffusion is already very slow at 400°C and effectively ceases at 200°C. [264].

The Curie point of the α-phase (Kamacite) is about $T_c^{bcc}(\mathrm{K}) = 1043 x_{Fe} + 456 x_{Ni} + 385.8 x_{Fe} x_{Ni}$ , with nickel and iron content $x_{Ni}$, $x_{Fe}$ in atomic fraction. The Curie temperature of the fcc (γ) ferromagnetic cubic phase varies with nickel content in a complicated fashion between less than 300 K (actually 0 for metastable phases) and close to 900 K [265]. If at least two main phases are present in meteoritic iron (α, γ), two magnetic phase transitions at different temperatures are expected, at ~1030 K for α and ~800 K for the γ phases.

Modelling the specific heat of FeNi alloys with arbitrary Ni content for the phases could possibly be done using the empirical CALPHAD approach described in [266] based on [267].

### 2.10.1 Metal carbides: Cohenite and cementite

Carbides are found associated to meteoritic iron FeNi: Cementite, iron carbide $Fe_3C$ and more general cohenite, iron-nickel carbide $(Fe,Ni)_3C$. For example, the Canyon Diabolo meteorite contains 5-8% c cohenite [257].

The limit of Ni solubility in cohenite $(Fe,Ni,Co)_3C$ is not known experimentally. Lunar cohenite containing 7.0 wt.% Ni has been observed in Apollo 17 soil fines. Meteoritic cohenite apparently has a Ni content of 2-3 wt.%. Terrestrial samples contain just over 3 wt.% Ni. However, since pure $Ni_3C$ has been synthesized it is reasonable to expect a continuous solid solution series between $Fe_3C$ and $Ni_3C$ [268] .

Pure iron carbide $Fe_3C$ is called cementite. It is opaque and strongly ferromagnetic below the Curie point, 485±5 K. For both carbides, the melting point ≈ 2110 K.  Cohenite decomposes >996 K.

While cementite is thermodynamically unstable for p<40 kbar, eventually and very slowly being converted to Austenite/Kamacite and graphite, it does not decompose on heating at temperatures below the eutectoid temperature (723 °C).

## 2.11 Phyllosilicates (Sheet silicates)

The nomenclature of phyllosilicates (and 'clay minerals') is complicated and changed over time. All phyllosilicate minerals are hydrated, with hydroxyl groups and water (in the case of clays) attached; they can form by aqueous alteration. However, carbonaceous chondrites frequently contain partly or completely dehydrated phyllosilicates, and such phases are likely present on asteroids Ryugu and Bennu [269]. In meteorites, the serpentines and montmorillonites (especially saponite) are the most frequently encountered phyllosilicates, micas are rare – mineral composition tables however often only list the constituents in some broad category, like 'saponite' or 'serpentine'.

Phyllosilicates are divided in a number of groups and subgroups, with mineral species making up the subgroups. We just describe the most relevant ones. Note that nomenclature is not fully consistent in the literature; presently recommended is [270]. In a phyllosilicate (sheet silicate), a 'sheet' or layer can be composed either of 1 tetrahedral : 1 octahedral sheet (1:1 layer), 2 tetrahedral :  1 octahedral sheet (2:1 layer) or the latter with a brucitic sheet in the interlayer (2:1:1 layer type).

Phyllosilicates, especially the smectite (clay) minerals are among the most complex inorganic compounds in nature. They display a complex and variable composition and (like amphiboles) often contain lattice vacancies; endmember compositions can often be not easily defined or the sheer number of ideal endmembers would be overwhelming. Thus, specific heat data are not available for any possible composition of a phyllosilicate we find naturally. We sometimes have to improvise, using analogue minerals as proxies (possibly scaled for mean atomic mass) or used predicted $c_P$-values from models (e.g., [155]) or ab-initio calculation (DFT). Measured specific heat data from the literature are often not directly usable; natural samples often don't have a 'reasonable' end-member composition, and impurities have to be determined and quantified such that the contribution of impurities can be subtracted from the measured properties. The hydration state of the mineral is another important point to consider, particularly for swelling clay minerals. Indeed, hydration energies are not negligible and depend on the nature of the clay mineral, interlayer cations, and relative humidity (RH). Calorimetric measurements have to be performed for a fixed and known hydration state. Blanc et al. propose a consistent suite of models for the prediction of (among other properties) $C_P(T)$ for anhydrous clay minerals, parameterized using calorimetric data from the literature [155]. Starting from the anhydrous state, the $C_P$ of a given hydration state can be calculated by adding the contribution of hydrate water to specific heat (details see chapter 2.12.1 below).

***The serpentine-kaolinite group*** has two subgroups, serpentines and kaolinites.

**The serpentines** describe a group of common rock-forming hydrous magnesium iron phyllosilicate $(Mg, Fe)_3Si_2O_5(OH)_4$) minerals. Common are iron-bearing cronstedtite, Mg-bearing antigorite/lizardite/chrysotile ('asbestos') $Mg_3Si_2O_5(OH)_4$), berthierine, and others. They decompose at ~750°C (chrysotile).

Cronstedtite is a complex, iron-rich serpentine, $Fe^{2+}{}_2Fe^{3+}(Si,Fe^{3+})O_5(OH)_4$, substitution between Si and $Fe^{3+}$ is variable, we assume Fe:Si 1:1 at least: $Fe^{2+}{}_2Fe^{3+}{}_2SiO_5(OH)_4$ but rather $Fe^{2+}{}_2Fe^{3+}(SiFe^{3+})O_5(OH)_4 = Fe^{2+}{}_2Fe^{3+}{}_2SiO_5(OH)_4$

**Kaolinite** $Al_2Si_2O_5(OH)_4$ has one tetrahedral sheet of silica ($SiO_4$) linked through oxygen atoms to one octahedral sheet of alumina ($AlO_6$) octahedra. Kaolinite undergoes a series of phase transformations upon thermal treatment in air at atmospheric pressure. Any $c_P$ must state the water content and is restricted to ≤ 550°C, where nonreversible dehydration begins.

***The pyrophillite/talc group*** contains **Talc** as its most relevant member.: $Mg_3Si_4O_{10}(OH)_2$ (iron-bearing: $Fe^{2+}{}_3Si_4O_{10}(OH)_2$ , Tschermak: $Mg_2Al_2Si_3O_{10}(OH)_2$). Pyrophillite is $Al_2Si_4O_{10}(OH)_2$.

***The chlorite[9] group.*** The name *chlorite* is from the Greek *chloros* (χλωρός), meaning 'green', in reference to its colour. They do not contain the element chlorine, also named from the same Greek root. Layer type 2:1:1. The typical general formula is: $(Mg,Fe)_3(Si,Al)_4O_{10}(OH)_2 \cdot (Mg,Fe)_3(OH)_6$. Most relevant are clinochlore, Mg-rich, and chamosite which is Fe-rich.

### *The mica group and subgroups*

**Muscovite** $KAl_2[(OH,F)_2|AlSi_3O_{10}]$ is the most common mica .

**Biotite** $K(Mg,Fe)_3AlSi_3O_{10}(F,OH)_2$ or $K(Mg,Fe^{2+},Mn^{2+})_3[(OH,F)_2|(Al,Fe^{3+},Ti^{3+})Si_3O_{10}]$, is another common mica, primarily a solid-solution series between the iron-endmember annite $KFe_3^{2+}AlSi_3O_{10}(OH)_2$, and the magnesium-endmember phlogopite $KMg_3AlSi_3O_{10}(OH)_2$ ; more aluminous end-members include siderophyllite $KFe^{2+}_2Al(Al_2Si_2)O_{10}(F,OH)_2$ (rare).

The chemical variability of biotites is dominated by Fe-Mg-exchange, and the Tschermak substitution $[(Fe,Mg)^{oct} + Si^{tet} = Al^{oct} + Al^{tet}]$. There are four endmember components used to describe such biotite compositions:

| | |
|---|---|
| $KFe_3[(OH)_2AlSi_3O_{10}]$ | annite (Ann) |
| $KMg_3[(OH)_2AlSi_3O_{10}]$ | phlogopite (Phl) |
| $KAlFe_2[(OH)_2Al_2Si_2O_{10}]$ | siderophyllite (Sid) |
| $KAlMg_2[(OH)_2Al_2Si_2O_{10}]$ | eastonite (Eas) |

***Smectites (often imprecisely called clay minerals)*** are hydrous aluminium phyllosilicates, sometimes with variable amounts of iron, magnesium, alkali metals, alkaline earths, and other cations, $A_{0.3}D_{2-3}[T_4O_{10}]Z_2 \cdot nH_2O$. Subgroups include montmorillonite, saponite, nontronite and vermiculite.

**Montmorillonite** $(Na,Ca)_{0.33}(Al,Mg)_2(Si_4O_{10})(OH)_2 \cdot n(H_2O)$ is a 2:1 phyllosilicate mineral (meaning that it has two tetrahedral sheets of silica sandwiching a central octahedral sheet of alumina) characterized as having greater than 50% octahedral charge.

**Saponite is** trioctahedral. $Ca_{0.25}(Mg,Fe)_3((Si,Al)_4O_{10})(OH)_2 \cdot n(H_2O)$

**Vermiculite**, $Mg_{0.7}(Mg,Fe,Al)_6(Si,Al)_8O_{20}(OH)_4 \cdot 8(H_2O)$ forms by the weathering or hydrothermal alteration of biotite or phlogopite. It undergoes significant expansio, then exfoliation when heated.

[9] There is some confusion concerning the nomenclature of serpentine and chlorite minerals: In the older literature, the term chamosite was generally applied to minerals in ironstone deposits which frequently contain both Fe serpentine and Fe chlorite. Following [270] we propose to use ``clinochlore" and ``chamosite" for the Mg and Fe end- members of the chlorite group minerals, and ``amesite" and ``berthierine" for their analogues in the serpentine subgroup minerals.

*Other important phyllosilicates*

**Palygorskite**, also known as *Attapulgite,* is $(Mg,Al)_4[OH|(Si,Al)_4O_{10}]_2 \cdot (4+4)\ H_2O$.

**Illite** is a is a group of closely related non-expanding clay minerals similar to micas. Structurally, illite is quite similar to muscovite with slightly more silicon, magnesium, iron, and water and slightly less tetrahedral aluminium and interlayer potassium. The chemical formula is given as $(K,H_3O)(Al,Mg,Fe)_2(Si,Al)_4O_{10}[(OH)_2,(H_2O)]$, but there is considerable ion (isomorphic) substitution. The iron rich member of the illite group is glauconite. A typical empiciral formula is $K_{0.65}Al_{2.0}Al_{0.65}Si_{3.35}O_{10}(OH)_2$.

In Table 6 we list the simplified reference formulae for some common phyllosilicates we assume if in modal analyses for example, no explicit empirical formula is given.

**Table 6 Assumed reference (simplified, idealized) formulae for common phyllosilicates in modal analyses of meteorites**

| | |
|---|---|
| Chrysotile | $Mg_3(Si_2O_5)(OH)_4$ |
| Berthierine | $(Fe^{2+},Fe^{3+},Al)_3(Si,Al)_2O_5(OH)_4$ |
| Cronstedtite | $(Fe^{2+}Fe^{3+})_3(Si,Fe^{3+})_2O_5(OH)_4$ iron end; but usually contains some Mg: $(Mg, Fe)_3Si_2O_5(OH)_4$ |
| Saponite-serpentine | $Mg_3(Al_xSi_{4-x}O_{10})(OH)_2 \cdot 4H_20$, x~0.33 |
| Saponite, general formula | $(Ca_{0.5}|Na)_{0.3}(Mg|Fe^{2+})_3(Si|Al)_4O_{10}(OH)_2 \cdot 4\ H_2O$ [223] or $Ca_{0.25}(Mg,Fe)_3((Si,Al)_4O_{10})(OH)_2{\cdot}n(H_2O)$ , typically 3:1 for Si:Al and 1:1 or 1:2 for Mg:Fe, thus (next line) |
| Saponite ‚typical‘ | $Ca_{0.25}MgFe_2Si_3AlO_{10}(OH)_2{\cdot}4(H_2O)$ , molar mass 451.2613+$n$·18.0153 g/mol |
| Saponite (Orgueil[10]) | $(Mg_{2.55}Fe^{2+}_{0.45})(Si_{3.46}Al_{0.54})O_{10}(OH)_2 \cdot nH_2O$ ($n$≈4 if fully hydrated) |
| Serpentine (Orgueil[11]) | $(Mg_{2.55}Fe_{0.45})Si_2O_5(OH)_4$ (no crystal water) |

[10] after Bland, Cressey et al. [271] who use the Orgueil average serpentine and saponite compositions of Tomeoka and Buseck [272], table 2.
[11] after Bland, Cressey et al. [271] who use the Orgueil average serpentine and saponite compositions of Tomeoka and Buseck [272], table 2.

## 2.12 Hydrated minerals

Dehydrated phyllosilicates are common among carbonaceous chondrites, and are probably present in the regoliths of many asteroids. For example, the carbonaceous chondrite material in HED meteorites include dehydrated materials, probably heated during impact into the regoliths. On the other hand, all possible hydration stages are found in certain natural phyllosilicates, so we need a correlation for the heat capacity of a given amount of hydrate (crystal) water (note, that e.g., serpentine does not have crystal- or hydrate water). Swelling clay minerals can incorporate, between their sheet silicate layers, much more hydration water, of the order of 10% by mass or more. The thermodynamics of hydration of clays is in fact an active research field and quite complex [144, 155, 273, 274]. Note that from meteorites, one can never be sure that the hydrate water found is only extra-terrestrial!

The 'water content' of a 'hydrated' mineral can be misleading [275]. It usually means 'all mass loss upon heating to ~770°C' which comprises molecular water (adsorbed), mesopore or crystal water, release of $H_2O$ from (oxy-)hydroxide minerals like ferrihydrite and goethite, but also hydroxyl groups (-OH) from phyllosilicates.

According to Garenne et al. [275] the loss of mass[12] between 25 and 200°C is due to the adsorbed water and the water in (2–50) nm pores (mesopores) and this range is most easily contaminated by terrestrial water[13].

The hydrogen quantity in carbonaceous chondrites can be inferred from TGA as different hosts:

(i) weakly-bonded $H_2O$ (loss between 25 and 200 °C),
(ii) $H_2O$ in hydroxides (200–400 °C),
(iii) -OH from phyllosilicates (400–770 °C) and
(iv) a high *T* loss (calcium carbonates and sulphates, 770–900 °C) .

For anhydrous clay minerals dihydroxylation begins at temperatures, depending on the mineral, of 300 to 600°C. For hydrated phases, loss of water may begin close to 300 K [144]. For the hydrous serpentine cronstedtite, MacKenzie and Berezowski [277] found only loss (i) of ~0.7% of surface adsorbed water up to 200°C; oxidation of $Fe^{2+}$ to $Fe^{3+}$ sets in between 200 and 320°C, and a further loss of hydroxyl water for >400 °C. Natural smectites like saponites start to metamorphose in the range

[12] 0.3% to 5%, sometimes up to 10%, depending on mineral, grain size, mesoporosity and heating rate [276]
[13] which can be removed by heating to 127°C, in high vacuum, for 3 days [276]

200 – 250 °C in chlorites and illite, while their mass loss below 200 °C (~10%) corresponds to the desorption of physically adsorbed water and interlayer water associated with interlayer cations [278].

We will focus on (i) here, weakly bonded $H_2O$.

Example for hydrate water mass fraction: a typical saponite is $Ca_{0.25}(Mg_{0.8}Fe_{0.2})_3Al_{0.5}Si_{3.5}O_{10}(OH)_2 \cdot n(H_2O)$;  M= 407.66+n*18.015. If n=4, we have 17.7% by mass water in the saponite, and with a typical saponite content of say 33% in a CI-meteorite, there is 5.8 weight-% 'saponite hydrate' water in the meteorite.

### 2.12.1 The specific heat contribution of water in hydrated minerals

Physisorbed, excess[14] or crystal water in minerals has a reproducible specific heat contribution $C_P$(water) so we can write:

$$C_P(anhydrous, X) = C_P(hydrated, X \cdot nH_2\mathrm{O}) - n \cdot C_P(water) \tag{18}$$

Water adsorbed at the 2 – 3 lower 'layers' behaves quite differently from bulk water. It is at least partially ordered, does not freeze and its molecular mobility was shown to depend largely on hydrogen bond interactions between the adsorbed water molecules and the -OH groups on the surface. The fourth layer is transitional, and further layers are similar to bulk water [279, 280]. Also proper crystal water is normally ordered, 'ice-like'. Thus, $C_P$(water) can be treated as $C_P$(ice Ih), just extrapolated for $T$>273.16 K which is not too problematic as the ice $C_P(T)$ curve is fairly linear there.

Majzlan et al.  [281] fitted the bulk Ih $C_P$ data to the following approximate correlation equation

$$C_P(\text{ice Ih}) = n_1 D(\theta_1/T) + n_2 D(\theta_2/T) + m_1 E(\mathrm{T}_{E,1}/T) + m_2 E(\mathrm{T}_{E,2}/T) \tag{19}$$

with $D$ and $E$ the Debye- resp. Einstein-functions, $\theta_1$ = 126.77 K, $n_1$ = 0.3103; $\theta_2$ = 392.77 K, $n_2$ = 0.60133; $T_{E,1}$ = 652.91 K, $m_1$ = 0.52, $T_{E,2}$ = 1388.98 K, $m_2$ = 2.105.

Note the rather high specific heat of water (ice) compared to silicates; adding water almost always increases $c_P$ of a sample!

Using these equations, the heat capacity of water ice Ih at 298.15 K is predicted to be 41.28 J/(mol·K). Note that there are possible phase transitions of water (ice) in larger pores which include a glass transition of amorphous ice (Ia) at 120 -140 K and subsequent crystallization to cubic ice (Ic);

[14] i.e., water that is retained even after prolonged evacuation and storage in argon

transformation of cubic ice to hexagonal ice (Ih) at 160 – 210 K; melting of ice Ih at 273.15 K [281]. None of these transitions are observed in the heat capacity of excess or hydrate water.

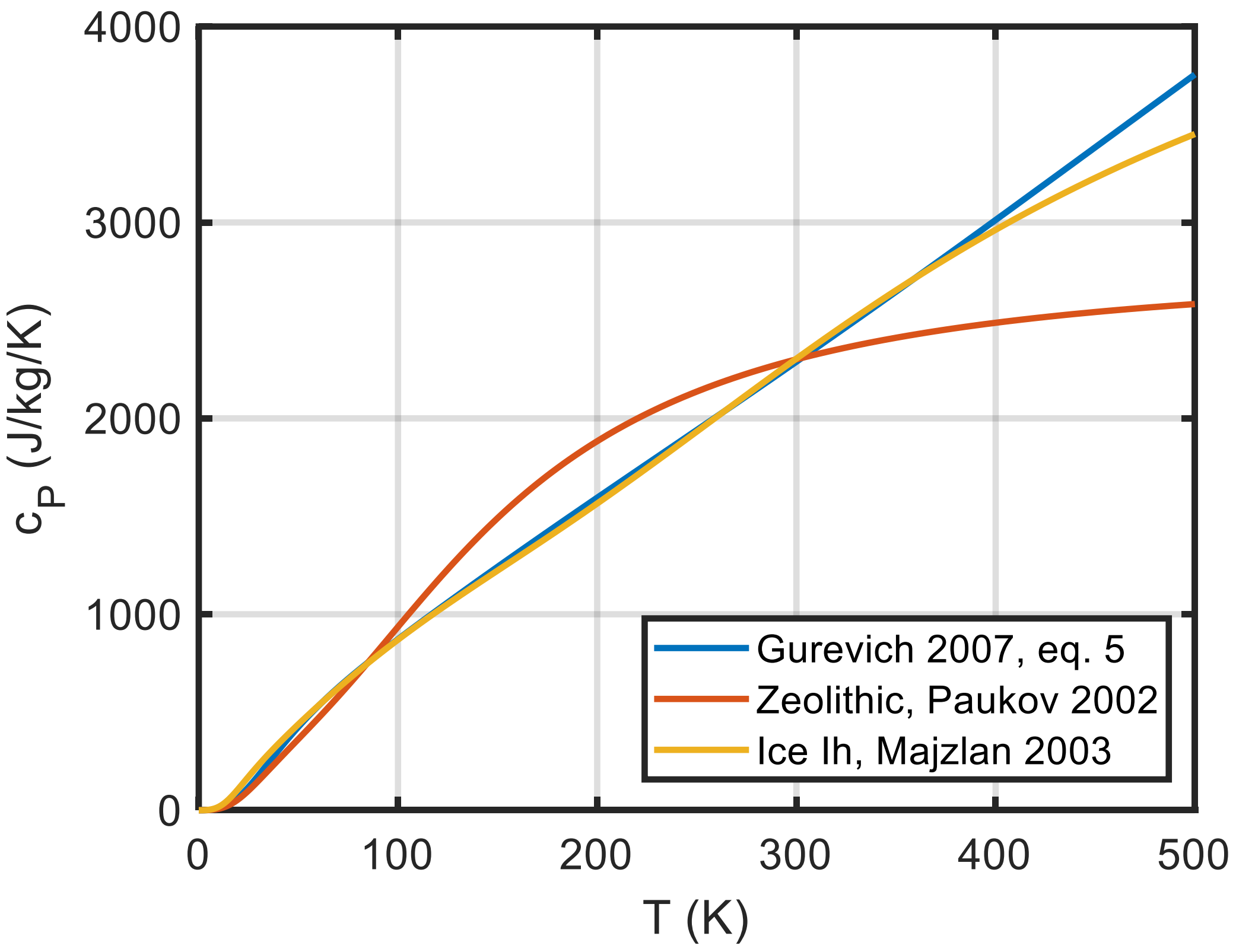


**Figure 13 bounds on crystal water $C_P$ contribution. It appears that the curve given by Gurevich 2007 is a good estimate for crystal water, we use it as our default.**

Experimentally, from the measured $C_P$ of epsomite and anhydrous magnesium sulphate, Gurevich et al. [282] derived the $C_P$ of crystal hydrate water in the range 0-303 K (and probably higher temperatures as extrapolation) as:

$$C_P^0(T)\{\mathrm{H_2O(cr)}\} = a_0 T (C_V)^2 + C_V$$
$$C_V = 3\left[(1/3)\sum_{j=1}^{3} a_j D_j(\theta_j/T) + a_4 E(\theta_E/T) + a_5 K(\theta_L/T, \theta_U/T)\right] \quad (20)$$

with the Debye, Einstein and Kieffer[15] functions $D(\theta_j/T)$, $E(\theta_E/T)$, $K(\theta_L/T, \theta_U/T)$ and the fitted constants:

| | | |
|---|---|---|
| $a_0$ | (J/mol) | $1.4689\cdot10^{-5}$ |
| $a_1$ | | 0.33333 |
| $a_2$ | | 0.33333 |
| $a_3$ | | 0.33333 |

[15] 'optical continuum' [62]. The integral, eqn. (21) can also be very precisely approximated by a new Padé approximant, see appendix 1.1

| | |
|---|---|
| $a_4$ | 0.33333 |
| $a_5$ | 0.33333 |
| $\theta_1$ (K) | 171 |
| $\theta_2$ (K) | 287 |
| $\theta_3$ (K) | 671 |
| $\theta_E$ (K) | 2047 |
| $\theta_L$ (K) | 484 |
| $\theta_U$ (K) | 1444 |

$$K(\theta_L/T, \theta_U/T) \equiv \frac{3R}{\theta_L/T - \theta_U/T} \int_{\theta_L/T}^{\theta_U/T} \frac{x^2 \exp(x)}{\left(\exp(x)-1\right)^2} dx \qquad (21)$$

This correlation for the heat capacity of crystal water gives a curve very similar to that for ice Ih of Majzlan et al. [281], see Figure 13. We use equation (20) as our standard curve for adsorbed, excess and hydrate water. It can probably be used with reasonable accuracy (better 10%) up to ~500 K. For even higher temperatures, one can use the general $H_2O$ curve of Robertson [100],

$C_P$[J/mol/K] $= 85.285 - 0.00155T - 537000/T^2 - 620.9/\sqrt{T} - 1.226 \cdot 10^{-6} T^2$, 298 to 1500 K.

More specifically, for all smectite end-members as well as Ca- and Mg-muscovite and -phlogopite the high-temperature $C_P$ , as a function of hydration, has been modelled by Vidal & Dubacq [273]. Viellard [274] gives detailed $C_P$ mostly for clays, but also other minerals, in the high-temperature range.

The heat capacity of water in zeolites (not a common mineral group in astro-material) or microporous minerals is different to the standard correlation (compare Figure 13), it has been measured and fitted to equation (22) by Paukov et al. [283] on the paranatrolite–tetranatrolite pair and the analcime–dehydrated analcime data of [284].

$$\Delta C_P^{zeo}(T) = \frac{1}{3} D(\theta_l/T) + \frac{2}{3} D(\theta_{tr}/T) + E(T_E/T) \quad \text{[J/mol/K]} \qquad (22)$$

This ‘zeolithic’ water heat capacity is, however, not well constrained above ~200 K where an anomalous behaviour resembling a glass transition appears in the data.

with the fitted constants for natrolite, $\theta_l$ = 175 K, $\theta_{tr}$ = 450 K, and $T_E$ = 583 K (and for analcime, $\theta_l$ = 230 K, $\theta_{tr}$ = 230 K, and $T_E$ = 525 K).

A more physical analysis has been performed by Geiger et al. [285] who also measured the water heat capacity in various minerals with microporous networks (see also [286]) .

A collection of results is shown in Figure 14**Fehler! Verweisquelle konnte nicht gefunden werden.**, from which we conclude that, at least for microporous minerals, the actual water heat capacity

depends on the mineral and can vary by roughly ± (4.3+0.014*T*) J/mol/K, 30≤*T*≤300 K.

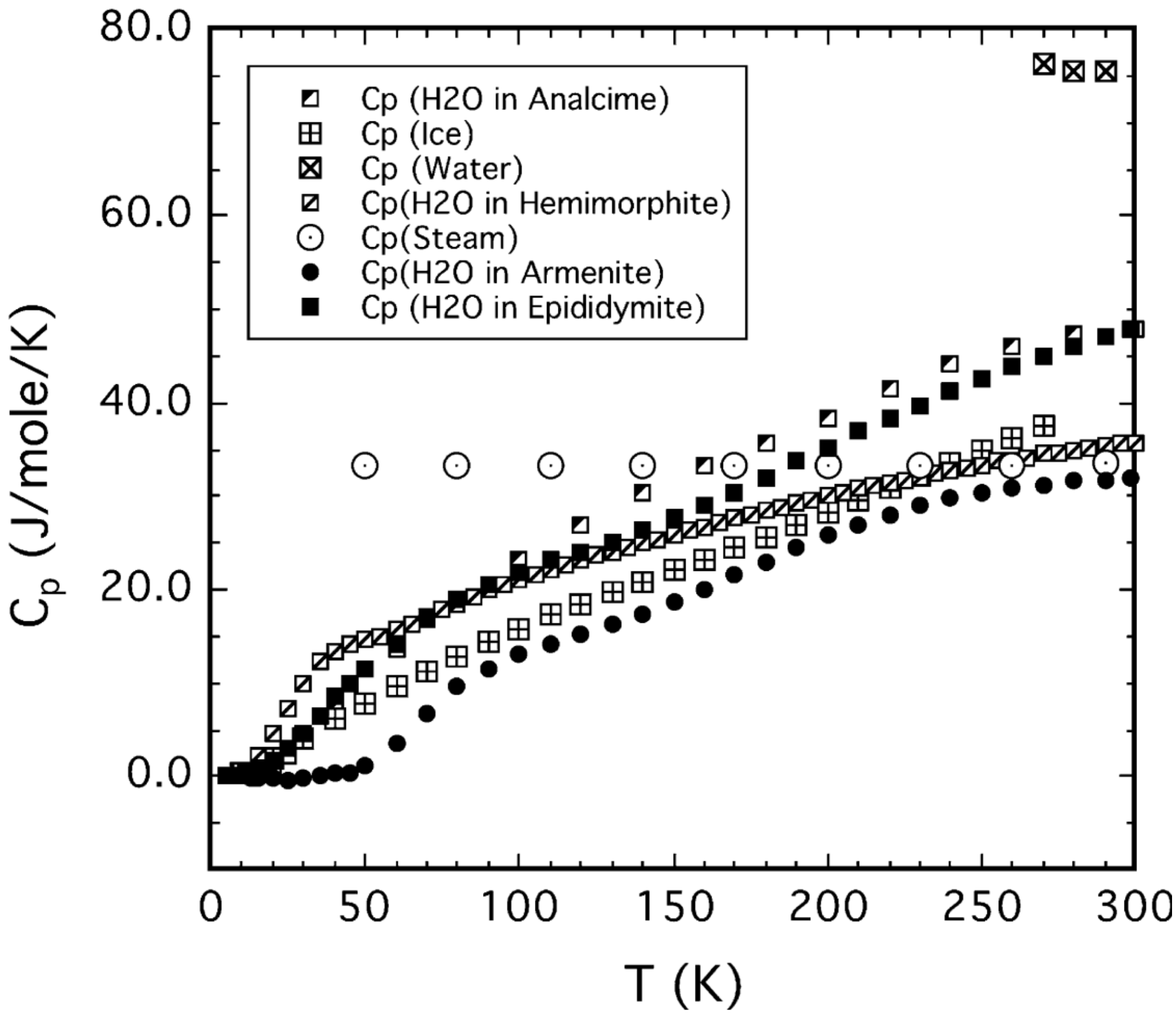


**Figure 14 (from [285], their fig.9) Heat capacity behaviour of confined $H_2O$ in armenite and epididymite as well as for hemimorphite [287] and analcime [284] at 0 K < T < 300 K. The squares with the + symbol are the $C_P$ of ice [288], the squares with the symbol the $C_P$ of super-cooled liquid water [289] and the circles with the symbol the $C_P$ of ideal $H_2O$ gas [290].**

## 2.13 Rare minerals

They do appear sometimes in astro-materials. We collected $C_P$ data on the following minerals (non-exhaustive list) [223, 225, 226].

### 2.13.1 Garnets

Garnets are hard, abrasive nesosilicates having the general formula $X_3Y_2(SiO_4)_3$. The X site is usually occupied by divalent cations (Ca, Mg, Fe, Mn)$^{2+}$ and the Y site by trivalent cations (Al, Fe, Cr)$^{3+}$. The garnet endmember minerals pyrope $Mg_3Al_2(SiO_4)_3$, almandine $Fe_3Al_2(SiO_4)_3$, spessartine

$Mn^{2+}{}_3Al_2(SiO_4)_3$; grossular $Ca_3Al_2(SiO_4)_3$ , rare uvarovite $Ca_3Cr_2(SiO_4)_3$ and andradite $Ca_3Fe_2Si_3O_{12}$ make up two solid solution series: pyrope-almandine-spessartine and uvarovite-grossular-andradite.

### 2.13.2 Spinels

Magnesio-iron spinel $(Mg,Fe)Al_2O_4$, is a common mineral in the Ca-Al-rich inclusions (CAIs) in some chondritic meteorites [291].

Chromite is iron chromium oxide, $FeCr_2O_4$. It is an oxide mineral belonging to the spinel structural group[16]. The element magnesium can substitute for iron in variable amounts as it forms a solid solution with magnesiochromite $MgCr_2O_4$. A substitution of the element aluminium can also occur, leading to hercynite $FeAl_2O_4$.

We consider the three most common endmember spinels in the substitution $(Mg,Fe)(Al,Cr)_2O_4$ :

Magnesiospinel ('spinel proper') $MgAl_2O_4$, chromite $FeCr_2O_4$, and hercynite $FeAl_2O_4$.

### 2.13.3 Other rare minerals

**Carlsbergite** is a nitride mineral that has the chemical formula CrN, or chromium nitride. It occurs in meteorites along the grain boundaries of kamacite or troilite in the form of tiny plates. It occurs associated with kamacite, taenite, daubreelite, troilite and sphalerite.

**Schreibersite**, $(Fe,Ni)_3P$, is generally a rare iron nickel phosphide mineral though common in iron-nickel meteorites. Even there it is a minor constituent, as there is only 0.50 to 1.3 wt.% P in iron meteorites [292]. Schreibersite and other meteoric phosphorus bearing minerals may be the ultimate source for the phosphorus on Earth.

**Tridymite** is a high-temperature polymorph of silica $SiO_2$ and usually occurs as minute tabular white or colorless pseudo-hexagonal crystals, or scales, in cavities in felsic volcanic rocks. It was found on Mars and probably is evidence for Martian silicic volcanism [293].

**Hibonite**, $CaAl_{12}O_{19}$ or, more generally, $(Ca,Ce)(Mg,Fe^{2+})Al_{10}(Ti^{4+},Al)O_{19}$ has been found in the Allende meteorite and in CAIs;

---

[16] Note the common usage of words like 'spinels', 'halites', 'perovskites', 'garnets' not for compositionally related minerals, but for having the same structure as the name-giving mineral. For example, chromite $FeCr_2O_4$ has a spinel structure.

**Melilite**, $(Ca,Na)_2(Al,Mg,Fe^{2+})(Si,Al)_2O_7$ in CAIs. Both hibonite and melilite are thought to have condensed very early during the cooling of the solar nebula, so they represent some of the most primordial minerals in the solar system [228]. Hibonite is even one of the minerals in presolar grains (besides silicate minerals (olivines and pyroxenes), corundum ($Al_2O_3$), spinel ($MgAl_2O_4$), graphite (C), diamond (C), titanium oxide ($TiO_2$), silicon carbide (SiC), titanium carbide (TiC) and other carbides within C and SiC grains, silicon nitride ($Si_3N_4$) [294].

**Moissanite** SiC, **bridgmanite** $(Mg,Fe)SiO_3$, **ringwoodite** γ-$(Mg,Fe)_2SiO_4$, **majorite** $Mg_3(MgSi)Si_3O_{12}$ and **wadsleyite** β-$Mg_2SiO_4$ are high-pressure polymorphs and rare minerals found only in meteorites but thought to be significant components of the deep Earth [228].

## 2.14 Carbon and carbon-rich / organic matter

Significant amounts of carbonaceous materials are contained in carbonaceous chondrites, mainly as solvent unextractable macromolecular matter, analogous to terrestrial kerogen or poorly crystalline graphite. During heating, these kerogen-type carbonaceous materials lose their labile fractions, and become more and more graphitized [295]. Note that ‘carbonaceous’ is a bit of a misnomer since the carbon content of some *carbonaceous* chondrites does not exceed 3 to 4% [296]; ureilite achondrites, on the other hand, tend to have a similar fraction of carbon (~3%) but in the form of graphite and trace amounts of nanodiamonds.

Elemental, ‘native’, stable carbon, graphite, has a well-known $c_P(T)$ curve up to temperatures of ~4000 K. Imperfect graphite, like ‘lamp black’ with numerous stacking faults and small crystallites exhibits an excess $C_P$ at low temperatures, significant only at <10 K. Just for the sake of completeness, we include diamond in the database; diamond has an extremely high Debye-Temperature, thus a smaller $C_P$ than most substances (except Be) over a wide range of temperatures.

For ill-defined, hydrogen-bearing, partly volatile and partly macromolecular carbonaceous material, we use two analogues: coal and ICOM (‘ill-defined complex organic matter’).

The specific heat capacity of coal is the highest of any mineral, being roughly 50% higher than that of graphite in the range 300 – 600 K. Typical ‘Sub-bituminous coal’ proposed as a kerogen substitute has the following composition [297]: total volatile matter 30-40%, ash 10%, moisture a few %. $c_P$ is referred to ‘daf’ composition, = dry, ash-free matter (i.e. the $c_P$ contribution of water and ash have been removed).

In the case of coal at elevated temperatures, irreversible changes of carbonaceous material associated with the release of volatile matter: *coal → solid carbonaceous material + released volatiles* takes

place. For this reason, the specific heat capacity of coal as a function of temperature is complex one; for example, our reference curve (which is the initial $c_P$ upon heating) has a maximum at ~900 K, decreasing at higher temperatures when the volatiles have mostly been liberated; it is then not reproducible but will become closer to the $c_P$ of graphite.

ICOM, or kerogen on Earth (a solid organic matter in sedimentary rocks), is insoluble in normal organic solvents because of the high molecular weight (upwards of 1,000 dalton) of its component compounds. It does not have a specific chemical formula. The soluble portion of kerogen is known as bitumen [295].

For most natural organic matter, at best the elemental composition (mass or atomic-% of C, H, O, N, S etc.) is known. Laštovka, Fulem et al. [298] have shown that $c_P(T)$ of most solid hydrocarbons (ICOM) can be well predicted from 0 K to the melting point based on a parametrization in 1/(mean atomic weight) = α, to an accuracy of ~6% rms (relative deviations max. 18%). The prediction is good for molar masses exceeding 200 g/mol and compounds with low mass fractions of hetero-atoms (O,N,S). On the high-$T$ end, $c_P$ predictions with values exceeding 2500 J/kg/K are not likely to be quantitative. The correlation equation has been trained with pure substances 0.10≤α≤0.22 mol/g, i.e., mean atomic weight from 4.5 to 10. The $c_P(T)$ curves of complex solid hydrocarbons look very different than those of silicates, almost linear with $T$, often with a slight bump at low temperatures, reminiscent of solid ammonia dehydrate, water ice Ih and polymers.
We set, a bit arbitrarily, α=0.138 (mean atomic mass 7.26 found in literature) and calculate $c_P(T)$ after Laštovka, Fulem et al. [298]. This will be the default $c_P(T)$ for ‘organic matter’ (if not dominated by elemental carbon, i.e., graphite).

There is also volatile organic matter. Some fresh carbonaceous chondrite meteorites smell of ‘tar’, so there is obviously some highly volatile organic fractioin in there, VOC, that is released already at room temperature in minute amounts. However, the mass fraction of VOC seems to be irrelevant for $c_P$ (not-macromolecular organic compounds <1000 ppm in total [296]; ~ 100 ppm VOC in Murchison (CM2) released from 20°C to 300 °C [299]).

Summarizing the thermal alteration for organic matter, decomposition or pyrolysis begins to be significant at ~200°C (coal: 250°C). It is complete at ~1000°C. Volatile organic matter desorbs already at room $T$, but in insignificant amounts for $c_P$ (<0.1% by mass).

## 2.15 Glasses

Some minerals in their amorphous state (structural glasses) form from quenched (silicate) melts, that is, cooled very quickly e.g., after impact events or volcanic eruptions. Lunar regolith (in most Apollo samples) contains a significant mass fraction of this type of glass. These glasses usually have a significantly lower density than their crystalline polymorphs.

Another type forms from impact shock, either as dense diaplectic glass (formed by a high-pressure solid-solid transition [300]), or also permanently densified glass, which can form by the quenching of dense mineral melts produced by high-pressure shock waves. The amorphous feldspar maskelynite (with a plagioclase composition $AbAn_{70\text{-}90}$) is abundant in Martian meteorites (shergottites) but it is not clear whether it is a densified glass from melt [301] or diaplectic [302].

Our terminology will only consider composition, thus for example an anorthite crystal, molten and subsequently quenched, becomes anorthite glass.

Glasses are in principle metastable, and can devitrify over geological timescales if they contain water (over time transforming to fine-grained mineral crystal fibres, observable in water-containing glasses like obsidian on the Earth's surface; no terrestrial specimens older than Cretaceous age are known). For the Moon, however, glass usually remains glass: the maximum water content of lunar volcanic glass is 100 ppm, for agglutinate glass it is effectively zero (Ryan Zeigler, pers. comm. 2019 to MZ). Thus, the lunar glasses do not really devitrify, because of the low water content. There is glass in meteorites that is 4.5 billion years old.

Glasses and other amorphous materials show typical anomalies in $c_P$ at (very) low temperatures though it is surprising that $c_P(T)$ of a solid lacking a crystal lattice is still quite similar to that of the crystallized variant. Obviously, the peaks in $c_P$ associated to lattice phase transitions are lacking in glasses.

Compared to the expected low-temperature Debye contribution, in glasses there are additional modes of vibration – one can be described by a two-level system (TLS), and some extra modes. In glasses, a quasilinear term in $C_P(T)$ comes from the contribution of the TLS, which results in an excess to the $T^3$ heat capacity $C_D(T)$ expected from the Debye theory. On top of the low-temperature quasilinear TLS term, often a so-called 'Boson peak' appears around 10 K with a long tail to higher temperatures (see below).

The specific heat of densified and diaplectic glasses seem to tend to the $C_P$ of the crystalline polymorph; $C_P$ differences between the normal and the high density glass exist, but seem to become significant only below ~90 K [303], reaching 50% below 10 K, that is, the densified glass shows less of the additional heat capacity found in normal glass.

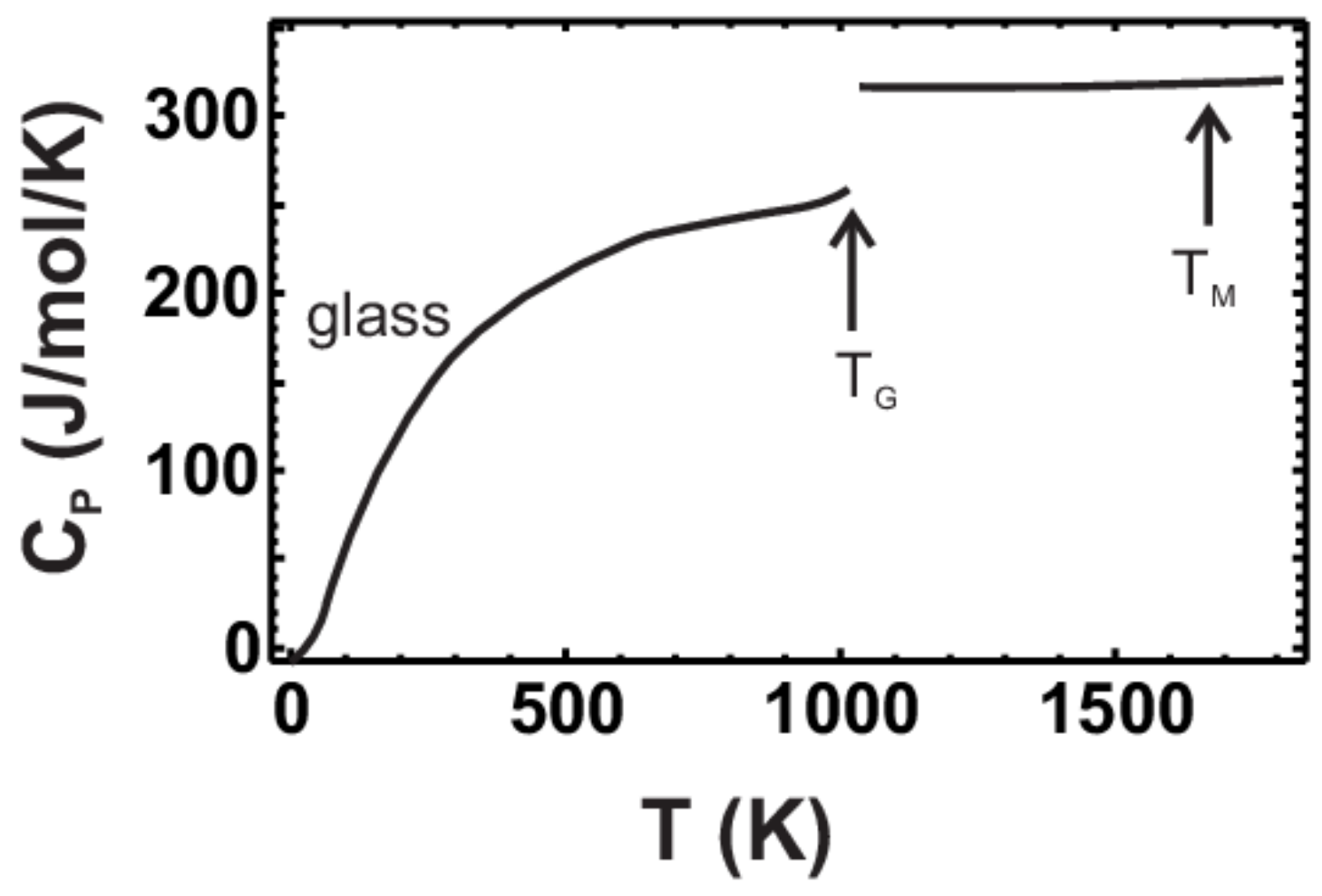


**Figure 15 Schematic $C_P$ curve of a glass, $T_G$ is the glass transition temperature, $T_M$ the melting point. After [183]. The heat capacity of the liquid, or the glass for $T>T_G$, is always greater than the heat capacity of the solid [Richet, P. in 90]**

Summarizing, it is found that, in all glasses similarly,

1) below 1-2 K: additive ~superlinear term $\propto T^{(1+\delta)}$, $0 \le \delta < 0.5$.
At very low temperatures $C_P$ can also depend on its cooling history and the number density of defects

2) 'Boson peak' excess $C_P$ at around ~10 K extending to the order of 90 K

3) Suppression of lambda peaks caused by phase transitions present in the crystalline form (example Hed)

4) $C_P$-$C_V$ allowed to be different for the glassy state compared to the crystalline polymorph, since thermal expansion and compressibility change, in general

5) At high temperatures, the onset of the glass transition (between 900 and 1000 K, typically well below the melting temperature of the crystalline variety) produces a broad $c_P$ peak or step (configurational heat capacity) of the order of ~8 J/g-atom/K, see Figure 15. Typical curves for hydrous basaltic glasses are given in [85] .

## 2.16 Solar system ices

Ices relevant in the solar system (comets, icy moons, TNOs) and reviewed for our database are: water ice Ih, carbon dioxide $CO_2$, carbon monoxide CO, methane $CH_4$, ethane $C_2H_6$, nitrogen $N_2$, ammonia dihydrate $NH_3 \cdot 2H_2O$, ethanol $C_2H_5OH$, and methanol $CH_3OH$. Except $CO_2$ and ethanol, all have transitions in their $c_P(T)$ curve. Figure 16 provides an overview; the curves end at the respective triple points.

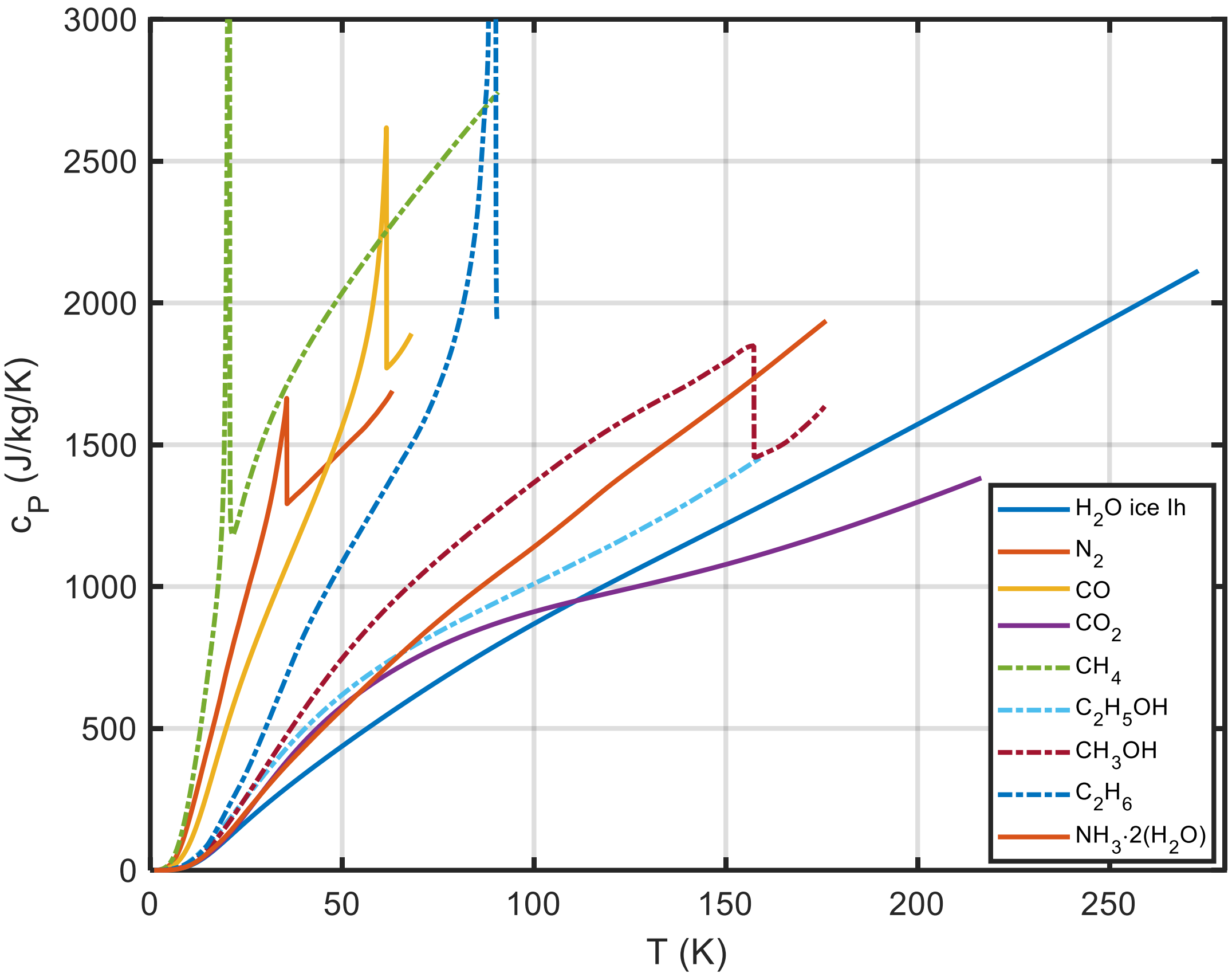


**Figure 16 Overview: c_p($T$) of common solar system ices. The curious dip for methanol at 157.34 K is real, it is the α/β conversion 'from crystal II to crystal I' just before melting.**

Note that the specific heat of most ices at low temperatures (say 40 K) is much higher (factor 3 to 20) than the $c_P$ of silicates at the same temperature! It is instructive to see why – There is two components: first, volatiles, having a much lower melting point than silicates, also have a much lower Debye temperature. Second, most volatiles have a lower average atomic mass $A_{av}$ (3 …14) than silicates (~22), and $c_P$ scales with $1/A_{av}$. More quantitatively, we recall the famous Lindemann formula, which can be written $T_m \cong cA\theta^2 a^2$ [304] with $T_m$ melting temperature in K, $\theta$ Debye temperature (taken, e.g., at a temperature $T^*$ when $C_v(T^*)=1/2\ C_v\ (T\rightarrow\infty)$), a typical interatomic distance (typically cube root of the volume per atom = $(M/\rho/N_A)^{1/3}$). Also, $c_P \propto 1/A_{av}$. Thus, the ratio of the heat capacities at some low, fixed temperature $T$ is approximately (D is the Debye function)

$$\frac{c_{P,1}}{c_{P,2}} \cong \frac{A_{av,2}\,\mathrm{D}(\theta_1/T)}{A_{av,1}\,\mathrm{D}(\theta_2/T)}; \quad \frac{\theta_1}{\theta_2} \cong \sqrt{\frac{T_{m,1}}{T_{m,2}}\frac{A_{av,2}}{A_{av,1}}\frac{a_2^2}{a_1^2}}$$

Typical volatiles have $T_m$=150 K, a=3.7 Å and $A_{av}$=8.5 u, while typical silicates have $T_m$=1500 K, a=4.7 Å and $A_{av}$=22 u. Thus, the ratio of Debye temperatures is expected to be approx. 0.65 (actually it is rather ~0.25).

At 40 K, silicates have typical Debye temperatures of 400 K; volatiles have typically 100 K. Again at 40 K, the D ratios is thus D(100/40)/D(400/40) ≈ 10. The ratio mean atomic mass is 22/8.5≈ 2.6; thus, at 40 K, volatiles typically have a $c_P$ that is predicted to be ~26 times higher than that of silicates at the same temperature (actually rather 34 times!).

Note that we here compile the $c_P$ of the crystalline state of cryocrystals. Many complex ices (methanol, ethanol) have a variety of polymorphs including amorphous and 'glassy crystal' states, with a $c_P$ that is typically 1.5-2× higher above the glass transition temperature and ~1-2% below compared to that of the thermodynamically stable crystal at the same temperature. It is not clear whether or under which conditions solar system ices exist in the glassy or crystalline state.

## 2.17 Tholins

Frequently, a very low thermal inertia $\Gamma(T) = \sqrt{\rho(T)k(T)c_P(T)}$ (as low as 0.1 to 3 J m$^{-2}$ K$^{-1}$ s$^{-½}$) is observed for comets and TNOs [305, 306, and references therein]. While surface material bulk densities are not so different than on other bodies, and amorphous ice with a lower thermal conductivity than crystalline ice may be present and, for granular media, the radiative part $\propto T^3$ of thermal conductivity is smaller at low temperatures, an important effect comes from the specific heat: directly, since $\Gamma \propto \sqrt{c_P}$ , and indirectly, since solid state thermal conductivity scales with $c_P$ at cryogenic temperatures (that is, below the maximum in *k* for well crystallized solids, e.g., [307]).

Objects beyond the ice line (comets, notably) and TNOs in particular are believed to contain a substantial fraction of frozen volatiles in their surface material. Small TNOs are thought to be low-density mixtures of rock and ice with some organic (carbon-containing) surface material such as 'tholins', detected in their spectra. The composition of some small TNOs could be similar to that of comets. The optical surfaces of small bodies are subject to modification by intense radiation, solar wind and micrometeorites. Consequently, the thin optical surface layer could be quite different from the icy regolith underneath, and not representative of the bulk composition of the body.

De Bergh et al, in [308], note that on the brightest TNOs and Centaurs (with VIS-NIR spectroscopy) several surface ices have been detected: $H_2O$, $CH_4$, $N_2$, $CH_3OH$, $C_2H_6$, $CO_2$, $NH_3$•n$H_2O$, and possibly HCN, in various combinations; water ice is by far the most common. Crystalline water ice, and possibly ammonia ice, have been found from spectroscopic observations of the TNO Orcus between 1.4

and 2.4 $\mu$m [309]. So, outer solar system body surfaces could be modeled as a mixture of ices ($H_2O$, $CO_2$, CO, $CH_4$,$N_2$, ethanol, methanol, ammonia dihydrate), maybe amorphous carbon, tholins and silicate 'dust' as a simple mechanical mixture.

The enigmatic tholins are believed to be created by intense radiation. Tholins are apparently found in great abundance on the surface of icy bodies in the outer solar system. Four major tholins have been proposed to fit the reddening slope [310]:

- Titan-tholin, believed to be produced from a mixture of 90% $N_2$ and 10% $CH_4$ (gaseous methane)
- Triton-tholin, as above but with very low (0.1%) methane content
- (ethane) Ice tholin I, believed to be produced from a mixture of 86% $H_2O$ and 14% $C_2H_6$ (ethane)
- (methanol) Ice tholin II, 80% $H_2O$, 16% $CH_3OH$ (methanol) and 3% $CO_2$

As an illustration of the two extreme TNO colour classes BB and RR, the following compositions have been suggested [311]

- for Sedna (**RR** very red): 24% Triton-tholin, 7% carbon, 10% $N_2$, 26% methanol, and 33% methane
- for Orcus (**BB**, grey/blue): 85% amorphous carbon, +4% Titan-tholin, and 11% $H_2O$ ice

Tholins ('complex abiotic organic gunk', [14]) are not one specific compound but rather are descriptive of a spectrum of molecules that give a reddish, organic surface covering on certain planetary surfaces. See, e.g., [312]. The typical composition of laboratory 'Titan tholins' is 35 at-% C, 15-30 at-% N, rest H. They are probably macromolecular and not soluble; no specific heat data are available.

McKay et al. [313] report that 'a detailed analysis of the organic compounds contained in tholin ... show that they include a complex organic mix of simple alkanes, aromatic compounds, heteropolymers, and amino acid precursors'. If that is so, tholins could be modelled like ill-defined complex organic matter based on elemental composition alone (α value).

**Table 7 Some lab tholins [313]. α is the inverse of the average atomic mass**

| Reference | Stoichiometry | C/N ratio | Conditions | α (mol/g) |
|---|---|---|---|---|
| Sagan et al. (1984) | $C_8H_{13}N_4$ | 1.9 | Low P | 0.151 |
| Coll et al. (1995) | $C_{11}H_{11}N$ | 11 | Low *T* | 0.146 |

| | | | | |
|---|---|---|---|---|
| McKay (1996) | $C_{11}H_{11}N_2$ | 5.5 | High $T,P$ | 0.140 |
| Coll et al. (1999) | $C_{11}H_4N_{14}$ | 2.8 | Low $T,P$ | 0.0873 |

## 2.18 Other compounds

**<u>For hydrate and crystal water</u>**: see section 2.12.1.

**<u>Calorimetric reference substances:</u>** we include some useful elements or compounds which are commonly used either as specific heat or temperature calibration materials or whose specific heat has to be known precisely for correcting $C_P$ measurements (copper Cu, aluminum Al, vacuum grease Apiezon® N etc.)

# 3 Some example applications

In this chapter, we demonstrate some applications of the $c_P$ database.

## 3.1 The $c_P(T)$ of tholin analogs

In this section, we present the calculated $c_P(T)$ curves of the tholin analogues discussed in section 2.17 above.

Figure 17 gives an overview of the $c_P(T)$ curves of our various model tholins we will discuss hereunder. One sees that there is a large variation between models, but that a silicate (e.g., lunar regolith) $c_P(T)$ curve is certainly not appropriate, it would be about an order of magnitude too low.

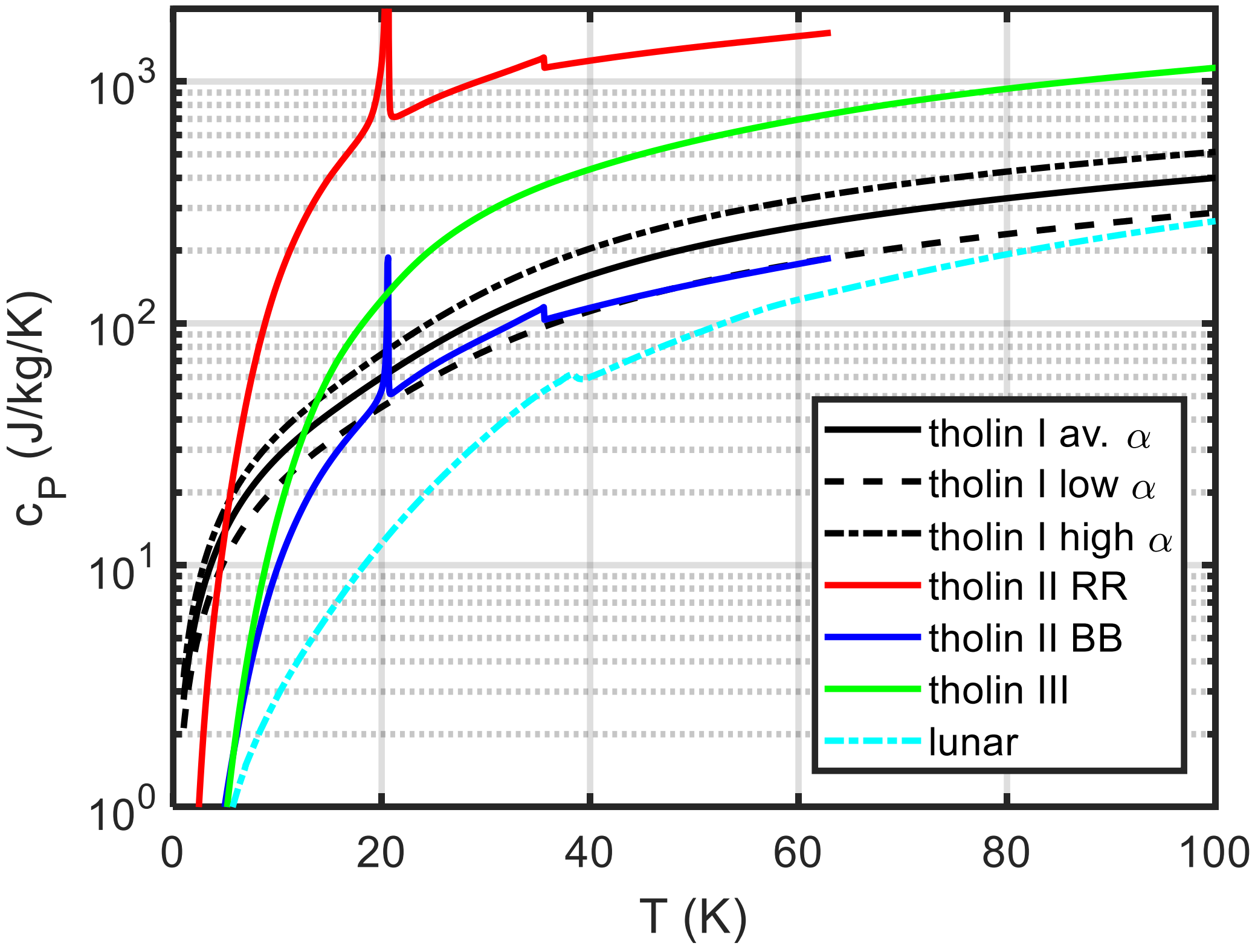


**Figure 17 Overview, specific heat of some model tholins. The range of specific heat at low temperatures is about one order of magnitude. The large λ peak at ~20 K is due to methane, the small anomaly near 35 K due to nitrogen. For comparison with common silicates, our lunar regolith curve is given**

### 3.1.1 Tholin model 1: ill-defined organic matter

We use the model of [298] with the parameter α (inverse of average atomic mass) varying between 0.087 and 0.151 (compare Table 7 and [312]), the result is shown in Figure 18. Calculated $c_P$ values are much lower than for model 3 (ammonia dihydrate).

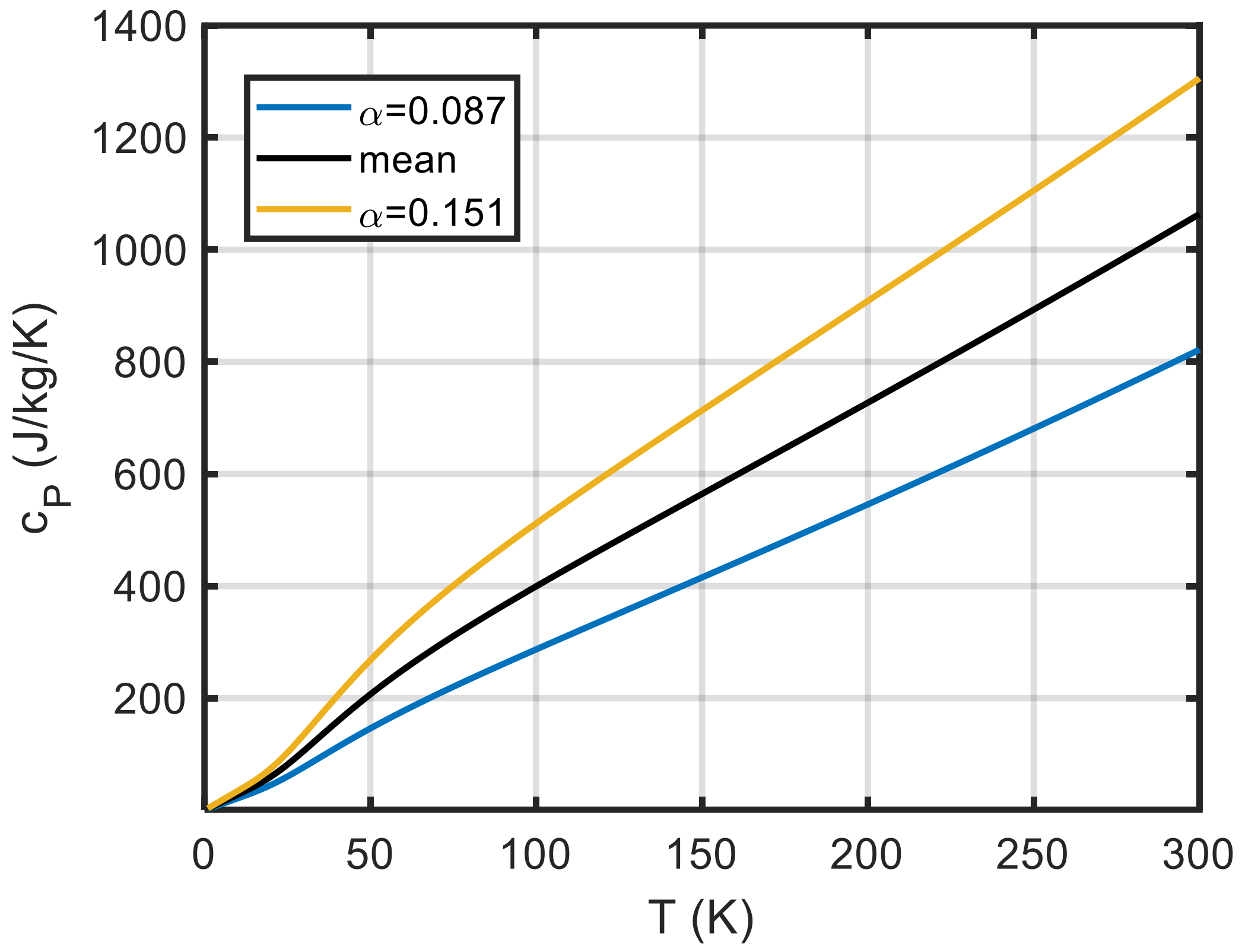


**Figure 18 $c_P$ of tholins, model 1. The blue and dark yellow curves indicate the likely range, the black curve is just the average of the blue and red ones.**

### 3.1.2 Tholin model 2: mix of ices and graphite

The following 'basic' tholins are defined (Table 8):

**Table 8 Basic tholins, composition [310]**

| Ice type abbreviation | Type | Composition |
|---|---|---|
| T1 | Titan | 90% $N_2$, 10% $CH_4$ |
| T2 | Triton | $N_2$ (0.1% $CH_4$ is negligible for $c_P$) |
| T3 | Ice tholin I | 86% $H_2O$ ice, 14% ethane $C_2H_6$ |
| T4 | Ice tholin II | 80% $H_2O$ ice, 16% $CH_3OH$, 3% $CO_2$ |

As an illustration of the two extreme classes BB and RR, the following (Table 9) compositions have been suggested

**Table 9 Composition for extreme spectral types tholins [311]**

| Spectral type | Composition | Example |
|---|---|---|
| RR, very red | 24% T1 ice, 7% graphite, 10% T2 ice ($N_2$), 26% methanol, 33% $CH_4$ | Sedna |
| BB, gray-blue | 85% graphite, 4% T1 ice, 11% $H_2O$ ice | Orcus |

and the resulting specific heat curves are shown in Figure 17 and for the basic tholins 1 – 4 in Figure 19.

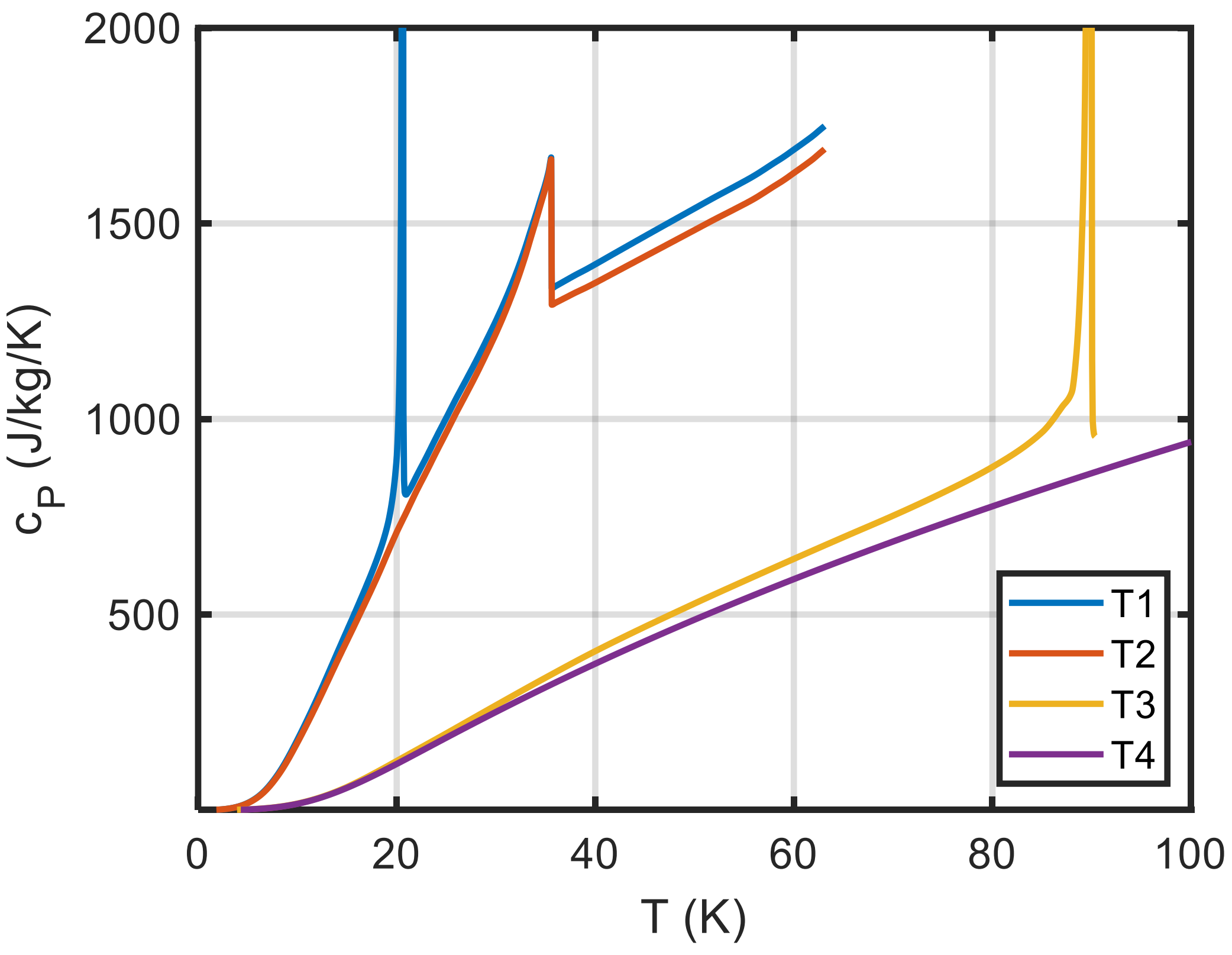


**Figure 19** $c_P(T)$ **of basic tholins, model 2**

### 3.1.3 Tholin model 3: Ammonia dihydrate $NH_3 \cdot 2H_2O$

Sometimes ammonia dihydrate has been taken as an analogue for 'tholins'. The specific heat for solid ammonia dihydrate is taken from [314] from near 0 K up to 176.2 K (melting temperature), Figure 20.

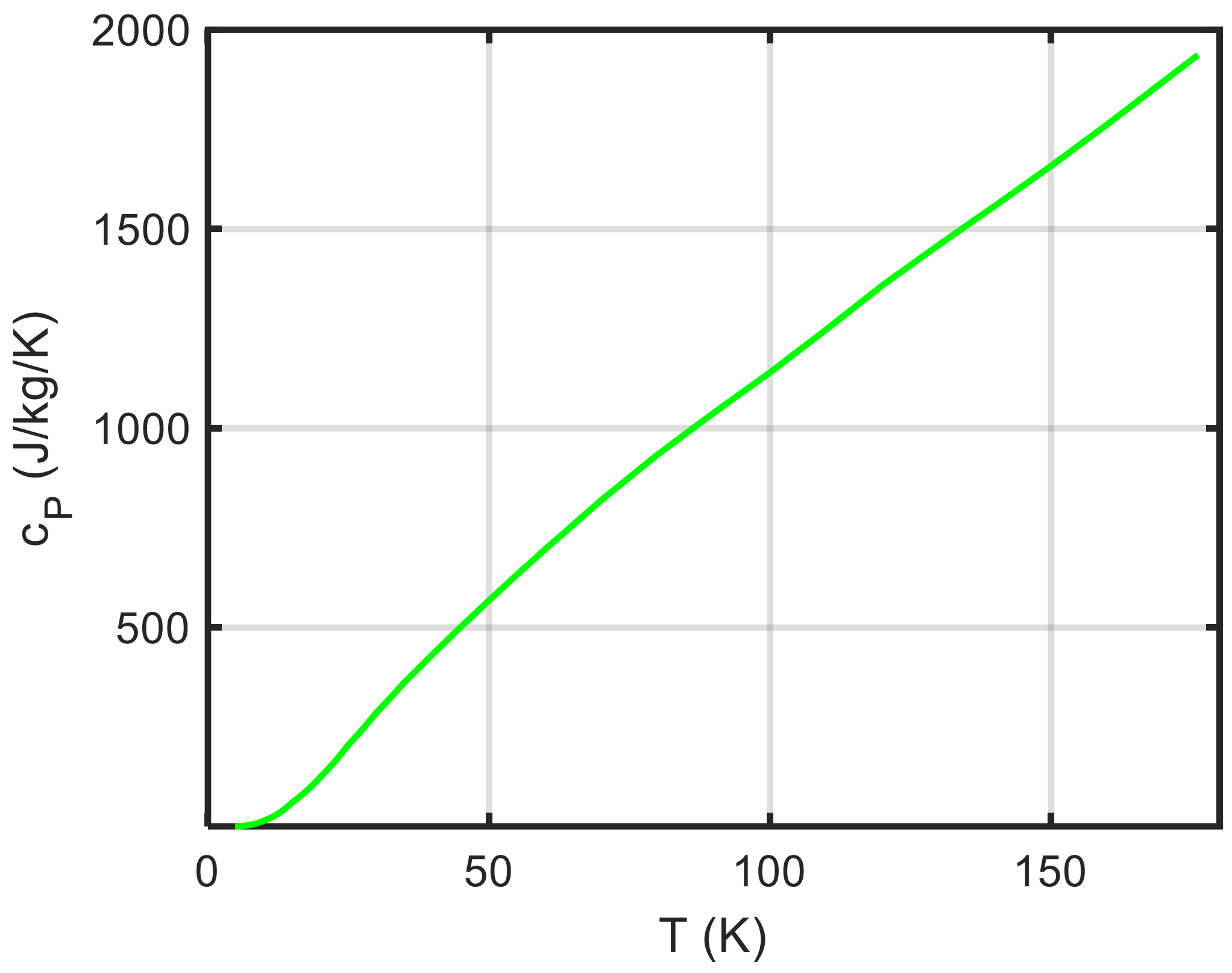


**Figure 20 Tholin model 3, specific heat of solid ammonia dihydrate, data of [314]**

## 3.2 The $c_P(T)$ of various regolith simulants

The most obvious 'forward' application is of course the construction of reference $c_P(T)$ curves for a material with known or assumed mineral composition. We start by calculating the specific heat of various regoliths simulants (detailed description see Appendix, section 5); compositions are given in Table 10.

Now applying the mixing model,

$c_P(T) = \sum_i w_i c_P^{(i)}(T)$ with $w_i$ the mass fractions of the constituents, $\sum w_i = 1$, we can immediately generate and plot the $c_P(T)$ curves, Figure 21.

**Table 10 Minerals and their mass fractions assumed for the $c_P(T)$ of DI regolith simulants. Some minerals appear more than once since they are part of different solid solutions. w is the mass fraction of the database mineral with**

abbreviation ‘abbr.’ (older nomenclature; Atg is antigorite, Plg Palygorskite=Attapulgite, Eps epsomite, Py pyrite, Vrm vermiculite, Sid siderite, Gy gypsum, Dol dolomite, Sms sodium metasilicate, Fo forsterite, Fa fayalite, Mag magnetite, En enstatite, Fs ferrosilite, Coal subbitumous coal, and FeNi meteoritic iron (10% Ni). For C2-1, antigorite has been substitute for its polymorph lizardite.

| CM-1 | | CM-2 | | CI-1 | | CI-2 | | C2-1 | | CR-1 | |
|---|---|---|---|---|---|---|---|---|---|---|---|
| **abbr.** | **w** | **abbr.** | **w** | **abbr.** | **w** | **abbr.** | **w** | **abbr.** | **w** | **abbr.** | **w** |
| Fa | 0.57 | Atg | 0.7 | Atg | 0.365 | Atg | 0.48 | Atg | 0.305 | Atg | 0.09 |
| Atg | 0.22 | Mag | 0.1 | Eps | 0.15 | Eps | 0.06 | Fo | 0.225 | En | 0.2325 |
| Fo | 0.0729 | Fo | 0.0675 | Mag | 0.115 | Mag | 0.135 | Fa | 0.025 | Fs | 0.0775 |
| Fa | 0.0081 | Fa | 0.0075 | Plg | 0.09 | Plg | 0.05 | Mag | 0.22 | Mag | 0.14 |
| Coal | 0.035 | Coal | 0.035 | Fo | 0.063 | Fo | 0.063 | Py | 0.085 | FeNi | 0.05 |
| Py | 0.025 | Py | 0.025 | Fa | 0.007 | Fa | 0.007 | Coal | 0.05 | Fo | 0.2475 |
| En | 0.015 | En | 0.015 | Py | 0.06 | Py | 0.065 | Vrm | 0.04 | Fa | 0.0825 |
| Fs | 0.005 | Fs | 0.005 | Vrm | 0.05 | Vrm | 0.09 | Plg | 0.04 | Py | 0.04 |
| Mag | 0.01 | Sms | 0.035 | Sd | 0.04 | Coal | 0.05 | Dol | 0.01 | Sms | 0.02 |
| Dol | 0.01 | Sd | 0.01 | Coal | 0.035 | | | | | Coal | 0.02 |
| Sms | 0.029 | | | Gp | 0.025 | | | | | | |

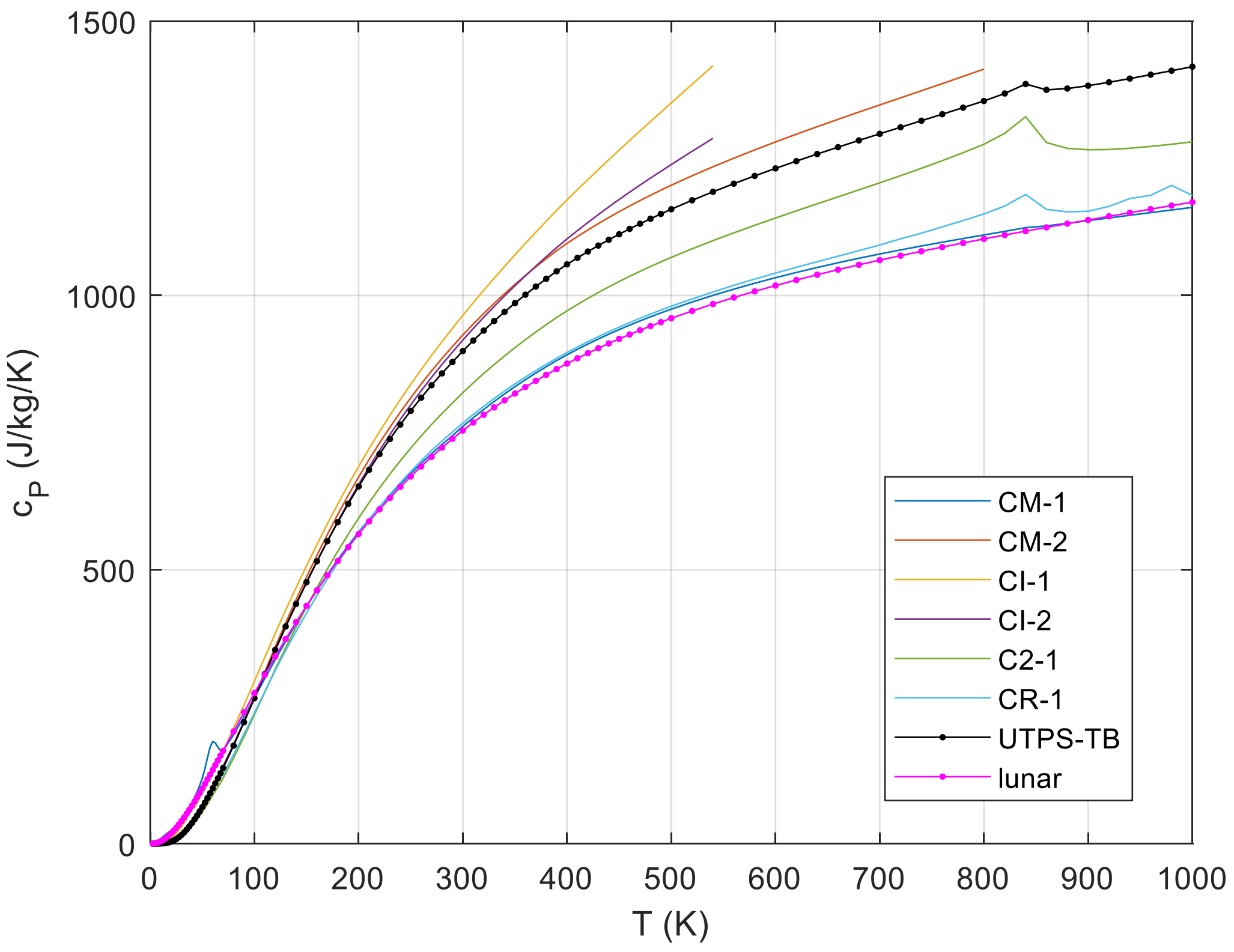


Figure 21 Calculated $c_P(T)$ of DI regolith simulants. For comparison, the standard lunar $c_P$-curve is given [315]. Note the ‘theoretical’ fayalite peak at ~60 K and the ‘theoretical’ magnetite peak at 840 K; the fayalite peak is expected to be smeared out in natural samples with the same mean fayalite content but from a range of olivine compositions. – For comparison, the standard lunar $c_P$-curve is given [315] .

## 3.3 Estimating the mineral composition from limited $c_P$ data and extrapolating

The basic idea is the following. Given a sample where the specific heat has been measured over a limited range of temperatures. Assume that we approximately know, or can guess, the mineral composition of this material, at least having an idea which minerals could be present in significant mass fractions (>1% or so). Then we can invert the mixing equation, equation (10), and solve for the mass fractions $w_i$ of the constituent minerals that produce a specific heat curve best fitting the data (Obviously, we need more data points ($T_i$, $c_{P,i}$) than the number of constituent minerals). Taking this composition as the best estimate of the truth, we can use equation (10) forward and calculate $c_P(T)$ for any temperature $T$ in which the constituent minerals are stable, that is, perform a physically meaningful extrapolation!

We have programmed and tested this method (for lunar data) with success. Here, we present only some main points; for more details (mathematics, figures), see appendix 1.2.
(In this section, to make notation more compact, $C$, $c$ designate specific heat of mixture and single endmember minerals, and $X$ stands for mass fractions)
Given the experimental $c_P$-data of a mineral mixture over a (wide as possible) temperature range and some idea about the main constituents, i.e., a list of endmember minerals. 'Main' means: mass fraction $X$ of a constituent $>X_{threshold} \approx 1\%$. We estimate the most likely mass fractions $X_i$ of the constituent minerals by (weighted) least squares solution of the constrained mixing equation (software function `cp_decompose`) and construct the model $c_P(T)$ curve over a wider temperature range (software function `cp_compose`). So far, this is simple and fast. More difficult and lengthy is the calculation of realistic uncertainties of the (extrapolated) model values, which we do by Monte Carlo, either adding random noise to the data or using a bootstrap method (bootstrap is preferred, since there are no assumptions about the form of the noise).
Note that the endmember mineral $c_P(T)$ curves are the base functions in our least-squares problem here; they are generally far from been orthogonal, and the problem only has a meaningful unique solution because of the constraint

$$X_i \geq 0, \quad \sum_i X_i = 1 \qquad (23).$$

Generally[17], low-temperature $c_P$ data are constraining the composition better than high-temperature ones, since the low temperature part is more 'diagnostic' of a compound, thus 'more orthogonal'. Given *M* experimental data points, $T_m, C_m(T_m), m = 1 \cdots M$ , we fit the *N*<*M* mass fractions $X_i$ using, as base functions, the $c_i(T)$ of *N* possible constituent minerals, since $C_m(T_m) = \sum_i X_i c_i(T_m)$ subject to the constraints (24). Let $\sigma$ be the uncertainties of the *C* data (weighting). This is a linear least-squares problem with bounds and linear constraints; for details, see appendix 1.2.
The first application of this method, for lunar surface material, is presented in the next section.

## 3.4 Construction of a lunar $c_P$ reference curve

Let us start by looking at all published (unsmoothed) Apollo $c_P$ data (see appendix). Although the 1σ uncertainty of the low-TAC Apollo specific heat data is only ~0.4%, it is clear (compare figures in Appendix, section 3) that the different Apollo samples have systematic differences among each other, within about ± 3%, likely due to compositional variations. This is also the range of relative differences between some previous lunar $c_P$ fit functions in the literature. We mention the polynomials in [316] and [317] , valid over the temperature range 90 K≤*T*≤350 K. Note that the Hemingway et al. [316] polynomial quickly diverges for *T*>350 K.

There are other correlation equations in the literature; Colozza [318] gives a crude extrapolation formula (logarithmic) of the lunar data up to melting temperatures. The expressions given by Ledlow et al. [319] are highly uncertain (possibly wrong) at high temperatures > 350 K and no improvement at low temperatures.

The work of Fujii and Osako [320] is often cited as a reference for 'basalt'. Actually, they measured thermal diffusivity of lunar samples and assumed the $c_P$ of this lunar basalt as a fit to the data on lunar crystalline rock 10057 by Robie et al. [321] . There is up to 5% deviation to the data [321] between 90 and 350 K.

A reasonable model for lunar regolith over wide temperature ranges, in particular for *T*>350 K up to the assumed melting temperature, ~1500 K, was given by Schreiner et al. [322]. It is notable that separate regression models are presented for high- and low-Ti Mare and Highlands regolith to demonstrate the effect of composition - after all, most Apollo samples are basaltic (nearside) mare material, and highlands regolith (including most of the far side) is under-represented. However, as can be seen

[17] complications arise if there are significant solid-solution composition-dependent transition peaks or other anomalies, as we have seen in the bronzite example in section 1.1.1.1

in Figure 22, the difference between Mare and Highlands is ~3.5% at most, and that between low- and high-Ti Mare <1%.

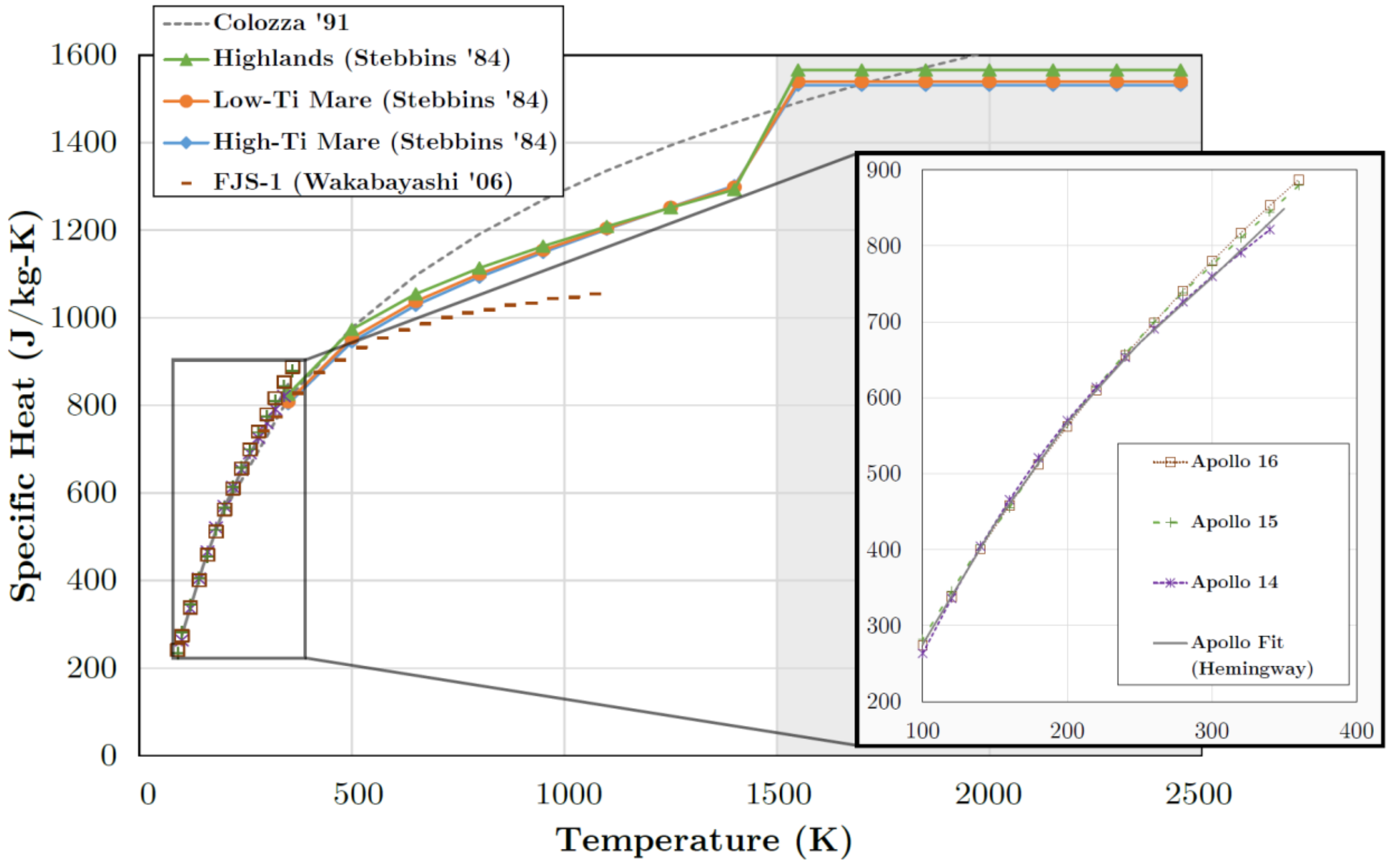


**Figure 22 The specific heat model for lunar regolith. In the lower temperature regime (≤ 350 K), a fit from [316] based on Apollo data is used. At higher temperatures (>350 K), a model by [323] is used. Melting temperature is 1500 K. [Reprinted from [322] Advances in Space Research, 57(5), Schreiner, S. et al., *Thermophysical property models for lunar regolith*, p. 1209-1222, 2016, with permission from Elsevier]**

We performed a preliminary exercise, using the Apollo data and extrapolation with pure anorthite $c_P$ data from our database for extrapolation to low temperatures, 10 to 80 K, and the Schreiner curve for higher temperatures (= Stebbins' (1984) model, [323] in [322], for 360 K to 1500 K). Biele et al. [315] found a convenient rational log-log fit function for this preliminary reference lunar average $c_P(T)$, equation (25). Besides the fact that rational functions often have better approximation properties than simple polynomials, we exploit the fact that a typical $c_P(T)$ curve looks simpler in log-log coordinates; a straight line (~$T^3$) at low temperatures, and no point of inflexion at medium temperatures.

$$\ln\left(\frac{c_p(T)}{1J/kg/K}\right)=\frac{p_1x^3+p_2x^2+p_3x+p_4}{x^2+q_1x+q_2} \qquad (25)$$
$$x=\ln(T/1\mathrm{K})$$

with just 5 fitted coefficients ($p_1$ is actually fixed)
$p_1 = 3$, $p_2 = -54.45$, $p_3 = 306.8$, $p_4 = -376.6$, $q_1 = -16.81$, $q_2 = 87.32$

This rational function has no poles; it correctly predicts zero heat capacity at 0 K and a ~$T^3$ dependence at $T$<5 K. It fits the mean, smoothed lunar sample data [316] with an absolute maximum

deviation of 3% and the high temperature Schreiner model to better than 1%. The estimated uncertainty of the low temperature portion rapidly increases below 50 K to ~5-10%.

But now, using the results of the previous section, we are in the position to construct an even more realistic lunar $c_P$ reference curve in a very wide temperature range. We use all the unsmoothed Apollo $c_P$ data, assuming a constant uncertainty of 2% and decompose it into a best-fit composition, using the following list of minerals (imposed mass fraction bounds in parentheses): enstatite (5-40%), diopside (1-20%), hedenbergite (1-20%), ferrosilite (6-24%), anorthite (12.5-100%), albite (2-14%), forsterite (1-8 %), fayalite (1-8%), orthoclase (0-5%), ilmenite (1-30%), troilite (0.3-2%), chromite (0.1-10%), magnesio-spinel (0.1-10%).

The result is the following best-fit composition (Table 11):

**Table 11 best-fit mineral composition for the Apollo $c_P$ data, with the given list of minerals and bounds and the $c_P$ database as of 2021.**

| **Mineral** | ***w*** | ***σ(w)*** |
|---|---|---|
| En | 0.05 | 0.02 |
| Di | 0.01 | 0.01 |
| Hd | 0.01 | 0.005 |
| Fs | 0.060 | 0.023 |
| An | 0.502 | 0.064 |
| Ab | 0.114 | 0.038 |
| Fo | 0.01 | 0.006 |
| Fa | 0.08 | 0.026 |
| Or | 0.05 | 0.015 |
| Ilm | 0.01 | 0.002 |
| Tro | 0.003 | 0.006 |
| Chr | 0.001 | 0.016 |
| Spl | 0.100 | 0.010 |

The normalized $\chi^2$ is 1.2, quite consistent with the estimated 2% uncertainties (intrinsic and caused by composition variations) in the data.

As expected, there is a dominating anorthite fraction, albite, spinel, olivine, some pyroxene, other feldspar, ilmenite but negligible troilite and chromite. There is a notable variation in the resulting best-fit composition depending which minerals are in the list and on their prescribed bounds, which was expected.

Now we can generate the synthetic curve (Figure 23) and compare with the data; preliminary uncertainty estimates for the synthetic curve (using many Monte Carlo realizations in composition, calculating the $c_P(T)$ curve for each and analysing the distribution for each temperature) showed a typical 1σ-uncertainty of ~2% from 90 to 1200 K, and exponentially increasing uncertainties below 90 K (relative uncertainty 0.87 at 10 K). This will be analysed in depth in paper II. Comparing with the preliminary model, equation (25), shows that the latter agrees to within 4% for temperatures from 90 to 1000 K, to ±8% for temperatures between 35 and 90 K and differing by >30% for T<20 K.

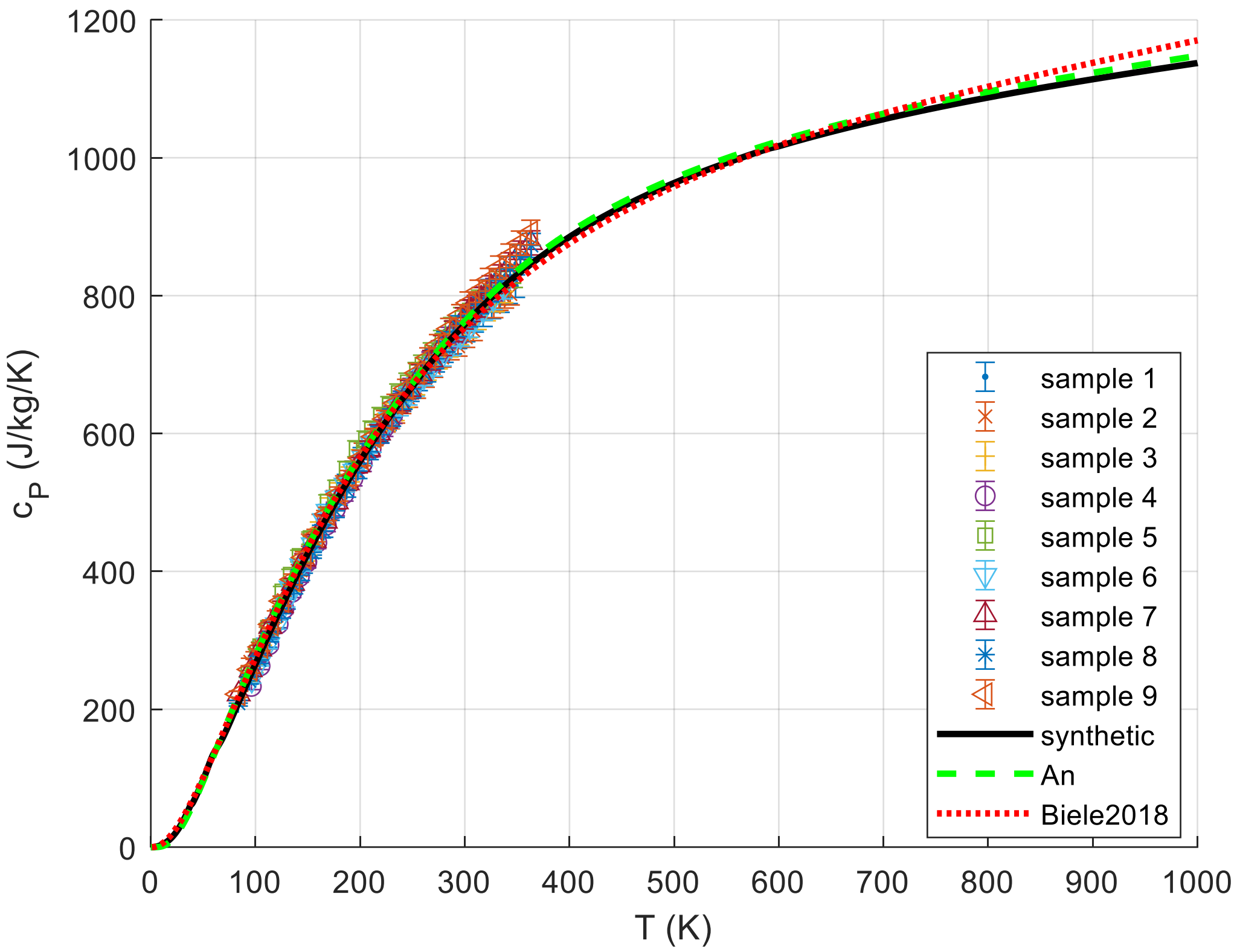


**Figure 23 Synthetic lunar $c_P$ curve, 0-1000 K, bold black line. Apollo data with error bars, separately for each of the 9 sample, are plotted with symbols as indicated in the legend; pure anorthite (An, green dashed) and analytical curve of Biele et al, 2018 [315] (red dotted line) for comparison**

It is obvious that just the precisely known heat capacity of anorthite explains most of the data and the curve. The region 90-350 K (where data exist) is enlarged in Figure 24, and the cryogenic temperature region is shown in Figure 25.

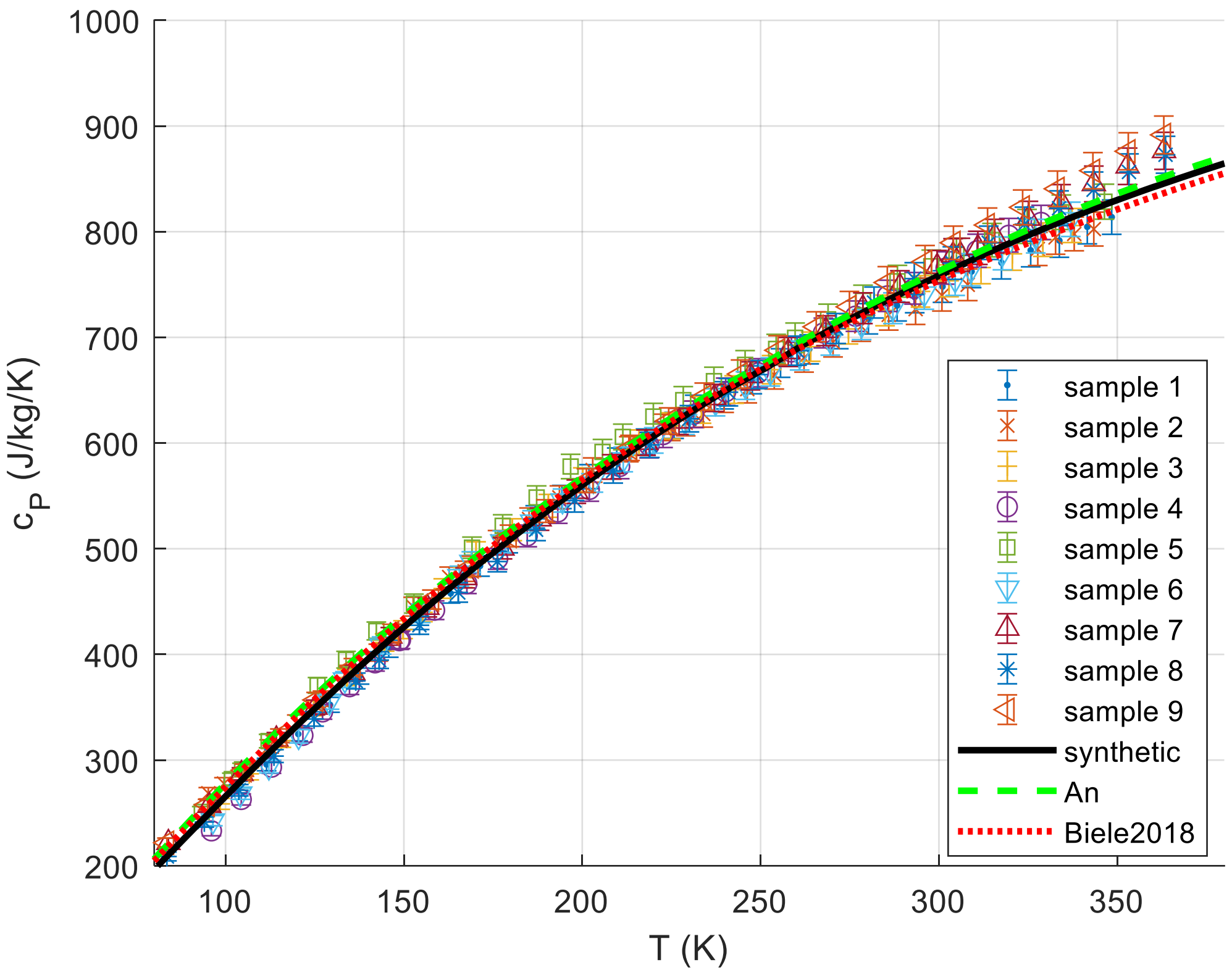


**Figure 24 Enlarged portion of Figure 23 showing all the well-known Apollo data points (numerical values: see appendix, chapter 4.1**

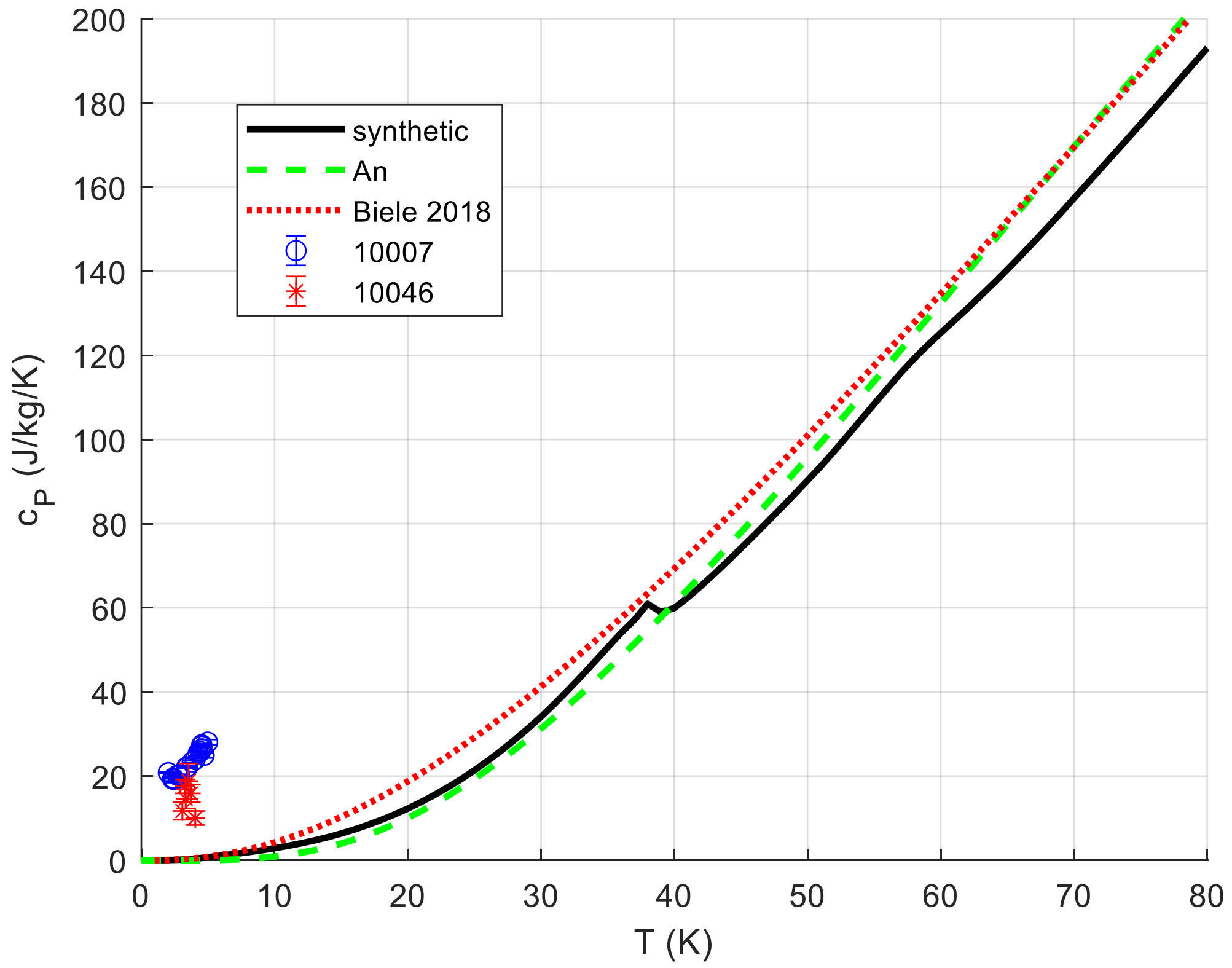


**Figure 25 Enlarged part of Figure 23, cryogenic temperatures. Added are the specific heats of the two samples measured at LHe temperatures [324] which are ~2 orders of magnitude larger than expected. This can't be explained**

**by a glass excess $c_P$ (factor 2 –3 only), maybe it indicates a Schottky anomaly in the liquid Helium temperature range or it is due to experimental errors.**

Finally, in Figure 26, the relative deviations of the synthetic curve to the datasets are shown. One can clearly see the systematic differences between the 9 datasets, which we believe are mainly due to compositional differences.

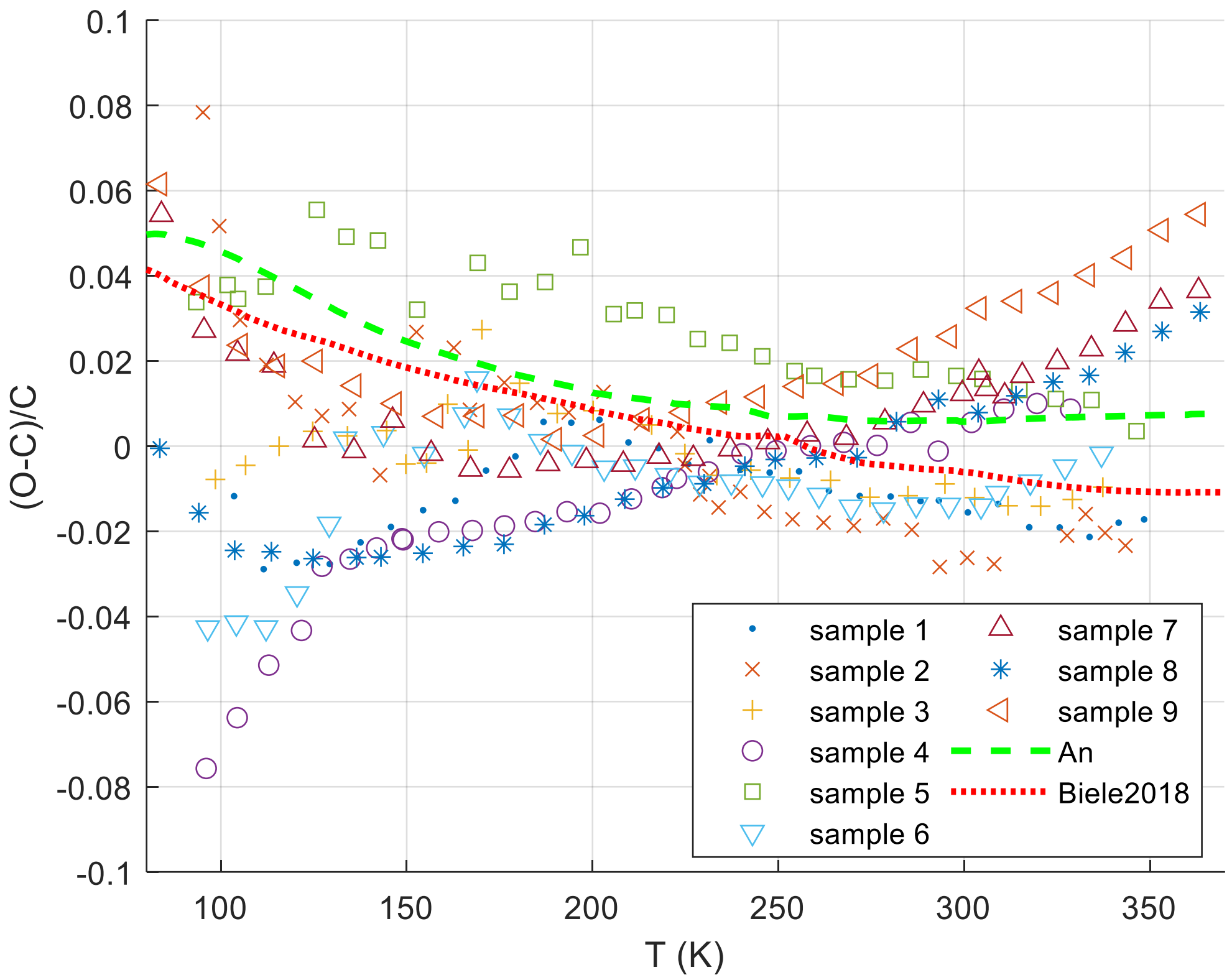


**Figure 26 Relative deviations of modelled 'synthetic' $c_P$ curve to Apollo data.**

## 3.5 Sensitivity of specific heat to composition-changing processes

We can also study the effects of composition on specific heat for the most important practical cases.

**Metal content.** It has long been known and understood that (meteorite) samples with a higher content of meteoritic iron FeNi (thus, also a higher density) have (at ~ room temperature) a smaller specific heat. This is easy to understand, since the $c_P$ of FeNi is smaller than that of most silicates, over wide range of temperatures. We can now quantify the difference vs. temperature, for a arbitrary mass fraction $w$ of FeNi in any mixed material X, $\Delta c_P(w) = c_P(\text{X, with } w \text{ FeNi}) - c_P(\text{X, no FeNi})$.

$$
\begin{aligned}
c_p &= c_{p,FeNi} w + c_{p,X}(1-w) \\
\Delta c_p &= c_p - c_{p,X} = w(c_{p,FeNi} - c_{p,X}) \\
\frac{\Delta c_p}{c_{p,X}} &= w\left[\frac{c_{p,FeNi}}{c_{p,X}} - 1\right]
\end{aligned}
\tag{26}
$$

**Weathering.** It is well-known [28] that weathered, originally metal-bearing meteorites (e.g., H-chondrite finds) have a specific heat higher than analogous pristine samples (falls). Terrestrial weathering has a number of effects, the most important being the oxidation of metal to various Fe-Ni oxyhydroxides, having a much higher specific heat than the native metals. Furthermore, in moist air, metal sulphides (e.g., troilite) oxidize to hydrated sulphates, mostly $FeSO_4$, which are usually so soluble that they are transported away, leaving the oxyhydroxides. Updating Mackes weathering model [28], we assume the following: replace a percentage of the FeNi metal with 1/3 of ferrihydrite, akaganéite and goethite each, and replace half of the same percentage of troilite with goethite, the other half runs off (not a closed system). Note that the degree of hydratisation, i.e. the number of moles of water per mole of akagenéite, goethite, ferrihydrite and sulphate also influences the weathering end product $c_P$ strongly.

Ordinary chondrite meteorite $c_P$ can be used to estimate the degree of weathering [24, 28].

In very strongly weathered specimens, silicate alteration (mainly olivines which react with water to serpentine or with ambient $CO_2$ to (Mg,Fe)-carbonates) or even massive replacement of silicates by clay and oxides may take place, leading to further changes in specific heat.

**Carbon (organics) content.** Analogously, we can study the effect of carbon (graphite) or 'organic matter' content on $c_P(T)$. Note that 'carbonaceous' in meteorites is essentially a misnomer since the carbon content of carbonaceous chondrites is very low (≤3.2%, after [325]), they just tend to look like coal. Thus, the effect is small: the ratio of $c_{P,Gr} / c_{p,lunar}$ varies from 0.3 to 1.3 in the temperature range 25 – 500 K, while the ratio $c_{P,IOM} / c_{p,lunar}$ varies from 1.5 to 3.2; thus, analogously to equation (27) the maximum change in $c_P$ to be expected is -2.2% to +7% at 25 K, but rather <±3% for temperatures >70 K. Still, carbonaceous chondrites tend to have a higher specific heat than ordinary chondrites (next paragraph).

**Phyllosilicate content.** More instructive is the effect of phyllosilicate content on $c_P(T)$. It turns out that (partially hydrated or dehydrated) phyllosilicates, which can be the dominant mineral species in some primitive carbonaceous chondrites, do have a significant effect on specific heat (they tend to increase $c_P$), compare Figure 27. We do not agree that heat capacity of ordinary and carbonaceous chondrites is similar claimed by Consolmagno et al. [24]. The hydration state also matters; this leads directly to the last item on our list,

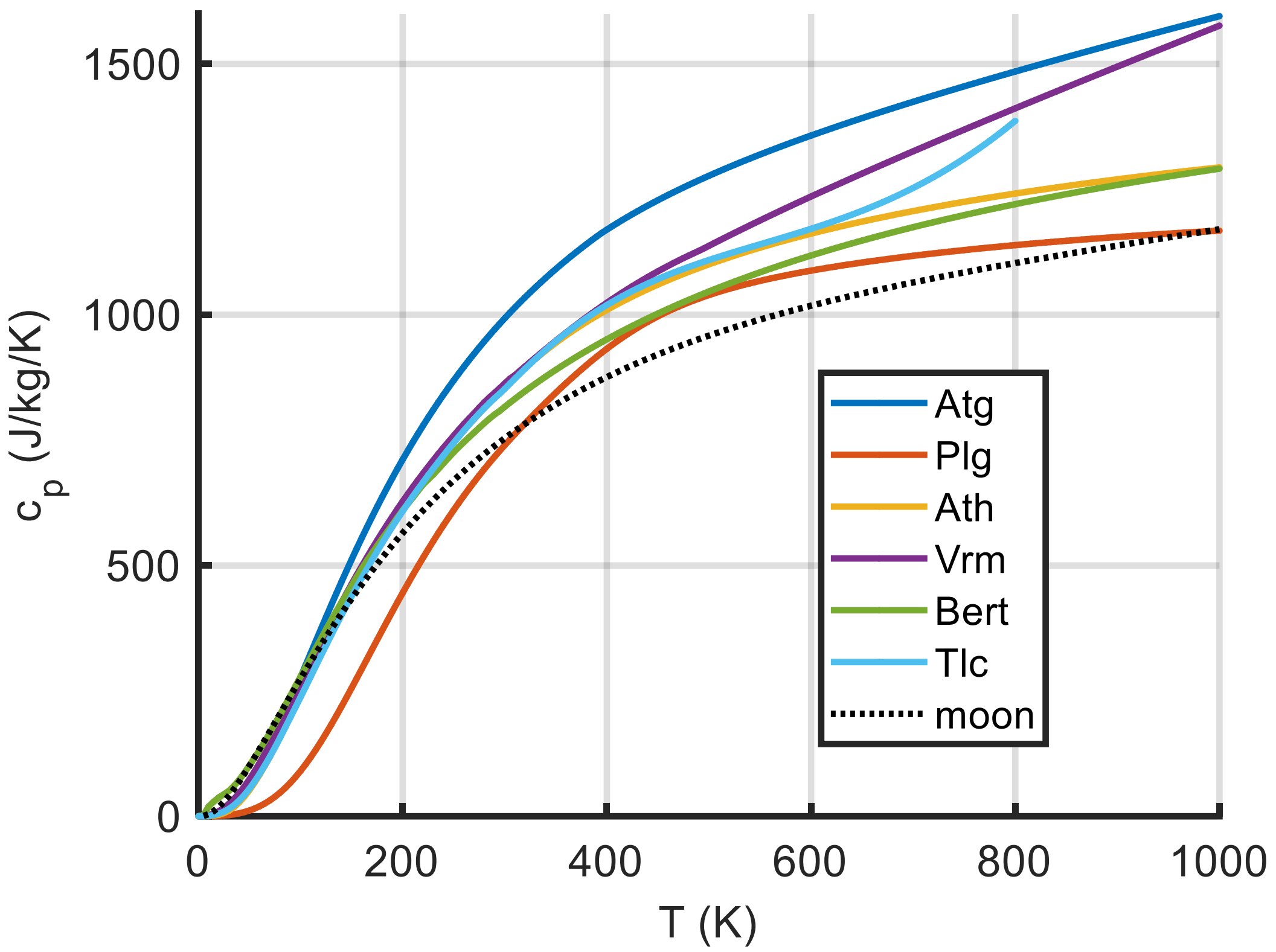


**Figure 27 comparison of $c_P(T)$ curves for the Moon ('basaltic', reasonable also for S-type asteroids) and for 4 different phyllosilicates or clay minerals. Talc decomposes for T>750 – 800 K, this is why the curve ends at 800 K.**

**Thermal metamorphism.** Here the idea is to estimate the variation of specific heat as a, say, primitive carbonaceous chondrite with petrologic type 1 (maximum hydrous alteration) to petrologic type 6 (maximum thermal metamorphism).

Dehydrated and thermally metamorphosed phyllosilicates are common among carbonaceous chondrites, and are probably present in the regoliths of many asteroids. For characteristic temperatures of dehydration, organic material degradation or decomposition and dehydroxylation, see e.g., section 2.12 and [276, 326]. The latter work also provides in depth-analysis of the loss of water by dehydration, comparing various proposed reaction pathways for thermal alteration and providing a wealth of new experimental data (thermal gravimetric analysis TGA, differential thermal analysis DTA) on the dehydration of common serpentine group minerals; it turns out that Fe-rich serpentines decompose at > (100 – 260) K lower temperatures that Mg-rich serpentines (550°C onset, 700°C for 5% mass loss).

We note that dehydration and dehydroxylation are very different processes. Dehydration is the removal of physisorbed or loosely bound 'crystal water' molecules. Dehydroxylation means -OH groups turn into molecular water; either two neighbouring $(OH)^-$ interact to produce $H_2O$ and $O_2^{2-}$, or a proton $H^+$ and a $(OH)^-$ combine into a water molecule. To preserve electrical neutrality, complicated cation diffusion is required which really alters the minerals involved.

What can be done with the database is to find a reasonable end state composition for a thoroughly heated, dehydrated/decomposed phyllosilicate and to calculate the $c_P$ of these (solid) end products, in

comparison to the $c_P$ of a the completely hydrated phyllosilicate, like a CI1. The water produced in the thermal alteration is of course not counted in the final petrologic type 6 composition, it is assumed to disappear.

# 4 Summary

In this paper (I), we summarize the theoretical and practically relevant background on the heat capacity of solids (in particular, minerals), its temperature dependence as well as useful approximations, and discuss the phase transitions and effects of pressure, crystallinity and particle size. The concept of endmember minerals and mechanical mixtures versus solid solutions is introduced, and the possibility to know the specific heat of astro-materials fairly accurately without measuring it - if only the mineral composition can be estimated. We are always talking about the 'complete' $c_P(T)$ curve, from ~10 K to ~1000 K.

For the non-mineralogist, we provide background on important minerals, their polymorphs and other relevant compounds; we discuss meteoritic iron, carbon-rich/organic matter, solar system ices and tholins in some detail.

A table with an overview of our database is given - the $c_P$ database itself will be subject of paper II, with a detailed description of methods and used input data (literature review), for each mineral and compound covered.
Important aspects of the specific heat like the influence of composition, (adsorbed/hydrate) water content, and thermal alteration are discussed. Put briefly, the carbon/organic matter content in e.g., carbonaceous chondrite meteorites, is insignificant for $c_P$ variation. However, the a high FeNi (meteoritic iron) fraction significantly decreases $c_P$ while a high content in phyllosilicates markedly increases specific heat, which can be expressed quantitatively with our $c_P$ database.
For hydrated minerals, but even for physisorbed water in porous silicates, the addition or loss (at elevated temperatures) of water also has a significant effect on specific heat.

The accuracy of composite $c_P$ curves is estimated to be of the order of 1% for $T$>70 K if the mineral composition is regarded as exact. For 10≤$T$≤70 K, the uncertainty can grow to the order of 5% (higher, but less relevant, in narrow temperature intervals near transition peaks) – in particular if there is a high proportion of solid solution minerals (other than olivine) present with high excess heat capacities (non-idealities). Anorthite (with mass fraction $w_{An}$) adds $w_{An}$ × 9% to the relative uncertainty near 510 K (480 – 520 K), due to a transition peak that is not predictable in its natural phase.
We give some quantitative example applications of the database already. First, looking at $c_P$ at cryogenic temperatures, we note that the temperature dependence of $c_P$, which traditionally has often been neglected, has a significant impact on thermal inertia. We now can calculate the specific heat at any

low temperature, not only for common (silicate) rocks and regolith materials, but also for solar system ices and some tholin analogues; the latter show a very large variation in $c_P$ between various models, but their specific heat is generally an order of magnitude higher than that of silicate rock at the same temperature.
An obvious application is the 'forward' prediction of $c_P$ curves for materials with known composition, like laboratory (asteroid) regolith analogues. We have done this calculation for 6 commercial analogue materials and for the Phobos simulant UTPS-TB of the University of Tokyo [327]. Result tables are in the appendix of this paper.

Turning to the extra-terrestrial material which has been studied best, lunar regolith, we show how to invert the measured $c_P$ data, which cover only the 90 – 350 K range, and construct a physically reasonable $c_P(T)$ curve from 0 to 1500 K. A very close predecessor of this curve was also cast into a very compact correlation equation, a rational function with only 5 fitted coefficients, which reproduces the measured and modelled values to ~4 %.

All published lunar sample $c_P$ data have been collected, for convenience, in the appendix. A brief data review on all (to date) published meteorite specific heat data is also in the appendix.

# 5 Outlook

This paper already being exceedingly long, we decided to end here (with a kind of cliff-hanger, some might say). Part II will be the database itself, that is, the data files and auxiliary software source code (on a repository), the explanation of using the database, of the methods used for data assimilation and a description of the input and final output $c_P(T)$ for each mineral and compound covered.

This will cover one of our goals, namely to supply the community with all the ingredients to calculate their own $c_P(T)$.

Finally, in order not to delay the publication of paper II, we might publish part III of the trilogy, on further applications and further standard reference curves (e.g., [328]) and in particular on the comparison with experiments (such experiments have recently started at the laboratory of one of us, MG). The applications could, for example, include the quantitative dependence (explicit correlations) of $c_P(T)$ with composition in terms of metal, organics /phyllosilicate content, and the effects of weathering and thermal metamorphosis; topics we have only touched, rather qualitatively, in the present paper. As for further specific heat reference curves, we plan to define up-to-date reference mineral compositions for the most important (~dozen) meteorite classes, calculate their specific heat curves and compare, if possible, to experimental data.

As of this writing, our $c_P$ dataset includes already more than 100 endmember mineral and compound $c_P(T)$ extracted and reviewed from the literature. It is a work in progress. We know that the database may, like any compilation, contain mistakes, misinterpretations, and omissions. We hope that those who publish $c_P$ data and/or use the database will help us to correct, improve, and extend it; don't hesitate to get in touch with us!

There is already a couple of minerals on our list where (new) specific heat data are sought, or just any because there is no data, e.g., hercynite for $T$>400 K, NiFe alloys ( of different compositions/phases) and meteoritic iron (kamacite, taenite) for 400-1200 K, pentlandite $(Fe,Ni)_9S_8$ for >300 K, some phyllosilicates, tholins (laboratory-made), amorphous variants of common minerals incl. diaplectic (or otherwise densified) glass; finally, more $c_P$ measurements on carbonaceous chondrites and iron meteorites with a well-characterized mineral composition would be very interesting, both low-*T* and high-*T*.

On the theoretical side, the plan is to model, if significant, excess heat capacities for feldspars and pyroxenes (simplified: only 'ideal' orthopyroxenes En-Fs and 'ideal' clinopyroxenes Di-Hed (or 'mean pigeonite' and 'mean augite', that is, with a fixed Ca content); then there is only 1 composition variable besides the *T* dependence). This might need a few more experimental data. Another issue is a study of transition peaks in $c_P$; in natural mineral mixtures with a spatial distribution of solid solution compositions, it is conceivable that narrow transition peaks are 'smeared out'; how to handle this is in $c_P$ models needs to be studied.

Finally, and this is relevant for the sample analysis community, we recommend to measure $c_P(T)$ of new samples returned from missions to asteroids (Ryugu, Bennu come to mind), new samples from the Moon (highland rocks, in particular) and other solar system bodies over wide temperature ranges. Nowadays, just 10-30 mg of a sample[18] suffices to determine the specific heat capacity accurately over the temperature range 2 – 900 K by PPMS and power-compensated DSC calorimetry. In particular, samples from primitive asteroids that contain a significant amount of phyllosilicates would be very interesting to compare to the models presented here.

[18] Are such measurements non-destructive besides homogenization and possibly grinding? Potentially yes, if the sample is heated beyond decomposition temperature (~200 – 300°C).

# Appendix (Supplementary Information)

**The specific heat of astro-materials:**
**Review of theoretical concepts, materials and techniques**


Jens Biele[(1)], Matthias Grott[(2)], Michael E. Zolensky[(3)], Artur Benisek[(4)], Edgar Dachs[(4)]

*[(1)] DLR – German Aerospace Center, RB-MUSC, 51147 Cologne, Germany, e-mail: Jens.Biele@dlr.de*

*[(2)] DLR – German Aerospace Center, Institute for Planetary Research, Berlin, Germany*

*[(3)] NASA Johnson Space Center (Houston, United States)*

*[(4)] Chemistry and Physics of Materials, University of Salzburg, 5020 Salzburg, Austria.*


# 1 Methods

## 1.1 Accurate Padé approximants to the Debye and Kieffer functions

For the **Debye function**[1], we approximate, following Goetsch et al. [1],

$$D(\theta_D/T) = 3\left(\frac{T}{\theta_D}\right)^3 \int_0^{\theta_D/T} \frac{x^4 e^x}{(e^x-1)^2}\,dx = 3\left(\frac{T}{\theta_D}\right)^3 \int_0^{\theta_D/T} \frac{x^4 e^{-x}}{(e^{-x}-1)^2}\,dx$$

with the rational function (Padé approximant) $g(T_n)$, $T_n=T/\theta_D$

$$g(T_n) = \frac{N_0 + \frac{N_1}{T_n} + \frac{N_2}{T_n^2} + \cdots}{D_0 + \frac{D_1}{T_n} + \frac{D_2}{T_n^2} + \cdots} \qquad (1)$$

The first one, two or three and the last one, two or three in each of the sets of coefficients $N_i$ and $D_i$ in $g(T_n)$ can be chosen to exactly reproduce both the low- and high-*T* limiting values and power-law dependencies in *T* and/or 1/*T* of the function it is approximating; this is a very important and powerful feature of the Padé approximant. Then the remaining terms in powers of $1/T_n$ in the numerator and denominator have freely adjustable coefficients that are chosen to fit the intermediate temperature range of the function. A physically valid approximant requires that there are no poles of the approximant on the positive real $T_n$ axis.

The constraints which assure the low-*T* and high-*T* limiting values and derivatives here are [1],

---

[1] as used in calorimetry

$$D_0 = \frac{N_0}{E}, \quad D_1 = \frac{N_1}{E}, \quad D_1 = \frac{-FN_0 + EN_2}{E}, \quad D_m = \frac{N_n}{G}$$
$$E = 3,\ F = -3/20, G = 12\pi^4/5 \qquad (2)$$
$$m = n+3$$

Result: The approximant in [1] has n=5 and m=8 and a maximum relative deviation of <0.3%; increasing the degrees to n=8, m=11, we obtain a maximum relative deviation to true Debye = $5.8 \cdot 10^{-6}$ at $T/\theta_D \approx 0.1$ (Table 2 and Figure 1)

**Table 1 Coefficients for Padé-approximant of the Debye function**

```
N0..N8:
                   3
         -0.17682974
         0.019953909
       0.00065686146
       4.0944374e-05
       1.3642663e-06
       4.4605133e-07
      -1.2957876e-08
       1.6782041e-09

D0..D11:
                   1
        -0.058943245
         0.056651303
       -0.0027325104
        0.0010646772
      -3.2134724e-05
       7.3396614e-06
       7.0397353e-08
       7.2090357e-09
       1.8991184e-09
       -5.540749e-11
       7.1785057e-12
```

The coefficients in bold numbers are not independent. Thus, 17 independent coefficients remain. The denominator-polynomial D has no zeros at positive temperatures. The Padé fit has been compared to the exact Debye function, calculated by precise numerical integration and with the function in the polylogarithmic form (the polylogarithm function, with double and optionally arbitrary precision is available in Matlab(TM)).

Debye function, polylogarithm form

x=θ<sub>D</sub>/T, $Li_s(x)$ is the polylogarithm of order *s* and argument *x*.

$$C_V/3R = \frac{4}{5}\frac{\pi^4}{x^3} + \frac{3x\exp(-x)}{\exp(-x)-1} + 12\ln\left[1-\exp(-x)\right] - \frac{36}{x}Li_2\left[\exp(-x)\right] - \frac{72}{x^2}Li_3\left[\exp(-x)\right] - \frac{72}{x^3}Li_4\left[\exp(-x)\right] \quad (3)$$

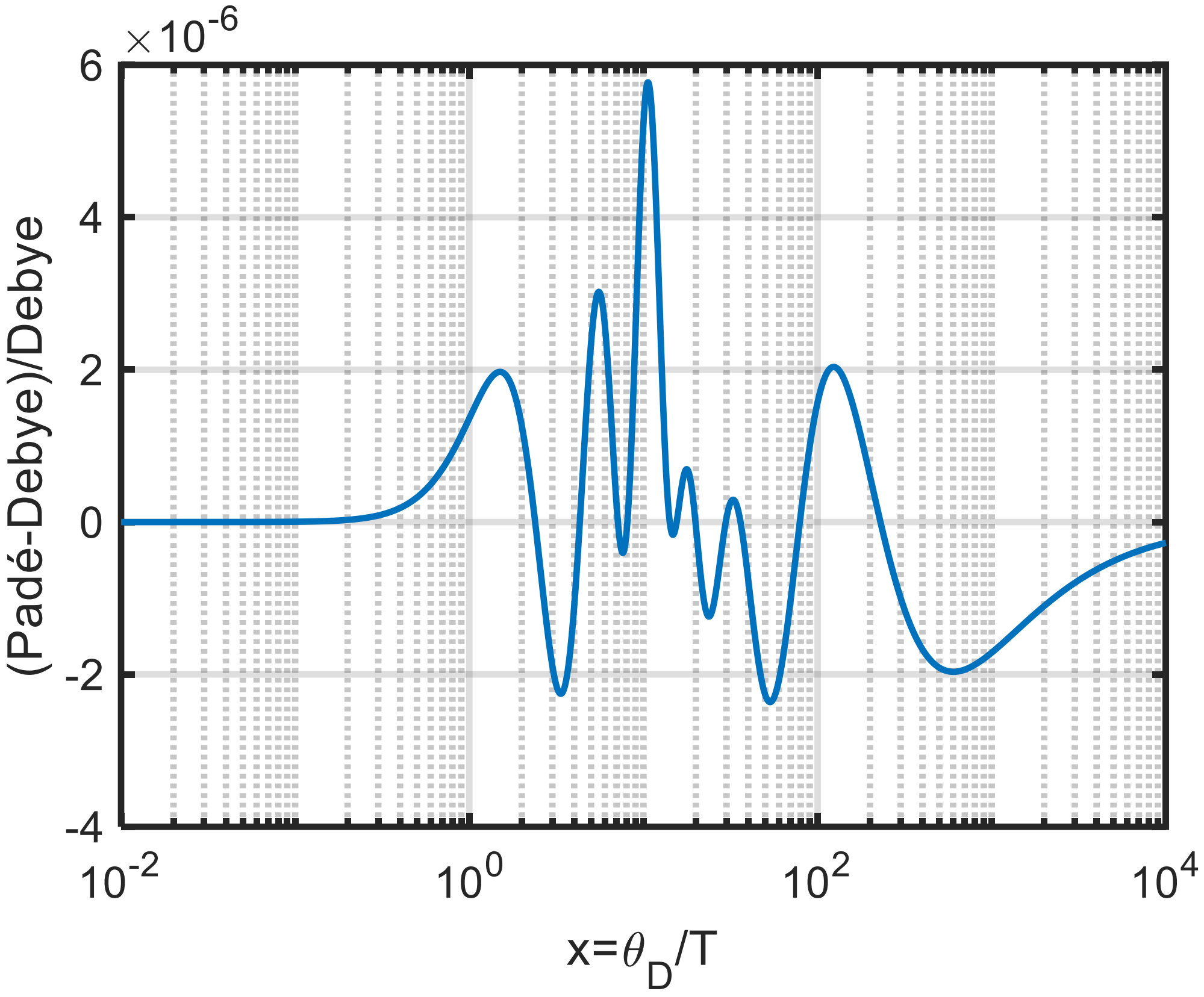

**Figure 1 Relative deviations of our Padé-fit from the true Debye-function**

### Padé-approximant for the Kieffer-function

Gurevich et al. [2], equation 5 for the $C_P$ of crystal hydrate water needs the Kieffer-function

$$K(\theta_L/T, \theta_U/T) \equiv \frac{3R}{\theta_L/T - \theta_U/T} \int_{\theta_L/T}^{\theta_U/T} \frac{x^2\exp(x)}{\left(\exp(x)-1\right)^2}dx \quad (4)$$

which has two arguments, lower and upper normalized temperatures.

The following Padé-approximant with numerical coefficients from Table 2 is accurate to <1·10$^{-6}$ in relative terms over arbitrary arguments (Figure 2):

$$k(x)=\frac{1+\sum_{n=1}^{8}p_n x^n}{1+\sum_{m=1}^{9}q_m x^m}$$

$$K(x_L,x_U)=\frac{3R}{x_U-x_L}\left[x_U k(x_U)-x_L k(x_L)\right] \qquad (5)$$

$$x_L=\theta_L/T,\quad x_U=\theta_U/T$$

**Table 2 Coefficients for Padé-approximant of Kieffer-function. Note that both** $q_9 = 1$ **(exactly) and** $p_8 = \pi^2/3$ **(exactly). Furthermore,** $p_{(1\ldots3)} = q_{(1\ldots3)}$**.**

| ***n, m*** | $p_n$ | $q_m$ |
|---|---|---|
| 1 | | **-291.060215932407** |
| 2 | | **1278381.54701115** |
| 3 | | **114269.722348969** |
| 4 | 22596.3244579461 | 58062.5772985723 |
| 5 | 2635.11630655455 | 5887.48838373051 |
| 6 | 337.259940890294 | 821.72512559065 |
| 7 | 7.47974537995418 | 102.316628131888 |
| 8 | **π²/3** | 2.27417852421094 |
| 9 | | **1** |

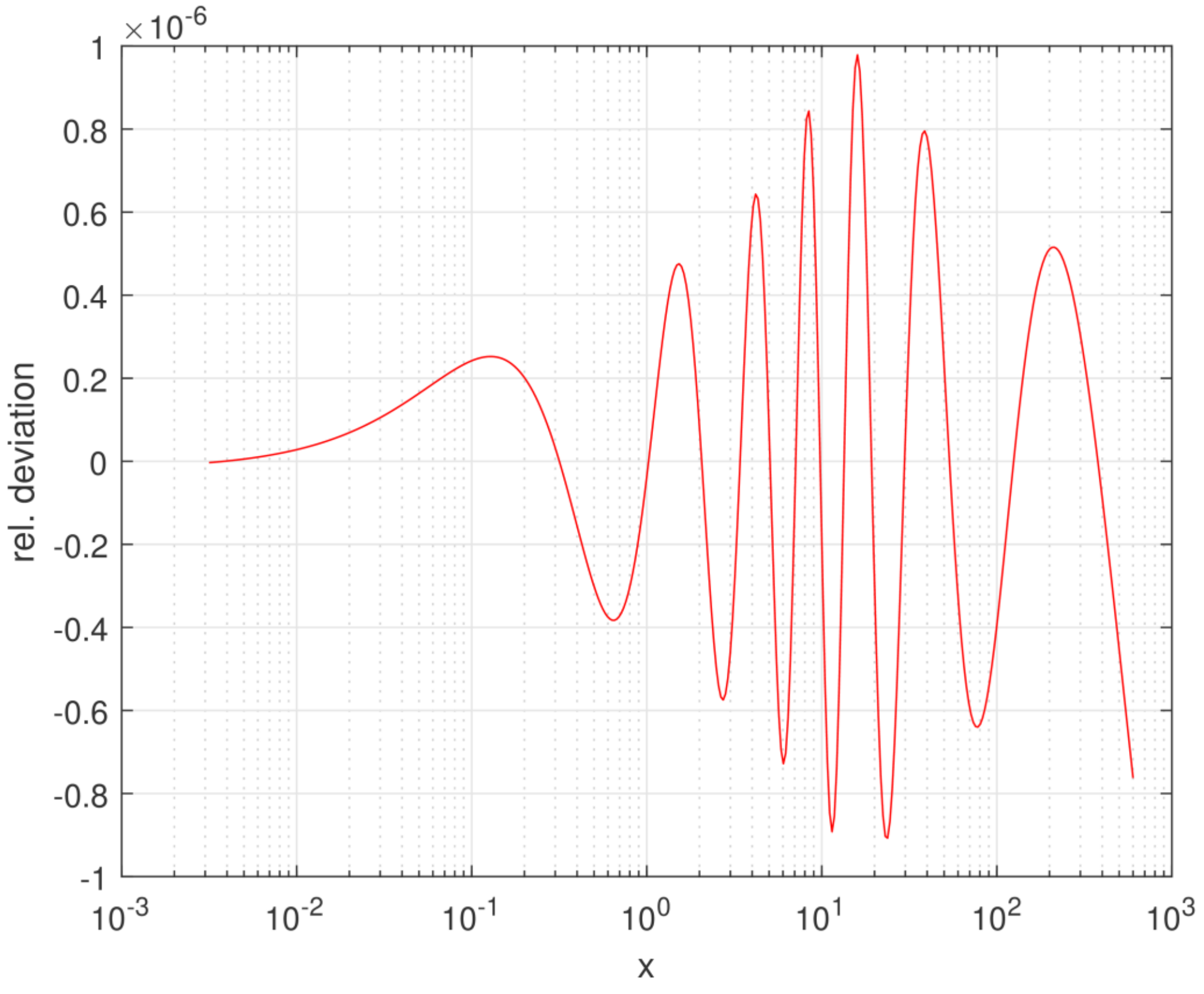

**Figure 2 Relative deviations of our Padé-fit from the true Kieffer-function k(x)**

## 1.2 Estimating the mineral composition from $c_p(T)$

The basic idea is the following (C, c here designate both specific heat capacities at zero pressure in J/kg/K):

Given the experimental $c_P$-curve ($T$; $c_P$) of a mineral mixture over a (wide as possible) temperature range and some idea about the main constituents (endmember minerals), i.e. a list of endmember minerals. Main means: mass fraction of a constituent >>1%. We estimate the 'most likely' mass fractions $X_i$ of the assumed or known constituent minerals by least squares solution of the constrained cp mixing equation (cp_decompose) and then, with the obtained mass fractions that are best (i.e., with the least $\chi^2$) reproducing the weighted cp data, construct the model $c_P(T)$ curve over a wider temperature range (cp_compose) and calculate the uncertainties of the fitted an extrapolated values both by Monte Carlo and a bootstrap [3] method. Note that the endmember mineral $c_P(T)$ curves are the base functions in our least-squares problem here; they are generally far from been orthogonal, and the problem is only solvable because of the constraint that $X_i \geq 0, \quad \sum_i X_i = 1$.

Given $M$ experimental data points, $T_m, C_m(T_m), m = 1 \cdots M$ , we fit the mass fractions $X_i$ using as base functions the $c_i(T)$ of $N$ possible constituent minerals, since $C_m(T_m) = \sum_i X_i c_i(T_m)$ subject to the constraints $X_i \geq 0, \quad \sum_i X_i = 1$ . Let $\sigma$ be the uncertainties of the $C$ data (weighting). This is a linear least-squares problem with bounds and linear constraints, in typical solver lingo (e.g,, Matlab's solver lsqlin) it is written:

$$\min_X \frac{1}{2} \left\| C \cdot X - d \right\|_2^2 \text{ such that } \begin{cases} A_{eq} X = b_{eq} \\ lb \leq X \leq ub \end{cases} \qquad (6)$$

with

$$C = \begin{bmatrix} c_1(T_1)/\sigma_1 & c_2(T_1)/\sigma_1 & \cdots & c_N(T_1)/\sigma_1 \\ c_1(T_2)/\sigma_2 & c_2(T_2)/\sigma_2 & \cdots & c_N(T_2)/\sigma_2 \\ \cdots & \cdots & \cdots & \cdots \\ c_1(T_M)/\sigma_M & c_2(T_M)/\sigma_M & \cdots & c_N(T_M)/\sigma_M \end{bmatrix}$$

$$\begin{aligned} X &= [X_1, X_2, \ldots, X_N]^T \qquad (7) \\ d &= [C_1/\sigma_1, C_2/\sigma_2, \ldots, \mathrm{C}_M/\sigma_M]^T \\ A_{eq} &= \mathbf{1} \quad (1 \times N \text{ vector}) \\ \mathrm{b}_{eq} &= 1 \\ lb &= \mathbf{0} \quad (1 \times N \text{ vector}) \\ ub &= \mathbf{1} \quad (1 \times N \text{ vector}) \end{aligned}$$

Because the problem being solved is always convex, the solver finds a global, although not necessarily unique, solution. It is highly recommended, if non-trivial constraints on the mineral composition are known, they should be furnished as bounds (*lb*, *ub*) for *X*.

**Uncertainty estimate of the solution:** By default we do both a Monte Carlo (varying the *d* vector with an assumed random noise + bias within error bounds of e.g., 1%) and a bootstrap resampling method to calculate the uncertainty distribution of the resulting $X_i$ . Sufficiently small $X_i$ can also be set identical to 0 and the process repeated. An alternative to Monte Carlo with an assumed random noise + bias is bootstrap resampling. Monte Carlo and bootstrap give rather similar results.

A number of $X_i$ can be grouped for convenience (belonging to the same mineral group or solution series, like fraction of fayalite and forsterite in olivine, or or-al-an in feldspars, or diopside-hedenbergite-enstatite-ferrosilite in pyroxene. We still have to solve the equation system above, but can then disentangle the $X_i$ into groups:

$$
\begin{aligned}
&C_m = \sum_{j=1:J} W_j \sum_{k=1:K(j)} w_k c_{j,k}(T_m) \\
&\sum W_j = 1, \quad W_j \geq 0 \\
&\sum w_k = 1, \quad w_j \geq 0 \qquad (8) \\
&X = [W_1 w_1, W_1 w_2, \ldots W_2 w_1, \ldots W_J w_K] \\
&\sum X_k = 1
\end{aligned}
$$

Convention: call *J* the number of *W*; *K* is a vector containing the number of *w* in each $W_j$.

**Example: lunar regolith data**

First we checked that a synthetic *T*, $c_P$, $\sigma(c_P)$ set (with known *X*) is correctly fitted and the result is identical to the known, true *X* (to within ~$1\cdot10^{-5}$ relative or $6\cdot10^{-7}$ in absolute mass fractions). The code can check after the first, nominal, fit whether any components of the result *X* are $<X_{\text{threshold}}$ (typically 0.001); it deletes the corresponding mineral(s) from the list, and re-performs the fit with the reduced minerals list. Then, we perform a Monte Carlo (~1000 runs) where each time we either do a bootstrap or add noise to the $d=c_P/\sigma(c_P)$-vector: nominally, our model noise is one third (0.58% if total uncertainty is the default 1%) of gaussian noise, one third of a linear bias with Gaussian random amplitude, one third of quadratic bias with Gaussian random amplitude. The result can be inspected in histograms (binned vs. mass fraction for each retained mineral), Figure 27.

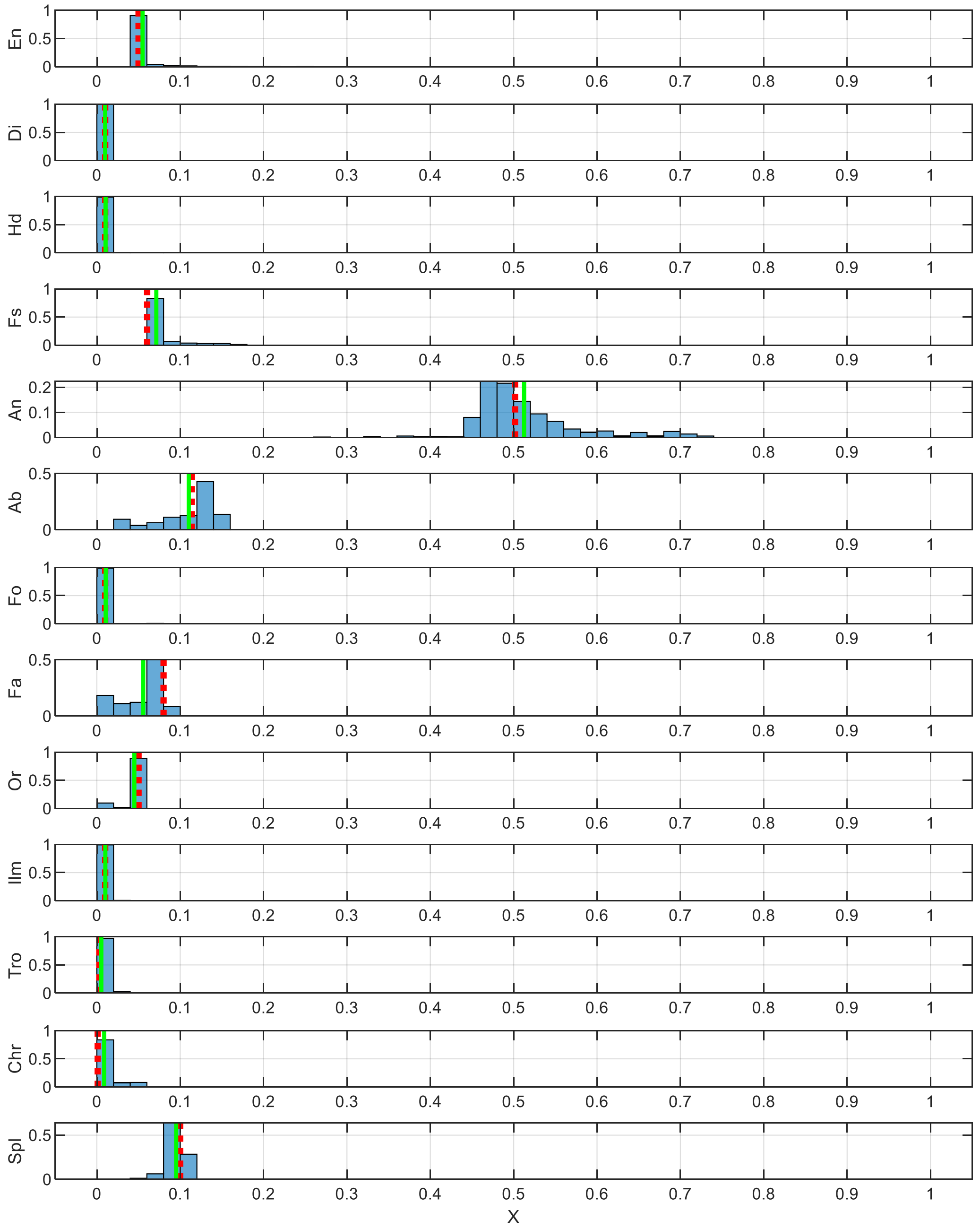

**Figure 3 Monte Carlo histogram of fitted mineral compositions for synthetic lunar regolith. Red dotted: nominal fit without noise. Magenta: mean; Black: median; blue: mode; bold green: true value. Calculated with lb=0 and ub=1 for each X, Xthreshold=0.001; MC=50000**

## 2 Data review for meteorite sample specific heat data

Only a handful of heat capacities of meteorites have been published until about ten years ago, when interest in astro-material thermophysical properties, including heat capacity, suddenly increased. The new meteorite $c_P$ data are, to our present knowledge, [4-13] and more are in preparation (e.g., Noyes, Macke, Opeil et al., "Novel Low Temperature Behavior in the Thermal Diffusivity and Thermal Inertia of Iron Meteorites" (submitted, 2022) including $c_P(T)$ over the range 2-300 K for four different iron meteorites [R. Macke, Jan 15, 2022, pers. comm.]).

The more interesting recent works start with Opeil et al. (2012) [4] who determined, besides thermal conductivity of various meteorite samples, also $c_P(T)$ for Martian shergottite meteorite Los Angeles Los Angeles: [14, 15] from ~2 to 380 K by PPMS.

Macke et al. (2014) [6, 7] measured heat capacity at 175 K for several HEDs (diogenites, eucrites, howardites). Macke et al. (2016) [8] provided PPMS data on heat capacities of ordinary chondrite falls (i.e., less likely weathered than finds) in the range (75–300) K, later extended to 5 - 300 K [11]. Opeil et al. (2020) determined thermal properties, including $c_P$ by PPMS, of five CM2 meteorites in the range 5 - 300 K [12].

Very interesting is the work of Piqueux et al. (2021) [13] who measured the specific heat of ~30 different meteorites of many classes from ~90 to 285 K by DSC, fairly accurately (~5% accuracy).

Virtually all of the older data were measured at temperatures at or above ~300 K, we describe them briefly hereunder.

Butler (1963) [16] reported the specific heat curve of octahedrite meteoritic iron from Canyon Diablo meteorite fall for 123 to 570 K (uncertainty unknown).

Matsui and Osako (1979) [17] directly measured the heat capacity of five Yamato meteorites (four ordinary chondrites and a howardite) at 300 K, 350 K, and 400 K, while Yomogida and Matsui (1983) [18] used laboratory data of the constituent minerals of ordinary chondrites to calculate their heat capacities; their calculated values, which they preferred, were 50% higher than their directly measured results. According to Yomogida & Matsui, this is very likely not due to any significant discrepancies between theoretical (linear mixing) and measured values but a bias introduced by sample preparation in the data of [17] (grinding leading to a bias in metal concentration).

Soini, Kukkonen et al. (2020) [19] compile the older $c_P$ data, including some modelled heat capacities, of a large number of meteorites plus new simultaneous contactless thermal conductivity and thermal diffusivity measurements, from which $c_P$ can be calculated to an accuracy of at best 5% (yet higher scatter from same samples measured in orthogonal directions with rather different $c_P$, up to 25% difference!). Even for class averages (at only 2 temperatures, 200 K and 300 K), the scatter is high (there are finds and fall meteorites in the list, i.e. potential weathering, as discussed above, and various petrographic grades), but the trend is $c_P$(LL)~$c_P$(C)>$c_P$(L)>$c_P$(H)>$c_P$(IAB), with the mesosiderite somewhere in between.

Beech et al. 2009 [20] measured the $c_P$ of a H5 chondrite Gao-Guenie fragment. Specific heat values have been determined over the temperature range between 296 and 773 K; furthermore, for Gao-Guenie (07C-TPRL) – H5, Gao-Guenie (08) – H5, Jilin-H5 and Sikhote-Alin-Iron IIAB the specific heat at ~350 K has been determined by 'A standard water immersion calorimetry technique', resulting in 732.0 ± 7.5, 739.7 ±27.5, 725.8 ±13.2 and 458.2 ±10.7 J/kg/K, respectively. Data might be problematic according to Flynn, Consolmagno et al. 2018 [21].

Szurgot et. al (2011) [22] determined, with DSC and a 'double-walled calorimeter' the specific heat at ~297 K of various samples of Brahin, Vaca Muerta, Allende, El Hammami, Gold Basin, Sahara 99471, DaG 610, Canyon Diablo, Gibeon, Sikhote Alin, Toluca and Morasko meteorites.

Opeil and Consomagno [23] noted that the heat capacity of chondritic meteorites is typically ~750 J/kg/K at 300 K, ~500 J/kg/K at 200 K.

(Łuszczek and Wach 2014) [24] determined $c_P$ of crust and interior of L-type ordinary chondrite NWA 6255 (petrographic type L5, shock stage S4, weathering grade W1) by DSC, 223-823 K. It contains a few % (1.0% in crust, 3.6% in interior) of troilite FeS, its transition peak is seen in the $c_P$ curves.

Flynn, Consolmagno et al. 2018 [21] report in their review on 'Physical properties of the stone meteorites' some problematic measurements, citing, for instance, Matsui and Osako (1979), Yomogida and Matsui, (1983) which we discussed above, but also Beech et al. (2009) [20].
We briefly add to this list, hereunder, a few instances where we found issues:

> The specific heat of the Morasko iron meteorite, from Szurgot et al. (2008) [25], from 263 to 303 K, must be grossly in error. They report, in a figure and in text, that the specific heat capacity $c_P$ varies from 50 to 590 J/kg/K in this temperature range, which we believe is impossible. The literature value is a variation from 428 to 449 J/kg/K over this small temperature range.
>
> Occasionally, the temperature of a specific heat value is not quoted correctly (or not at all), probably stemming from the assumption that temperature is not important; for example, (Yu and Ji 2015) [26] estimate $c_P$ of Bennu (B-type, analogue likely CM meteorites as ~500 J/kg/K or 560 J/kg/K, no temperature given, referring to Opeil 2012 [27] . These values appear very small! In the Opeil (2012) paper, we find ~750 J/kg/K at 300 K; the 560 J/kg/K value comes actually from Gundlach and Blum (2013) [28] ; 500 J/kg/K is a realistic CM value for 175 K, the average temperature[2] in Consolmagno & Macke's LN2 experiments.

Ghosh and McSween (1999) [29] studied, theoretically, the temperature dependence of specific heat capacity and its effect on asteroid thermal models. Their $c_P(T)$ curves, 300-1100 K (their fig. 1) are grossly in error. They note that 'there are no measurements on ferrosilite, albite, orthoclase, or diopside' - they must have overlooked it, precise data on said minerals exist at least since the 1980ies.

Henke (2012) [30] calculate cp of H and L ordinary chondrites based on an assumed composition taken from the literature ([18, 31] and heat capacities from Barin (1995) [32]. However, the calculated curves might be in error and the said references are outdated.

Szurgot and Wach (2012) [33] give the $c_P$ of the ordinary chondrite L6 Sołtmany meteorite with ~5% troilite (an equilibrated ordinary chondrite L6 class) and the Gao-Guenie chondrite; the data look problematic, at least for Sołtmany.

[2] Applying the curvature correction to the mean Consolmagno & Macke LN2 values, the real $c_P$ at the average temperature of 185.5K is about 4.1% higher (see section 1.4); the average $c_P$ rather corresponds to a notional temperature of 177.3 K (this latter value is only exact for the lunar $c_P(T)$ curve; Consolmagno and Macke cite 175K at the notional temperature, quite close). The $c_P$ data measured with the LN2 drop calorimetry have random and systematic uncertainties of ~2% and ~2%, respectively.

## 3 Data review for the 9 lunar samples where $c_P$ was measured

Note that the sample number prefix for Apollo samples is 10 for Apollo-11, 12 for Apollo-12, 14 for Apollo-14, 15 for Apollo-15, 6 for Apollo-16 and 7 for Apollo-17. Major geologic features and rock types sampled are listed in Table 3.

**Table 3 Context: Apollo missions, major geologic features and rock types (source: NASA)**

| Apollo | Major Geologic Features and Rock Types |
|---|---|
| 11 | Mare (Sea of Tranquillity), basaltic lava |
| 12 | Mare (Ocean of Storms), rocks are basaltic lava; ray from Copernicus Crater crosses the site. |
| 14 | Highlands (Fra Mauro formation) - thought to be ejecta from the Imbrium Basin |
| 15 | Mare (Hadley Rille in a mare area near the margin of Mare Imbrium) and highlands (Apennine Mountains, a ring of the Imbrium basin); rocks are breccia and basalt |
| 16 | Highlands (Descartes formation and Cayley Plains); rocks are anorthosite and highlands soil. |
| 17 | Mare (Sea of Serenity) and Highlands; rocks are mare soil, orange soil, basaltic lava, anorthosite. |

Detailed descriptions of all lunar regolith samples have been published online by the Lunar and Planetary Institute (LPI), see https://www.lpi.usra.edu/lunar/samples/#petrographic.

Table 4 gives an overview about the lunar samples for which cP (or thermal diffusivity) has been measured.

**Table 4 Overview lunar cp samples**

| Sample | Description | Reference, notes |
|---|---|---|
| 14321,153 | Breccia, Big Bertha, Apollo 14. | [34] |
| 15555,159 | Soil, Apollo 15 | [34] |
| 14163,186 | Soil, Apollo 14 | [34] |
| 15301,20 | Soil, Apollo 15 | [34] |
| 60601,31 | Soil, Apollo 16 | [34] |
| 10057 | Type A vesicular basalt, Apollo 11 | [35] |
| 10084 | Regolith fines, Apollo 11, | [35] |
| 10021,41 | Breccia, Apollo 11 | [36] |
| 12018,84 | Olivine-dolerite, Apollo 12 | [36] |
| 10020 | Only composition: Pyroxene 45.4%, plagioclase 24.6%, olivine 3.9%, ilmenite 22.7%, troilite 0.9%, other 0.7%, void 1.8% (volume fractions) | [37] (thermal diffusivity of four Apollo 11 lunar specimens measured over the temperature range -130°C to + 150°C) |
| 10046 | Only composition: Pyroxene 16.8%, Opaquest (~ilmenite) 8.6%, plagioclase 4.7%, unidentified 3.5%, glass 2.9%, matrix <40 µm | [37] (thermal diffusivity of four Apollo 11 lunar specimens measured over the temperature range -130°C to + 150°C) |

| Sample | Description | Reference, notes |
|---|---|---|
| | (approximated by 30% glass, 55% pyroxene, 15% plagioclase) 63.5% (volume fractions) | |
| 10017 and | (type A) | [40] LHe temperatures |
| 10046 | (type C) | [40] LHe temperatures |

The average mineral composition of lunar regolith material can be seen in **Table 5** [38]:

**Table 5 Adopted average mineral composition of lunar surface materials**

| $\rho_{mix}$=3.14 g/cm³ | **Mineralogy**<br>**(if solid solition, mole-% fractions)** | **Vol-%** | **Density g/cm³** | **Mass-%** |
|---|---|---|---|---|
| Plagioclase | 10:90 albite/anorthite | 32 | 2.736 | 27.9 |
| Pyroxene | 40Mg60Fe mix of enstatite and ferrosilite, but ~25% Ca. i.e. a half towards 40diopside-60hedenbergite | 39 | 3.4 | 42.3 |
| Glass | Plagioclase glass with Fe:<br>51% $SiO_2$, 24% $Al_2O_3$, 11% FeO, 14% CaO<br>Or 2 types:<br>• Mare glass, mostly basalt (pyroxenes)<br>• Highland glass, mostly anorthosite | 19<br><br><br>8.5<br>11 | 2.9 | 17.5<br><br><br>8<br>10 |
| Olivine | ~45:55 forsterite (Mg-endmember: $Mg_2SiO_4$) and fayalite (Fe-endmember: $Fe_2SiO4$) | 7 | 3.5 | 7.8 |
| Ilmenite | $FeTiO_3$ | 3 | 4.74 | 4.5 |
| Troilite | FeS | | | 1 |
| Native iron | elemental iron-nickel metal with typically 5.7% Ni | | | 0.3±0.15 |

Cremers, 1974 [39] reported the Apollo $c_P$ data and also further data from lunar samples. Unfortunately, because of lacking temperature information, a single measurement on a third breccia sample (10065) by Bastin *et al.* and measurements of the Soviet Luna-16 sample reported by Vinogradov and Avduevskii *el al.*. are not really useful for us.

Morrison and Norton (1970) [40] measured the specific heat of Apollo 11 sample 10017 (type A) and 10046 (type C) at liquid helium temperatures (Table 9). The specific heat of sample 10017 increases monotonically from 19 to 28 J/kg/K in the temperature range between 2.34 K and 4.97 K. The specific heat of sample 10046, measured on the temperature range between 3.08 and 4.05 K, ranges from 10 J/kg/K to 19 J/kg/K with a maximum at 3.54 K. These values of specific heats are two orders of magnitude larger than those expected from elastic properties of these samples (Debye crystals) - they can be explained by the high fraction of amorphous constituents (glass) in the samples. The mineral composition of 10017 is rather well known, see Table 6 below.

**Table 6 approximate mineral composition of lunar sample 10017 [derived from Apollo 11 Lunar Sample Information Catalog Publication: JSC-12522, https://curator.jsc.nasa.gov/lunar/catalogs/apollo11/10017.pdf]**

| Mineral | volume fraction, normalized to 100%, in % | mass fraction (calculated), in % |
|---|---|---|
| En | 8.9 | 8.1 |
| Fs | 2.2 | 2.5 |
| Di | 35.5 | 33.2 |
| Hd | 8.9 | 9.3 |
| An | 18.4 | 14.5 |
| Ab | 4.7 | 3.5 |
| Or | 0.5 | 0.4 |
| Ilm | 20.4 | 27.9 |
| Tro | 0.4 | 0.6 |

and the resulting $c_P$ is well described, up to 5.5 K, by $c_P$(cryst) = $0.00266 \times T^3$ [J/kg/K]. The amorphous excess, $\Delta c_P = c_P - c_P(cryst.)$ is shown in Figure 4, it is indistinguishable from the measured $c_P$.

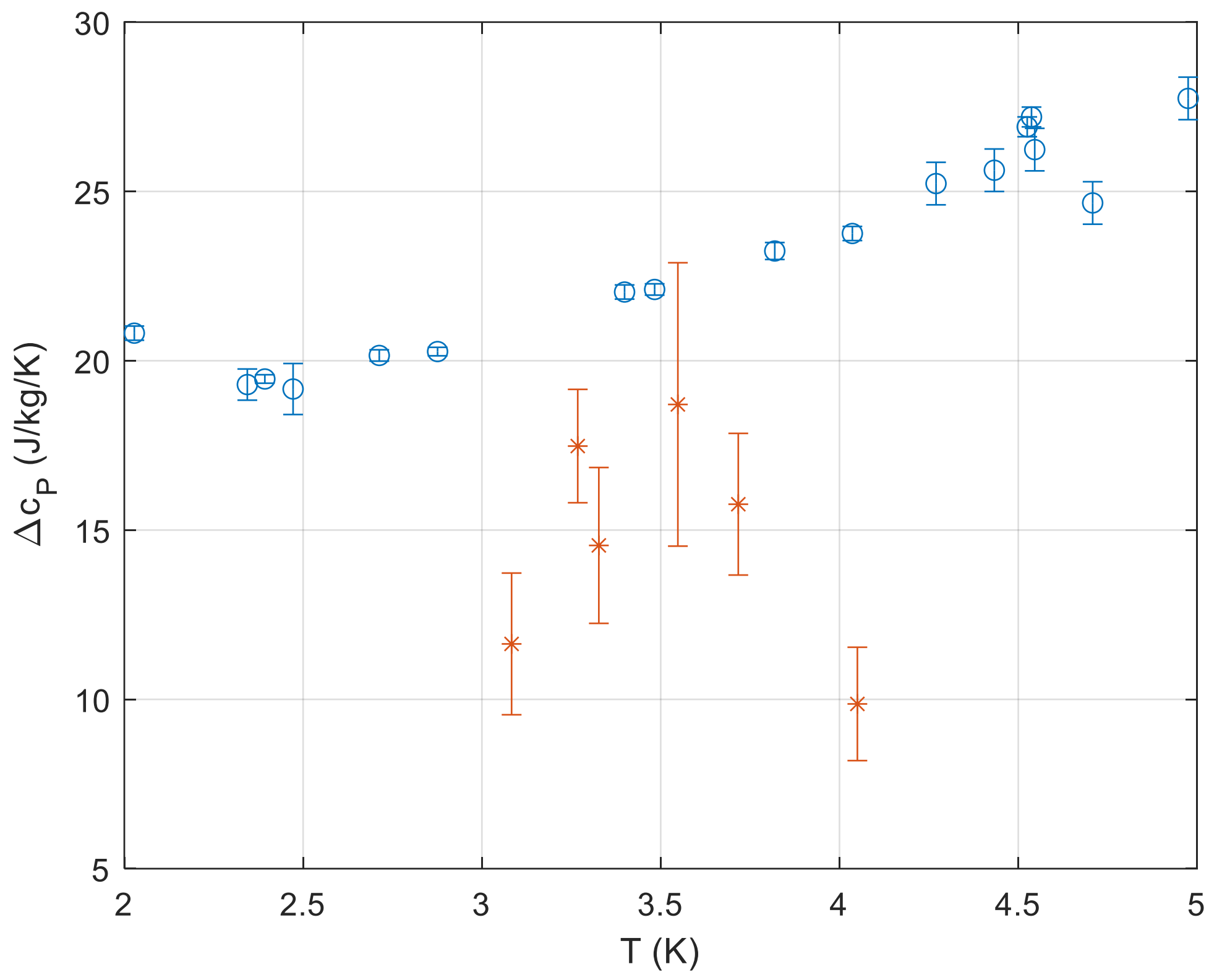

**Figure 4 Excess $c_P$ of lunar samples 10017 (blue circles) and 10046 (red stars) after Morrison & Norton (1970). 10046 is a typical breccia rock with a relatively high glass content; 10017 is a vesicular basalt 10017. The results for the rock 10017 are more extensive and more accurate because the equilibrium time after heating was much shorter for that rock (<0.5 min compared with 10 to 15 min for rock 10046), consistent with their bulk structures (10046 being more porous and cracked, its thermal conductivity 10 to 100 times smaller than that of rock 10017)**

All Apollo $c_P$ data for the temperature range ~90 K to ~340 K are listed in Table 7 and graphically shown in Figure 5.

If we subtract a rather arbitrary fit through each of the data sets individually, we get the scatter plot of relative differences `(fit-measured)/measured` in **Figure 6**, indicating experimental uncertainties of the order of max ±2%.

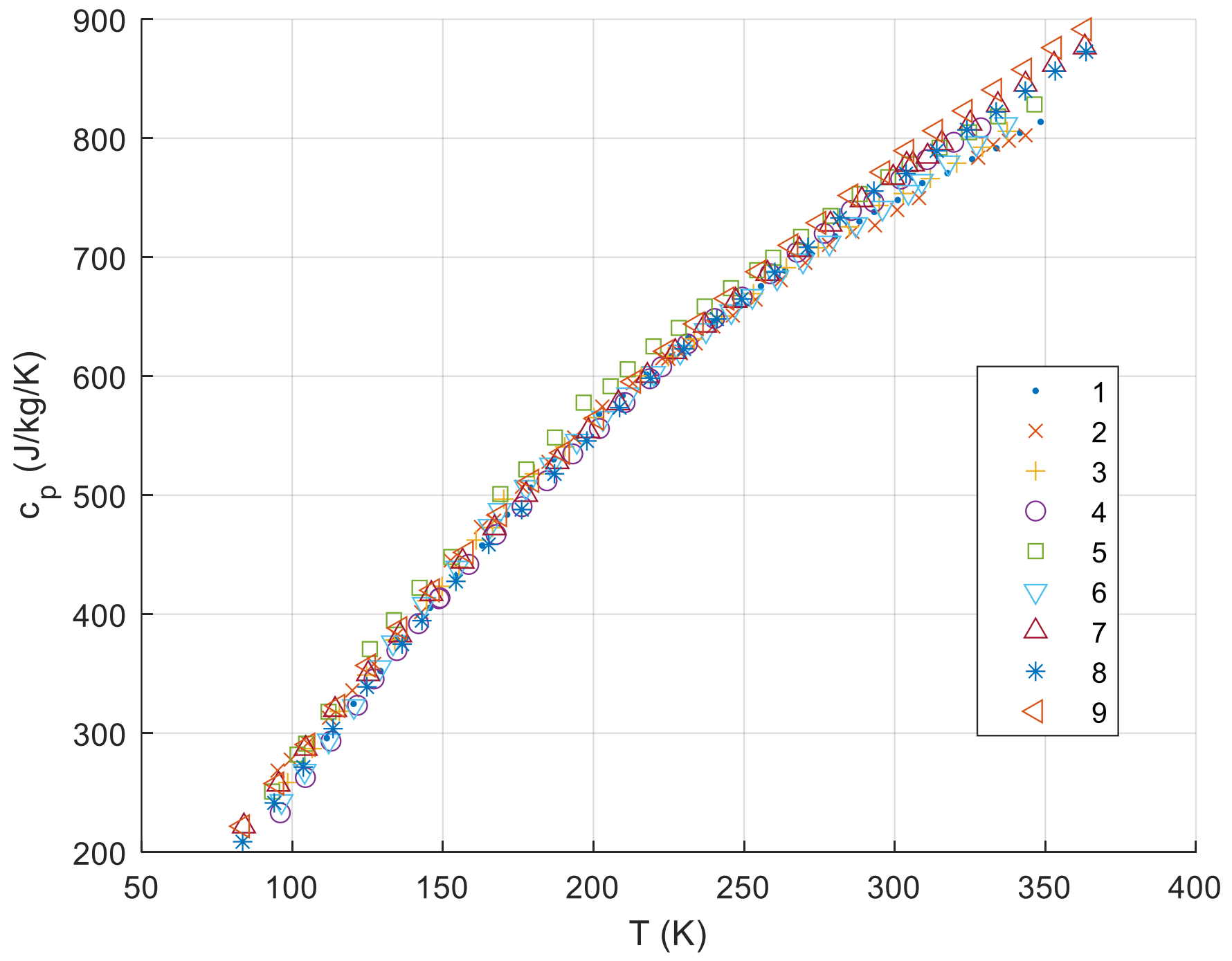

**Figure 5 Data overview: all published Apollo specific heat data. Marker indicates data set number used in this paper.**

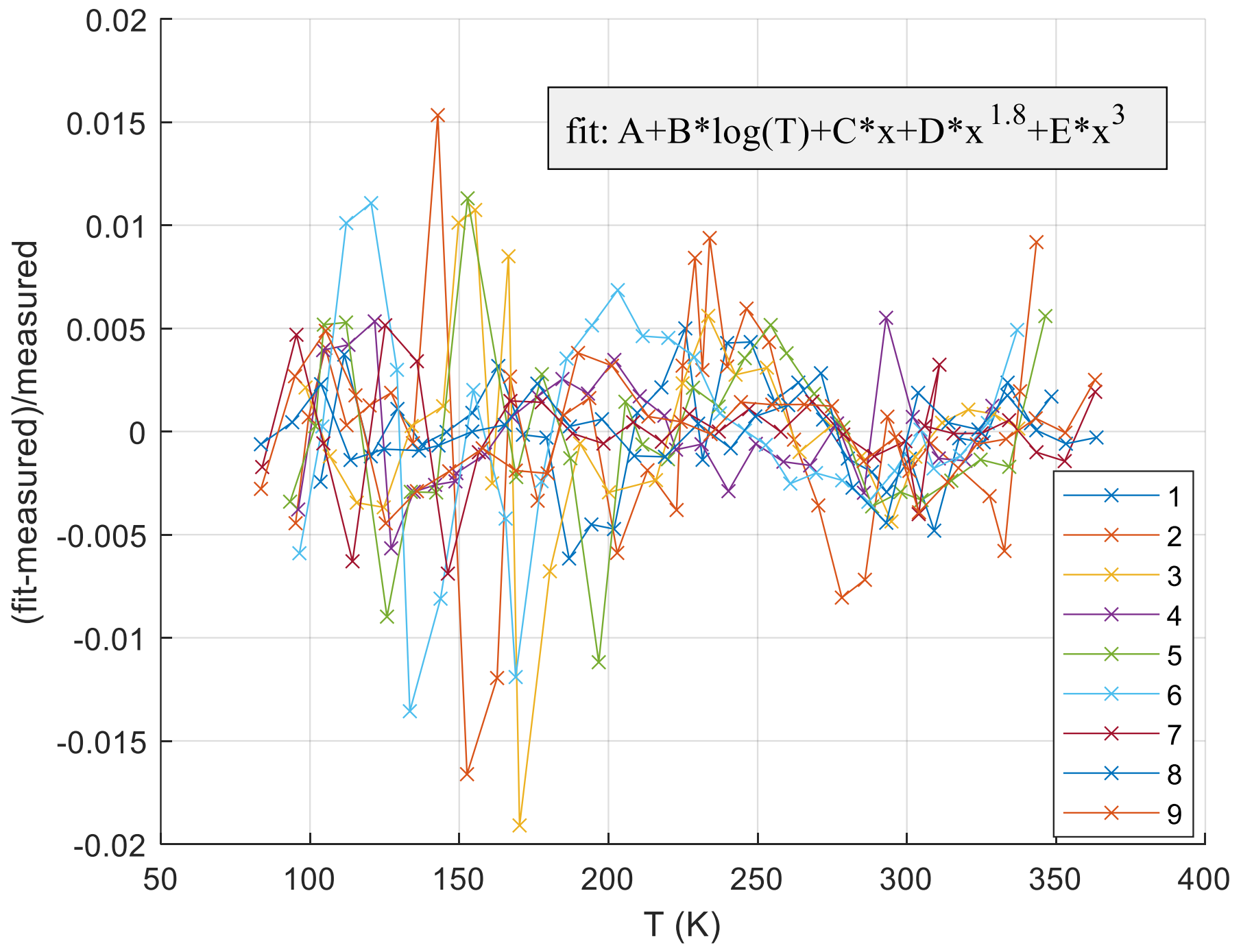

**Figure 6 scatter of lunar $c_P$ data wrt. a smooth fit (equation indicated in figure) for each set. Legend indicates data set number used in this paper. Standard deviation is ~0.5%.**

The data on specific heat of Apollo samples are scattered in several publications, some are smoothed, some are not. To facilitate convenient access, we reproduce the published 9 datasets (unsmoothed) hereunder.

**Table 7 Overview: Apollo samples specific heat datasets. n is the dataset number used in this paper. 1 cal = 4.1840 J (thermochemical calorie)**

| n | Reference | Notes |
|---|---|---|
| 1 | Robie&Hemingway, 1970 [35] | unsmoothed; *T*(K), cp (cal/g/K)<br>10057 (vesicular basalt) |
| 2 | Robie&Hemingway, 1970 [35] | unsmoothed; *T*(K), cp (cal/g/K)<br>10084 (regolith) |
| 3 | Robie&Hemingway, 1971 [36] | unsmoothed; *T*(K), cp (cal/g/K)<br>10021,41 (breccia) from Tranquillity Base. |
| 4 | Robie&Hemingway, 1971 [36] | unsmoothed; *T*(K), cp (cal/g/K)<br>12018,84 olivine dolerite from the Sea of Storms |
| 5 | Hemingway&Robie, 1973 [34] | 14163,186 (>1 mm fines) from Fra Mauro, cp in cal/g/K |
| 6 | Hemingway&Robie, 1973 [34] | 14321,153 (breccia) from Fra Mauro, cp in cal/g/K |
| 7 | Hemingway&Robie, 1973 [34] | 15301,20 (soil) from Hadley-Apennine Base, cp in cal/g/K |
| 8 | Hemingway&Robie, 1973 [34] | 15555,159 (basalt) from Hadley-Apennine Base, cp in cal/g/K |
| 9 | Hemingway&Robie, 1973 [34] | 60601,31 (soil) from Lunar Highland, cp in J/g/K |

[34-36] also reported specific heat data (Figure 30) that were smoothed using a form of least-squares orthogonal polynomials. The accuracy of the data is quoted as ±0.4 percent which of course does not include $c_P$ differences from one sample to the other due to compositional variations. The latter are seen more clearly in **Figure 7**.

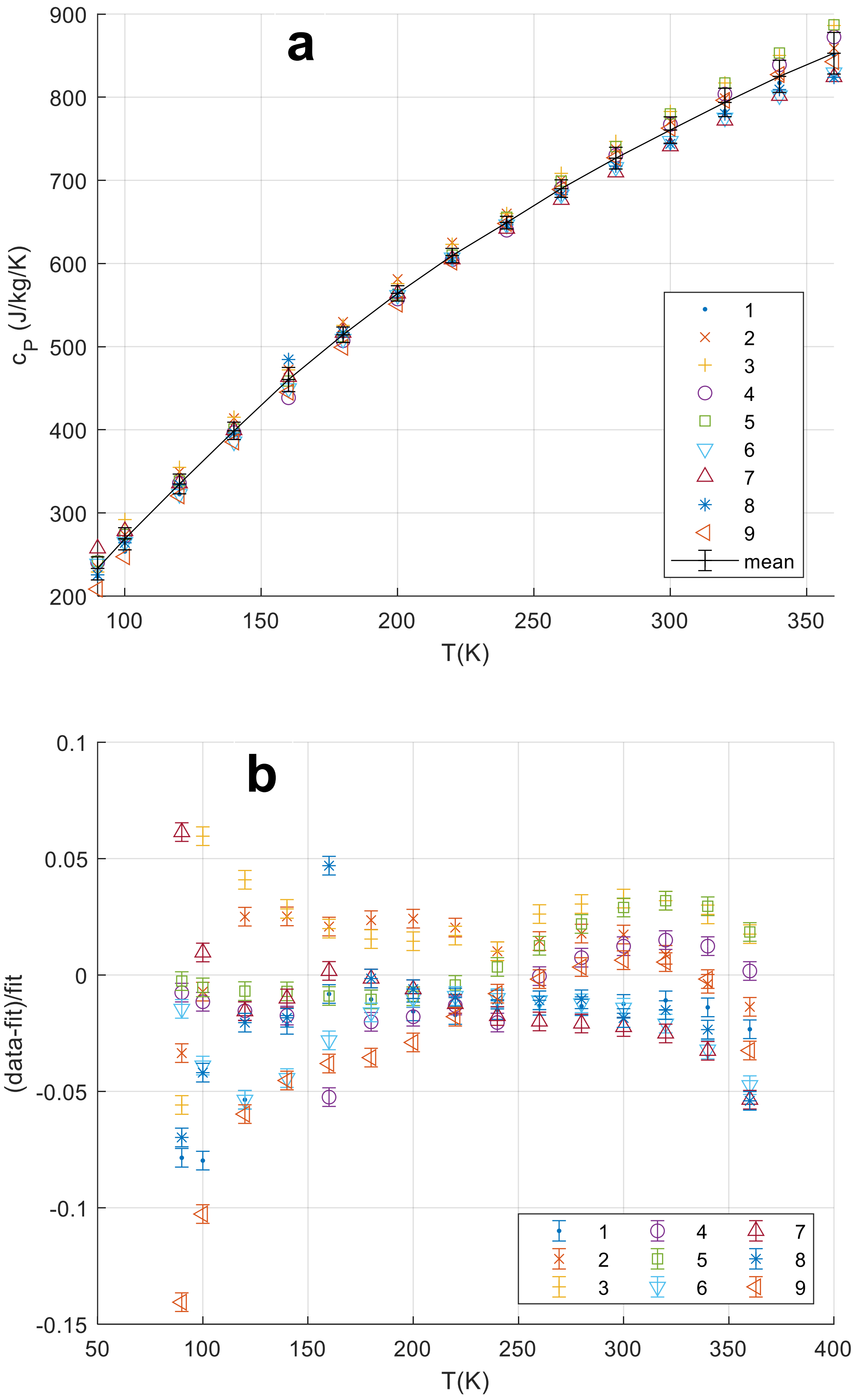

**Figure 7** Lunar $c_P$. (a) Data from 9 smoothed datasets (symbols) and mean (line) with 1σ standard deviation bars. (b) relative deviations of 9 smoothed datasets (symbols) from the polynomial fit of Hemingway&Robie [34], with 0.4% stated uncertainty of each data point errorbar. Fit (J/kg/K) = polyval([9.6552e-8 -7.3699e-5 1.5009e-2 2.127 -23.173],*T*) Note that datasets 1-3, 5-6 values at 90 and 360 K are extrapolated thus uncertain; dataset 4 values at 90, 340 and 360 K are extrapolated thus uncertain..

# 4 Data tables

## 4.1 Apollo lunar samples specific heat

Table 8 Numerical data, specific heat (J/kg/K). Reference for dataset number: see Table 7. cal converted into J.

| T (K) | cP | T (K) | cp | T (K) | cp | T (K) | cp |
|---|---|---|---|---|---|---|---|
| **dataset #1** | | **dataset #3** | | 196.75 | 577.8 | 303.96 | 777.8 |
| 103.54 | 274.4704 | 98.52 | 258.5712 | 205.69 | 591.6 | 310.91 | 784.7 |
| 111.50 | 295.8088 | 106.64 | 287.0224 | 211.38 | 605.8 | 306.01 | 778.2 |
| 120.38 | 324.6784 | 115.66 | 318.4024 | 219.94 | 625.1 | 315.65 | 795.8 |
| 129.30 | 352.2928 | 124.71 | 348.9456 | 228.32 | 640.6 | 325.09 | 812.5 |
| 137.57 | 379.4888 | 134.06 | 378.2336 | 236.92 | 658.6 | 334.25 | 828.0 |
| 145.76 | 405.4296 | 144.60 | 411.2872 | 245.66 | 674.0 | 343.42 | 845.2 |
| 154.41 | 432.2072 | 149.77 | 423.4208 | 254.37 | 689.1 | 352.89 | 861.9 |
| 163.10 | 457.7296 | 155.34 | 439.7384 | 259.68 | 699.6 | 363.10 | 876.5 |
| 171.35 | 483.6704 | 161.04 | 462.3320 | 268.92 | 717.1 | **dataset #8** | |
| 179.25 | 506.2640 | 166.52 | 472.7920 | 278.72 | 734.7 | 83.56 | 208.8 |
| 186.82 | 530.1128 | 170.23 | 496.6408 | 288.37 | 753.1 | 94.05 | 241.4 |
| 194.29 | 548.9408 | 180.38 | 517.9792 | 297.86 | 767.3 | 103.71 | 271.5 |
| 201.87 | 568.1872 | 190.50 | 540.5728 | 304.91 | 778.2 | 113.58 | 303.8 |
| 209.68 | 583.6680 | 200.17 | 565.2584 | 314.89 | 792.0 | 124.82 | 338.9 |
| 217.74 | 601.2408 | 215.92 | 600.4040 | 324.70 | 805.0 | 136.51 | 374.9 |
| 225.75 | 616.7216 | 224.86 | 616.3032 | 334.30 | 818.4 | 143.07 | 394.6 |
| 231.55 | 632.6208 | 233.44 | 631.3656 | 346.40 | 828.4 | 154.35 | 427.6 |
| 239.78 | 645.1728 | 242.6 | 650.6120 | **dataset #6** | | 165.24 | 458.6 |
| 247.68 | 659.8168 | 253.10 | 669.4400 | 96.45 | 242.7 | 176.18 | 487.9 |
| 255.59 | 675.7160 | 264.05 | 691.1968 | 104.15 | 268.2 | 187.06 | 518.0 |
| 263.63 | 688.6864 | 274.64 | 707.9328 | 112.15 | 293.7 | 197.83 | 545.6 |
| 271.93 | 703.3304 | 284.9 | 725.5056 | 120.49 | 322.6 | 208.68 | 573.6 |
| 280.23 | 717.5560 | 294.84 | 743.4968 | 129.14 | 355.2 | 218.95 | 598.3 |
| 288.28 | 730.1080 | 302.83 | 753.5384 | 133.46 | 376.1 | 230.05 | 623.0 |
| 293.21 | 738.0576 | 311.81 | 766.0904 | 143.74 | 408.4 | 240.98 | 648.1 |
| 301.01 | 748.0992 | 320.58 | 779.0608 | 154.73 | 438.9 | 249.20 | 664.8 |
| 309.16 | 762.3248 | 329.14 | 792.4496 | 165.55 | 474.0 | 260.22 | 687.4 |
| 317.53 | 770.6928 | 337.33 | 805.8384 | 168.91 | 487.4 | 271.21 | 708.4 |
| 325.72 | 782.4080 | **dataset #4** | | 177.58 | 506.7 | 281.80 | 733.0 |
| 333.76 | 791.6128 | 96.05 | 233.0488 | 185.96 | 525.5 | 293.09 | 755.6 |
| 341.59 | 804.5832 | 104.40 | 262.7552 | 194.47 | 545.6 | 303.77 | 770.3 |
| 348.44 | 813.7880 | 112.88 | 293.2984 | 203.13 | 564.8 | 313.96 | 789.5 |
| **dataset #2** | | 121.69 | 323.4232 | 211.49 | 584.5 | 323.97 | 807.1 |
| 95.17 | 268.6128 | 127.19 | 345.5984 | 220.06 | 602.5 | 333.68 | 822.2 |
| 99.56 | 277.8176 | 134.76 | 369.4472 | 228.81 | 620.5 | 343.36 | 839.7 |
| 105.16 | 291.6248 | 141.92 | 392.0408 | 237.36 | 638.5 | 353.28 | 856.5 |
| 112.25 | 312.9632 | 148.73 | 412.9608 | 245.74 | 654.4 | 363.53 | 872.8 |
| 119.98 | 335.9752 | 149.06 | 413.7976 | 252.74 | 667.3 | **dataset #9** | |
| 127.19 | 358.1504 | 158.61 | 441.8304 | 260.94 | 682.8 | 83.57 | 221.8 |
| 134.50 | 381.9992 | 167.66 | 466.9344 | 269.55 | 697.1 | 94.97 | 257.7 |
| 142.83 | 401.6640 | 176.30 | 490.3648 | 278.32 | 712.1 | 105.29 | 290.4 |
| 152.61 | 445.1776 | 184.58 | 512.1216 | 287.10 | 727.6 | 115.24 | 323.0 |
| 162.69 | 473.2104 | 193.14 | 534.7152 | 295.90 | 741.4 | 125.39 | 356.9 |
| 167.02 | 478.6496 | 201.98 | 556.0536 | 304.55 | 754.8 | 135.83 | 388.3 |
| 176.27 | 507.1008 | 210.50 | 577.8104 | 308.97 | 764.0 | 146.65 | 420.1 |
| 185.10 | 528.0208 | 218.75 | 597.8936 | 317.81 | 779.5 | 157.86 | 451.9 |
| 193.60 | 548.5224 | 222.65 | 607.9352 | 327.13 | 795.8 | 168.95 | 483.3 |
| 202.94 | 574.4632 | 231.20 | 627.1816 | 336.96 | 811.7 | 179.61 | 512.1 |
| 213.07 | 594.1280 | 240.23 | 648.5200 | **dataset #7** | | 189.82 | 535.6 |
| 222.84 | 615.0480 | 249.38 | 666.5112 | 83.98 | 221.8 | 201.17 | 564.4 |
| 224.85 | 614.6296 | 258.64 | 686.1760 | 95.46 | 256.9 | 213.39 | 595.4 |
| 229.00 | 619.2320 | 267.60 | 704.1672 | 104.47 | 287.0 | 224.23 | 620.9 |
| 231.49 | 627.1816 | 276.60 | 720.0664 | 114.25 | 319.7 | 234.21 | 643.9 |
| 233.93 | 627.6000 | 285.60 | 739.3128 | 125.20 | 349.8 | 244.40 | 665.3 |
| 239.78 | 641.8256 | 293.03 | 746.4256 | 135.82 | 382.4 | 254.90 | 687.8 |
| 246.27 | 651.0304 | 301.98 | 765.6720 | 146.20 | 417.1 | 265.60 | 710.0 |
| 253.83 | 664.4192 | 310.70 | 781.9896 | 156.58 | 444.3 | 274.91 | 728.9 |
| 262.18 | 680.7368 | 319.59 | 796.6336 | 167.16 | 472.4 | 285.53 | 751.9 |
| 270.31 | 695.3808 | 328.64 | 808.7672 | 177.64 | 500.4 | 296.10 | 771.5 |
| 278.22 | 710.4432 | **dataset #5** | | 188.03 | 528.0 | 304.06 | 789.5 |
| 285.93 | 721.3216 | 93.34 | 251.0 | 198.34 | 554.0 | 313.63 | 806.3 |
| 293.46 | 726.7608 | 101.77 | 282.0 | 208.28 | 577.4 | 323.47 | 823.0 |
| 300.84 | 739.7312 | 104.64 | 291.2 | 217.89 | 600.4 | 333.23 | 840.6 |
| 308.07 | 749.7728 | 112.07 | 318.0 | 227.13 | 620.5 | 343.17 | 857.7 |
| 327.71 | 783.6632 | 125.80 | 370.7 | 237.00 | 642.7 | 353.09 | 876.1 |
| 332.70 | 794.5416 | 133.79 | 395.0 | 247.16 | 663.6 | 363.04 | 891.6 |
| 337.93 | 797.8888 | 142.28 | 422.2 | 257.76 | 686.2 | | |
| 343.42 | 802.4912 | 152.82 | 448.1 | 268.32 | 706.3 | | |
| | | 169.10 | 501.0 | 278.68 | 727.6 | | |
| | | 177.70 | 521.7 | 289.07 | 748.1 | | |
| | | 187.22 | 548.5 | 299.50 | 766.9 | | |

### Table 9 LHe cP of two lunar samples, [40]

| $T$ (K) | $c_P$ (J/kg/K) | $\sigma(c_P)$ (J/kg/K) |
|---|---|---|
| **sample 10017** | | |
| 2.344 | 19.33 | 0.46 |
| 2.393 | 19.50 | 0.13 |
| 2.472 | 19.20 | 0.75 |
| 2.713 | 20.21 | 0.17 |
| 2.876 | 20.33 | 0.13 |
| 2.028 | 20.84 | 0.21 |
| 3.399 | 22.13 | 0.21 |
| 3.483 | 22.22 | 0.17 |
| 3.819 | 23.39 | 0.25 |
| 4.036 | 23.93 | 0.21 |
| 4.27 | 25.44 | 0.63 |
| 4.433 | 25.86 | 0.63 |
| 4.525 | 27.15 | 0.29 |
| 4.537 | 27.45 | 0.29 |
| 4.546 | 26.48 | 0.63 |
| 4.708 | 24.94 | 0.63 |
| 4.975 | 28.07 | 0.63 |
| **sample 10046** | | |
| 3.083 | 11.72 | 2.09 |
| 3.268 | 17.57 | 1.67 |
| 3.327 | 14.64 | 2.30 |
| 3.548 | 18.83 | 4.18 |
| 3.717 | 15.90 | 2.09 |
| 4.05 | 10.04 | 1.67 |

## 4.2 Analogs and lunar reference specific heat

**Table 10 Specific heat capacities of regolith simulants, calculated from $c_P(T)$ of minerals and composition, assuming mechanical mixture or ideal solid solutions. Jens.Biele@DLR.de, April 25, 2022**
**Note that the values at high temperatures may not be reproducible, since decomposition of phyllosilicates starts well below 1000K and the notorious epsomite (magnesium sulfate heptahydrate) in the CI simulants starts to loose crystal water already at T>293K. Accuracy is believed to be of order of a few %, except for T<90K where magnetic transition peaks occur; these depend sensitively on whether the minerals are only mechanically mixed (then the curves are correct) or in solid solution (then the endmember peaks tend to blur and shift to lower temperatures). Linear interpolation of the data in this file should reproduce the curves well.**

| | cp (J/kg/K) | | | | | | | |
|---|---|---|---|---|---|---|---|---|
| **T (K)** | **CM-1** | **CM-2** | **CI-1** | **CI-2** | **C2-1** | **CR-1** | **UTPS-TB** | **lunar** |
| 0 | 0 | 0 | 0 | 0 | 0 | 0 | 0 | 0 |
| 2.5 | 0.088199 | 0.0086767 | 0.024995 | 0.019043 | 0.012423 | 0.092583 | 0.006204 | 0.14838 |
| 5 | 0.70191 | 0.065641 | 0.18427 | 0.12613 | 0.084825 | 0.66188 | 0.046509 | 0.83601 |
| 7.5 | 3.8755 | 0.19593 | 0.61896 | 0.40001 | 0.33069 | 1.6614 | 0.17142 | 2.1934 |
| 10 | 8.8118 | 0.41192 | 1.4222 | 0.88048 | 0.74601 | 2.8362 | 0.40628 | 4.2409 |
| 12.5 | 13.312 | 0.73866 | 2.6483 | 1.5861 | 1.2607 | 3.9898 | 0.77091 | 6.9591 |
| 15 | 17.242 | 1.266 | 4.3969 | 2.5916 | 1.9353 | 5.201 | 1.3523 | 10.312 |
| 17.5 | 20.647 | 2.1233 | 6.794 | 3.9942 | 2.8592 | 6.5719 | 2.2939 | 14.256 |
| 20 | 23.971 | 3.4372 | 9.9283 | 5.8889 | 4.133 | 8.2388 | 3.7621 | 18.746 |
| 22.5 | 27.587 | 5.314 | 13.849 | 8.3491 | 5.8337 | 10.341 | 5.8014 | 23.737 |
| 25 | 31.649 | 7.893 | 18.612 | 11.455 | 8.0412 | 12.983 | 8.4913 | 29.185 |
| 27.5 | 36.463 | 11.225 | 24.2 | 15.216 | 10.793 | 16.258 | 11.851 | 35.049 |
| 30 | 42.044 | 15.286 | 30.566 | 19.592 | 14.084 | 20.246 | 15.852 | 41.29 |
| 32.5 | 48.545 | 20.043 | 37.655 | 24.53 | 17.896 | 25.101 | 20.465 | 47.871 |
| 35 | 55.985 | 25.529 | 45.566 | 30.011 | 22.231 | 30.843 | 25.681 | 54.758 |
| 37.5 | 64.453 | 31.775 | 54.321 | 35.989 | 27.062 | 38.297 | 31.45 | 61.92 |
| 40 | 73.447 | 37.848 | 62.349 | 42.416 | 32.359 | 39.656 | 37.721 | 69.329 |
| 42.5 | 83.592 | 43.522 | 66.375 | 49.223 | 38.071 | 45.175 | 44.416 | 76.957 |
| 45 | 94.681 | 50.615 | 74.702 | 56.363 | 44.172 | 51.564 | 51.482 | 84.78 |
| 47.5 | 106.55 | 58.065 | 83.154 | 63.829 | 50.653 | 58.471 | 58.918 | 92.775 |
| 50 | 119.57 | 65.988 | 92.173 | 71.618 | 57.512 | 65.847 | 66.734 | 100.92 |
| 52.5 | 136.7 | 74.305 | 101.52 | 79.727 | 64.849 | 74.034 | 74.926 | 109.2 |
| 55 | 157.46 | 82.913 | 110.95 | 88.114 | 72.631 | 83.128 | 83.431 | 117.59 |
| 57.5 | 176.18 | 91.763 | 120.46 | 96.712 | 80.603 | 92.281 | 92.18 | 126.08 |
| 60 | 186.26 | 100.79 | 130.06 | 105.45 | 88.476 | 100.35 | 101.1 | 134.65 |
| 62.5 | 183.85 | 109.93 | 139.68 | 114.26 | 96.051 | 106.84 | 110.1 | 143.29 |
| 65 | 175.77 | 119.21 | 149.35 | 123.18 | 103.6 | 112.86 | 119.22 | 151.99 |
| 67.5 | 169.57 | 128.72 | 159.16 | 132.31 | 111.44 | 119.25 | 128.58 | 160.73 |
| 70 | 169.61 | 138.53 | 169.16 | 141.71 | 119.77 | 126.77 | 138.23 | 169.5 |
| 80 | 198.58 | 179.92 | 210.51 | 181.21 | 156.16 | 162.14 | 178.91 | 204.75 |
| 90 | 231.96 | 223.44 | 253.06 | 222.7 | 194.92 | 199.64 | 221.62 | 239.88 |
| 100 | 266.6 | 268.22 | 296.35 | 265.49 | 235.22 | 237.98 | 265.45 | 274.51 |
| 110 | 301.42 | 313.48 | 339.95 | 308.95 | 276.38 | 276.68 | 309.58 | 308.39 |
| 120 | 335.83 | 359.04 | 383.85 | 353.01 | 318.34 | 315.17 | 353.77 | 341.35 |
| 130 | 369.18 | 402.9 | 426.16 | 395.37 | 357.7 | 351.34 | 396.32 | 373.28 |
| 140 | 401.43 | 445.24 | 467.08 | 436.23 | 395.25 | 385.87 | 437.33 | 404.09 |
| 150 | 432.52 | 486.29 | 507.01 | 476.05 | 431.82 | 419.25 | 476.98 | 433.75 |
| 160 | 462.3 | 525.91 | 545.8 | 514.66 | 467.16 | 451.2 | 515.12 | 462.25 |
| 170 | 490.74 | 563.97 | 583.34 | 551.93 | 501.06 | 481.65 | 551.64 | 489.59 |
| 180 | 517.85 | 600.42 | 619.56 | 587.78 | 533.47 | 510.5 | 586.56 | 515.79 |
| 190 | 543.65 | 635.28 | 654.48 | 622.23 | 564.44 | 537.88 | 619.88 | 540.87 |
| 200 | 568.28 | 668.57 | 688.07 | 655.25 | 593.93 | 564.19 | 651.64 | 564.87 |
| 210 | 591.71 | 700.33 | 720.39 | 686.88 | 622.04 | 588.85 | 681.9 | 587.81 |
| 220 | 614.01 | 730.6 | 751.48 | 717.16 | 648.76 | 612.3 | 710.71 | 609.75 |
| 230 | 635.3 | 759.47 | 781.39 | 746.15 | 674.2 | 634.85 | 738.16 | 630.71 |
| 240 | 655.61 | 787.03 | 810.27 | 773.99 | 698.54 | 656.26 | 764.52 | 650.75 |
| 250 | 675 | 813.29 | 838.06 | 800.6 | 721.64 | 676.79 | 789.6 | 669.9 |
| 260 | 693.51 | 838.39 | 864.94 | 826.2 | 743.74 | 696.64 | 813.5 | 688.2 |
| 270 | 711.21 | 862.3 | 890.82 | 850.66 | 764.73 | 715.6 | 836.28 | 705.7 |
| 280 | 728.09 | 885.08 | 915.88 | 874.21 | 784.76 | 733.22 | 857.96 | 722.43 |
| 290 | 744.22 | 906.8 | 940.03 | 896.66 | 803.77 | 749.78 | 878.62 | 738.43 |
| 300 | 761.24 | 927.84 | 963.81 | 918.66 | 822.65 | 766.21 | 898.62 | 753.74 |
| 310 | 776.97 | 948 | 986.92 | 939.8 | 840.64 | 781.89 | 917.73 | 768.39 |
| 320 | 792.03 | 967.15 | 1009.4 | 960.15 | 857.61 | 796.63 | 935.91 | 782.42 |
| 330 | 806.47 | 985.54 | 1031.4 | 979.93 | 873.97 | 810.77 | 953.34 | 795.86 |
| 340 | 820.29 | 1003.2 | 1053 | 999.11 | 889.67 | 824.29 | 970.03 | 808.74 |
| 350 | 833.53 | 1020.1 | 1074.2 | 1017.7 | 904.79 | 837.17 | 986.05 | 821.08 |
| 360 | 846.22 | 1036.3 | 1095 | 1035.9 | 919.33 | 849.47 | 1001.4 | 832.92 |
| 370 | 858.37 | 1052 | 1115.4 | 1053.5 | 933.32 | 861.25 | 1016.2 | 844.29 |
| 380 | 870.02 | 1067 | 1135.5 | 1070.6 | 946.79 | 872.59 | 1030.4 | 855.2 |
| 390 | 881.18 | 1081.5 | 1155.3 | 1087.3 | 959.75 | 883.46 | 1044.1 | 865.68 |
| 400 | 891.73 | 1094.9 | 1174.5 | 1103.2 | 971.99 | 893.84 | 1056.8 | 875.76 |
| 410 | 901.72 | 1107.4 | 1193.1 | 1118.4 | 983.59 | 903.73 | 1068.7 | 885.45 |
| 420 | 911.28 | 1119.4 | 1211.5 | 1133.2 | 994.73 | 913.23 | 1080.1 | 894.78 |
| 430 | 920.43 | 1130.9 | 1229.6 | 1147.5 | 1005.4 | 922.34 | 1091 | 903.76 |
| 440 | 929.2 | 1142 | 1247.5 | 1161.5 | 1015.7 | 931.14 | 1101.5 | 912.42 |
| 450 | 937.61 | 1152.7 | 1265.2 | 1175.1 | 1025.5 | 939.62 | 1111.7 | 920.77 |
| 460 | 945.67 | 1163 | 1282.7 | 1188.4 | 1035 | 947.78 | 1121.4 | 928.82 |
| 470 | 953.41 | 1172.9 | 1300 | 1201.4 | 1044.1 | 955.63 | 1130.8 | 936.59 |
| 480 | 960.83 | 1182.5 | 1317.2 | 1214.1 | 1052.8 | 963.19 | 1139.9 | 944.1 |
| 490 | 967.97 | 1191.8 | 1334.3 | 1226.5 | 1061.3 | 970.48 | 1148.7 | 951.36 |
| 500 | 974.84 | 1200.9 | 1351.3 | 1238.8 | 1069.5 | 977.49 | 1157.3 | 958.39 |
| 520 | 987.84 | 1218.1 | 1385.2 | 1263 | 1085.1 | 990.72 | 1173.6 | 971.78 |
| 540 | 999.96 | 1234.5 | 1419.2 | 1286.6 | 1100 | 1003.2 | 1189.1 | 984.36 |
| 560 | 1011.3 | 1250.2 | | | 1114.2 | 1015.4 | 1203.9 | 996.22 |
| 580 | 1022 | 1265.2 | | | 1127.9 | 1026.8 | 1218.1 | 1007.4 |
| 600 | 1032.2 | 1279.8 | | | 1141.2 | 1037.7 | 1231.8 | 1018.1 |
| 620 | 1041.8 | 1293.9 | | | 1154.3 | 1048.3 | 1245 | 1028.2 |
| 640 | 1050.9 | 1307.7 | | | 1167.1 | 1058.5 | 1257.9 | 1037.8 |
| 660 | 1059.5 | 1321.1 | | | 1179.9 | 1068.6 | 1270.5 | 1047 |
| 680 | 1067.6 | 1334.3 | | | 1192.6 | 1078.9 | 1282.7 | 1055.9 |
| 700 | 1075.4 | 1347.3 | | | 1205.4 | 1089.2 | 1294.8 | 1064.4 |
| 720 | 1082.9 | 1360.3 | | | 1218.5 | 1100.4 | 1306.8 | 1072.6 |
| 740 | 1090 | 1373.2 | | | 1231.9 | 1110.9 | 1318.7 | 1080.6 |
| 760 | 1096.9 | 1386.2 | | | 1245.9 | 1122.3 | 1330.6 | 1088.3 |
| 780 | 1103.6 | 1399.4 | | | 1260.4 | 1133.7 | 1342.6 | 1095.8 |
| 800 | 1110.1 | 1412.8 | | | 1275.8 | 1145.8 | 1354.7 | 1103.1 |
| 820 | 1116.7 | | | | 1295.8 | 1160.2 | 1368.4 | 1110.3 |
| 840 | 1123.5 | | | | 1326 | 1181.1 | 1385.7 | 1117.3 |
| 860 | 1126.7 | | | | 1278.9 | 1153.7 | 1375 | 1124.2 |
| 880 | 1131.3 | | | | 1268.2 | 1149 | 1377.4 | 1131 |
| 900 | 1136.2 | | | | 1265.9 | 1149.9 | 1382.7 | 1137.7 |
| 920 | 1141.1 | | | | 1266.2 | 1158.4 | 1388.8 | 1144.3 |
| 940 | 1146 | | | | 1268.5 | 1172.3 | 1395.6 | 1150.9 |
| 960 | 1150.8 | | | | 1271.9 | 1177.9 | 1402.7 | 1157.4 |
| 980 | 1155.6 | | | | 1275.8 | 1195.6 | 1409.9 | 1163.9 |
| 1000 | 1160.2 | | | | 1280.2 | 1176.7 | 1417.3 | 1170.3 |

# 5 Mineral composition of analogue materials

Commercial asteroid simulants from the now defunct private company Deep Space Industries, USA [https://deepspaceindustries.com/simulants/ - Deep Space Industries was acquired on January 1, 2019 by Bradford Space[3] .

An iron/nickel mixture consisting of 93% Fe, 7% Ni is used. Sub-bituminous coal is a kerogen substitute.

Note that in order to find the correct mass fractions of endmember minerals in solid solutions (like olivine, or (ortho-)pyroxenes listed here, one need to convert the given atomic fractions into mass fractions (equations (1.14),(1.16)).

## 5.1 CI simulants

Orgueil-type CI. Available are UCF/DSI-CI-**1** (better elemental fidelity, especially volatiles) and UCF/DSI-CI-**2** (better mineral fidelity, appearance and stability) with the following spec sheets (Table 11-Table 16)

Table 11 UCF/DSI-CI-1 Orgueil simulant mineralogical composition

| **Mineral** | Weight % | Notes |
|---|---|---|
| **Antigorite** | 48.0% | A serpentine mineral, $(Mg,Fe^{++})3Si_2O_5(OH)_4$ |
| **Epsomite** | 6.0% | Magnesium sulfate heptahydrate $MgSO_4 \cdot 7H_2O$ |
| **Magnetite** | 13.5% | Iron Oxide – $Fe_3O_4$ (actually present 14.5%) |
| **Attapulgite** | 5.0% | AKA palygorskite, $(Mg,Al)_2Si_4O_{10}(OH) \cdot 4(H_2O)$ This clay binds strongly, without swelling/shrinking |
| **Olivine** | 7.0% | Magnesium Iron Silicate – $(Mg_{0.9}\,Fe_{0.1})_2SiO_4$ |
| **Pyrite** | 6.5% | Iron Sulfide ($FeS_2$) substituted for troilite (FeS) |
| **Vermiculite** | 9.0% | A smectite-group clay $(Mg,Fe,Al)_3(Al,Si)_4O_{10}(OH)_2$ - $4H_2O$ |
| **Coal** | 5.0% | Sub-bituminous coal is a kerogen substitute |

Table 12 UCF/DSI-CI-2 Orgueil simulant mineralogical composition

| **Mineral** | Weight % | Notes |
|---|---|---|
| **Antigorite** | 36.5% | A serpentine mineral, $(Mg,Fe^{++})_3Si_2O_5(OH)_4$ |
| **Epsomite** | 15.0% | Magnesium sulfate heptahydrate – $MgSO_4 \cdot 7H_2O$ |

[3] https://www.bradford-space.com; https://spacenews.com/deep-space-industries-acquired-by-bradford-space/

| | | |
|---|---|---|
| **Magnetite** | 11.5% | Iron Oxide – $Fe_3O_4$ (actually present 14.5%) |
| **Attapulgite** | 9.0% | AKA palygorskite, $(Mg,Al)_2Si_4O_{10}(OH)\cdot 4(H_2O)$<br>This clay binds strongly without swelling/shrinking |
| **Olivine** | 7.0% | Magnesium Iron Silicate – $(Mg_{0.9}Fe_{0.1})_2SiO_4$ |
| **Pyrite** | 6.0% | Iron Sulfide ($FeS_2$) substituted for troilite (FeS) |
| **Vermiculite** | 5.0% | A smectite-group clay $(Mg,Fe,Al)_3(Al,Si)_4O_{10}(OH)_2$ - $4H_2O$ |
| **Siderite** | 4.0% | Iron Carbonate – $FeCO_3$ |
| **Coal** | 3.5% | Sub-bituminous coal is a kerogen substitute |
| **Gypsum** | 2.5% | Calcium Sulfate Di-hydrate – $CaSO_4\cdot 2H_2O$ |

## 5.2 CM simulants

Like the CI-type carbonaceous chondrites, the CMs are dominated by phyllosilicates. The major difference is the predominance of iron rich serpentine polymorph cronstedtite. Available are UCF/DSI-CM-1 (best appearance and physical characteristics) and UCF/DSI-CM-2 (improved mineralogy and volatiles content).

**Table 13 UCF/DSI-CM-1 Murchinson simulant mineralogical Composition**

| **Mineral** | Weight % | Notes |
|---|---|---|
| **Ferrous Silicate** | 57.0% | An iron rich silicate ($Fe_2SiO_4$) aka Fayalite or "copper slag grit" substituted for the serpentine mineral Cronstedtite, $(Fe^{++}{}_2,Fe^{+++})(Si,Fe^{+++})O_5(OH)_4$ |
| **Antigorite** | 22.0% | A serpentine mineral, $(Mg,Fe^{++})_3Si_2O_5(OH)_4$ |
| **Olivine** | 8.1% | Magnesium Iron Silicate – $(Mg_{0.9}Fe_{0.1})_2SiO_4$ |
| **Coal** | 3.5% | Sub-bituminous coal is a kerogen substitute |
| **Pyrite** | 2.5% | Iron Sulfide ($FeS_2$), substituted for troilite (FeS) |
| **Pyroxene** | 2.0% | $Mg_{0.75}Fe_{0.25}SiO_3$ |
| **Magnetite** | 1.0% | Iron Oxide $Fe_3O_4$ |
| **Dolomite** | 1.0% | $(CaMg)[CO_3]_2$ |
| **Sodium Silicate** | 2.9% | Note that 5.0% of sodium silicate pentahydrate is added, but the water is driven out by the lithification process |

**Table 14 UCF/DSI-CM-2 Murchison simulant mineralogical composition**

| **Mineral** | Weight % | Notes |
|---|---|---|
| **Antigorite** | 70.0% | A serpentine mineral, $(Mg,Fe^{++})_3Si_2O_5(OH)_4$ |
| **Magnetite** | 10.0% | Iron Oxide – $Fe_3O_4$ |
| **Olivine** | 7.5% | Magnesium Iron Silicate – $(Mg_{0.9}Fe_{0.1})_2SiO_4$ |
| **Coal** | 3.5% | Sub-bituminous coal is a kerogen substitute |
| **Pyrite** | 2.5% | Iron Sulfide ($FeS_2$), substituted for troilite (FeS) |
| **Pyroxene** | 2.0% | $Mg_{0.75}Fe_{0.25}SiO_3$ |
| **Sodium Silicate** | 3.5% | Note that 6.0% of sodium silicate pentahydrate is added, but the water is driven out by the lithification process |
| **Siderite** | 1.0% | $FeCO_3$ |

## 5.3 C2 type simulant

Tagish Lake-type C2 Carbonaceous Chondrite simulant. This material sampled a different mixture of source materials than the CIs.  It is more olivine and magnetite rich while being a little depleted in serpentine relative to CIs. Available as UCF/DSI-C2-1.

**Table 15 UCF/DSI-C2-1 Tagish Lake Simulant mineralogical composition**

| **Mineral** | Weight % | Notes |
|---|---|---|
| **Lizardite** | 30.5% | A serpentine mineral, $Mg_3Si_2O_5(OH)_4$ |
| **Olivine** | 25.0% | Magnesium Iron Silicate – $(Mg_{0.9}Fe_{0.1})_2SiO_4$ |
| **Magnetite** | 22.0% | Iron Oxide – $Fe_3O_4$ |
| **Pyrite** | 8.5% | Iron Sulfide ($FeS_2$) |
| **Coal** | 5.0% | Sub-bituminous coal is a kerogen substitute |
| **Vermiculite** | 4.0% | A smectite-group clay $(Mg,Fe,Al)_3(Al,Si)_4O_{10}(OH)_2 \cdot 4H_2O$ |
| **Attapulgite** | 4.0% | aka palygorskite, $(Mg,Al)_2Si_4O_{10}(OH)\cdot 4(H_2O)$<br>This clay binds strongly without swelling/shrinking |
| **Dolomite** | 1.0% | Calcium Magnesium Carbonate – $CaMg(CO_3)_2$ |

## 5.4 CR simulant

CRs are less “primitive” than CMs and CIs, have lower clay contents and thus lower volatile contents, more FeNi free metal, and have more mafic silicates. As such they seem to represent an intermediate group in the spectrum of volatile rich to volatile poor carbonaceous chondrites. The most pristine CR falls are also the most anomalous members of this group (and may not really belong in the group). DSI uses a collection of five low-to-moderate weathering Antarctic finds that have mineralogies roughly average for the overall group as guide for the CR recipe.

Table 16 UCF/DSI-CR-1 Simulant Mineralogical composition

| Mineral | Weight % | Notes |
|---|---|---|
| **Antigorite** | 9.0% | A serpentine mineral, $(Mg,Fe^{2+})_3Si_2O_5(OH)_4$ |
| **Pyroxene** | 31.0% | $Mg_{0.75}Fe_{0.25}SiO_3$ |
| **Magnetite** | 14.0% | Iron Oxide – $Fe_3O_4$ |
| **Iron-Nickel** | 5.0% | An iron/nickel mixture consisting of 93% Fe, 7% Ni |
| **Olivine** | 33.0% | Magnesium Iron Silicate – $(Mg_{0.9}Fe_{0.1})_2SiO_4$ |
| **Pyrite** | 4.0% | Iron Sulfide ($FeS_2$), substituted for troilite (FeS) |
| **Sodium Silicate** | 2.0% | Note that 3.5% of sodium silicate pentahydrate is added, but the water is driven out by the lithification process |
| **Coal** | 2.0% | Sub-bituminous coal is a kerogen substitute |

## 5.5 UTPS-TB Phobos simulant

This is one of the the University of Tokyo Phobos simulant, Tagish Lake Variant by Hideaki Miyamoto and Takafumi Niihara, [41].
The mineral composition is approximately (Miyamoto, priv. comm.),

```
Serpentine         62.5%
Magnetite           7.9%
Pyrite              9.4%
Olivine             7.6%
Limestone           4.6%
Dolomite            4.7%
Organic materials   3.3%
```

We took antigorite as the serpentine; an olivine with $Fo_{90}Fa_{10}$ and sub-bitumous coal as "organic material".

In Figure 8 we plot the estimated UTPS-TB specific heat together with the measured $c_P$ of a Martian shergottite (basaltic).

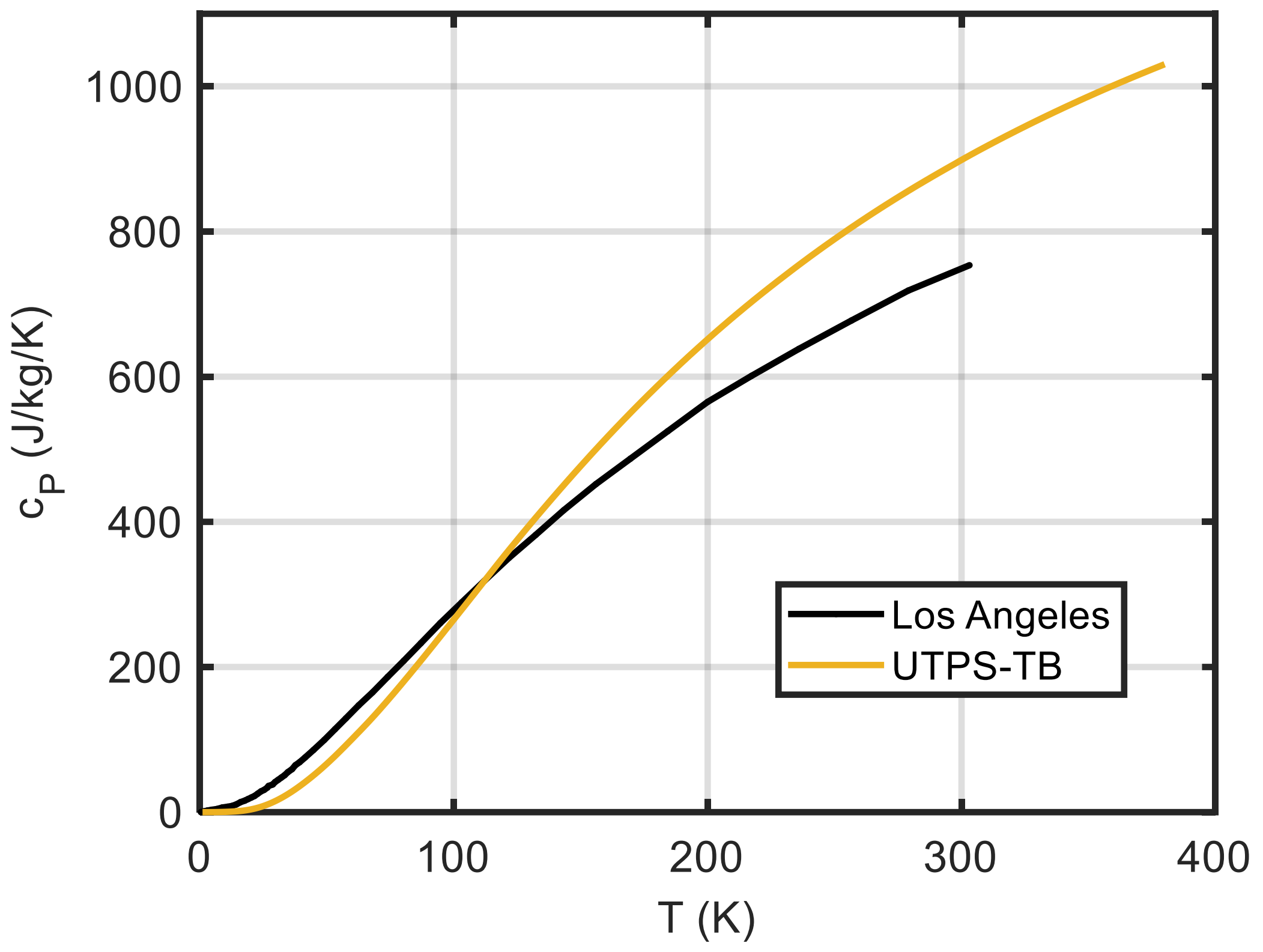

**Figure 8 Comparison of specific heat capacity of possible Phobos surface materials. Blue curve, calculated $c_P$ of Phobos surface simulant UTPS-TB, based on Tagish Lake meteorite (which most likely originated from 773 Irmintraud, a D-type asteroid). Black curve, measured $c_p$ of Martian meteorite Los Angeles after [27].**

## 6 Supporting data and figures for the bronzite mixing example

The bronzite sample was described by [42] as follows: “We measured three perfect, gem-quality, centimeter-sized orthopyroxene single crystals that are similar except for their minor element chemistries, principally the chromium and aluminum concentrations (Table1). Multiple spot analyses of each chip revealed no chemical inhomogeneities within any single chip. The crystals ranged mass from 0.19 to 1.65 g. The crystals were carefully characterized by optical, microscopy, X ray diffraction, and electron microprobe techniques. Unit cell dimensions are included in Table 1. Each crystal is of a single phase, free of exsolution lamellae, and chemically homogeneous (Figure 1). .. Electron microscopy reveals that the crystals are remarkably devoid of defects and contain no Gunier-Preston zones”.

From the chemical analysis given in [43], together with the compositions of the three chips given in table 1 of [42] we estimated (mole fractions normalized to Mg+Fe+Ca=1 **and** Si=1; see Table 17 and Table 18) the following empirical formula of the orthopyroxene, the remaining impurities and the corresponding 1-σ uncertainties of the stoichiometric coefficients:

$$Mg_{x1}Fe_{x2}Ca_{x3}SiO_3 + Al_{0.014}Mn_{0.003}Cr_{0.001}O_{0.026}$$

x1=0.843±0.028, x2=0.151±0.021, x3=0.006±0.007. Molar mass 105.23 g/mol.

The impurities correspond to about 0.008 $Al_2O_3$ + 0.003 MnO + 0.0008 $Cr_2O_3$ and are neglected.

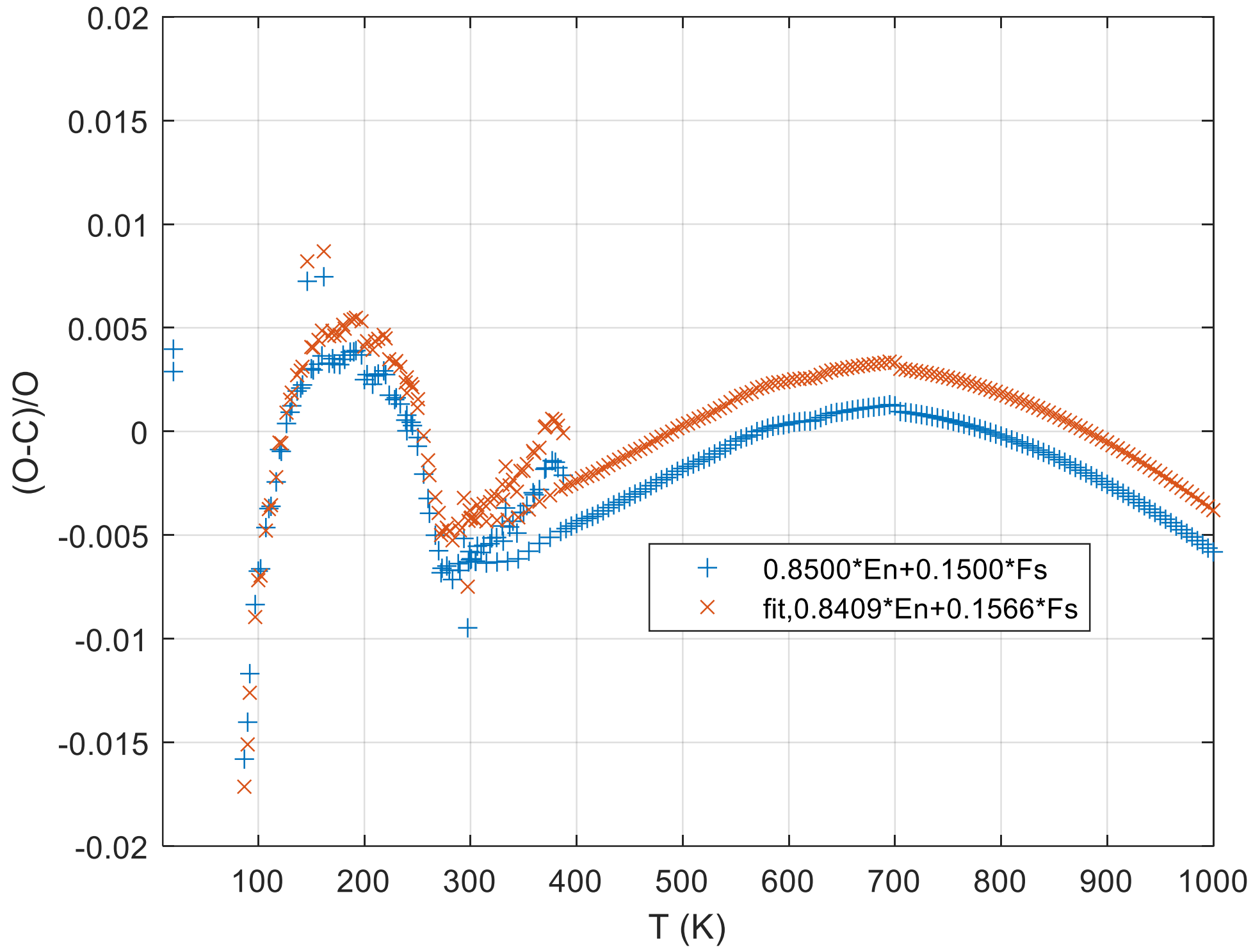

**Figure 9 Bronzite [43] relative deviations observed Cp minus calculated Cp; improved fit allowing Mg and Fe (En and Fs) mole fractions to be free. Data <100 K were excluded from the fit. Relative deviations are less than 1% outside of the Schottky peak regions (10-100 K). The fit used weighting with 1% relative uncertainty for all data points; the fit equation is Cp(T) = (x1*cp_mineral(T,'En')*0.100389+x2*cp_mineral(T,'Fs')*0.131931) where the numerical coefficients are just the molar masses of En and Fs. At 12 K, the relative deviation is 46%, at 38 K, 67%.**

**Table 17 Bronzite compositions, chemical analysis [42, 43]**

3 chips; OxideWeightPercent | CationsPerSixOxygens
Table 1 of HUEBNER ET AL. 1979

| | MF-1 | MF-2 | MF-3 |
|---|---|---|---|
| **SiO2** | 55.34 | 55.71 | 57.07 |
| **$Al_2O_3$** | 1.89 | 1.3 | 0.55 |
| **$Cr_2O_3$** | 0.51 | 0.07 | 0.02 |
| **FeO** | 10.56 | 11.57 | 8.51 |
| **MgO** | 30.61 | 30.45 | 33.18 |
| **MnO** | 0.18 | 0.2 | 0.16 |
| **CaO** | 0.23 | 0.21 | 0.19 |
| **TiO2** | 0.02 | 0.04 | 0 |
| **Na2O** | 0.01 | 0 | 0 |

Composition Krupka 1985a, oxide weight percents

| | |
|---|---|
| **SiO2** | 55.86 |
| **$Al_2O_3$** | 0.66 |
| **$Cr_2O_3$** | 0.1 |
| **FeO** | 10.11 |
| **MgO** | 31.61 |
| **MnO** | 0.19 |
| **CaO** | 0.29 |
| **Total** | 98.82 |

## Table 18 Bronzite empirical formula results, various normalisations

| Empirical formulae, normalized to .. | | | | | 3O | | 1Si | | Mg+Fe+Ca=1 | | (Mg+Fe+Ca=1) -Mg*En-Fe*Fs | | Normalize rest to 1O |
|---|---|---|---|---|---|---|---|---|---|---|---|---|---|
| | **Krupka** | **Huebner MF1** | **Huebner MF2** | **Huebner MF3** | **Mean** | **Std** | **Mean** | **Std** | **Mean** | **Std** | **Mean** | **Std** | |
| **Element** | | **mole fraction normalized to 3O** | | | | | | | | | | | |
| Si | 0.9913 | 0.9795 | 0.9874 | 0.9951 | **0.9883** | **0.0066** | **1.0000** | **0.0067** | **1.0059** | **0.0068** | 0.0059 | 0.0068 | 0.1100 |
| O | 3.0000 | 3.0000 | 3.0000 | 3.0000 | **3.0000** | 0.0000 | **3.0355** | 0.0000 | **3.0532** | 0.0000 | **0.0532** | 0.0000 | 1.0000 |
| Al | 0.0138 | 0.0394 | 0.0272 | 0.0113 | **0.0229** | **0.0130** | **0.0232** | **0.0132** | 0.0233 | 0.0133 | **0.0233** | **0.0133** | 0.4384 |
| Cr | 0.0014 | 0.0071 | 0.0010 | 0.0003 | 0.0024 | 0.0032 | 0.0025 | 0.0032 | 0.0025 | 0.0032 | 0.0025 | 0.0032 | 0.0468 |
| Fe | 0.1500 | 0.1563 | 0.1715 | 0.1241 | **0.1505** | **0.0198** | **0.1523** | **0.0200** | **0.1532** | **0.0201** | 0.0000 | 0.0201 | 0.0000 |
| Mg | 0.8362 | 0.8077 | 0.8045 | 0.8625 | **0.8277** | **0.0272** | **0.8375** | **0.0275** | **0.8424** | **0.0277** | 0.0000 | 0.0277 | 0.0000 |
| Mn | 0.0029 | 0.0027 | 0.0030 | 0.0024 | 0.0027 | 0.0003 | 0.0028 | 0.0003 | 0.0028 | 0.0003 | **0.0028** | **0.0003** | 0.0522 |
| Ca | 0.0055 | 0.0044 | 0.0040 | 0.0035 | **0.0044** | **0.0008** | **0.0044** | **0.0009** | **0.0044** | **0.0009** | 0.0000 | 0.0009 | 0.0000 |
| Ti | 0.0000 | 0.0003 | 0.0005 | 0.0000 | 0.0002 | 0.0003 | 0.0002 | 0.0003 | 0.0002 | 0.0003 | 0.0002 | 0.0003 | 0.0038 |
| Na | 0.0000 | 0.0002 | 0.0000 | 0.0000 | 0.0000 | 0.0001 | 0.0000 | 0.0001 | 0.0000 | 0.0001 | 0.0000 | 0.0001 | 0.0008 |
| sum | 5.0011 | 4.9976 | 4.9991 | 4.9991 | **4.9992** | **0.0014** | | | | | | | |
| M | 105.3657 | 105.6447 | 105.9951 | 104.4292 | **105.3587** | **0.6710** | | | | | | | |

# Table of Contents